\documentclass[aps,prmaterials,reprint,groupedaddress,longbibliography]{revtex4-2}
\usepackage[english]{babel}
\makeatletter
\@ifundefined{captionsen}{%
  \expandafter\let\csname captionsen\endcsname\captionsenglish
  \expandafter\let\csname dateen\endcsname\dateenglish
  \expandafter\let\csname extrasen\endcsname\extrasenglish
  \expandafter\let\csname noextrasen\endcsname\noextrasenglish
  \expandafter\let\csname l@en\endcsname\l@english
}{}
\makeatother
\usepackage{graphicx}%
\usepackage{siunitx}
\usepackage{textgreek}
\usepackage{xspace}
\usepackage{hyperref}
\usepackage{multirow}%
\usepackage{bm}
\usepackage{threeparttable}%
\usepackage{booktabs}%
\usepackage{miller}
\usepackage{placeins}
\usepackage{caption}%
\usepackage{subcaption}%
\usepackage{listings}%
\usepackage{adjustbox}
\usepackage{enumitem}
\usepackage[version=4]{mhchem}%
\usepackage{xr}

\DeclareSIUnit\angs{\text{Å}}
\DeclareSIUnit\atp{at.\%}

\newcommand{\myqtyrange}[3]{\qtyrange[range-units=single,range-phrase=\text{--}]{#1}{#2}{#3}}

\newcommand{\aFe}{\textalpha-Fe\xspace}

\begin{document}

\title{Fast machine learned interatomic potential for hydrogen-induced embrittlement in \aFe}

\author{Eetu Makkonen}
    \email[Contact author: ]{eetu.makkonen@helsinki.fi}
    \affiliation{Department of Physics, University of Helsinki, Finland}

\author{Alvaro Lopez-Cazalilla}
    \affiliation{Barcelona Supercomputing Center, Pla{ç}a Eusebi G{ü}ell 1-3, 08034 Barcelona, Spain}

\author{Flyura Djurabekova}
    \affiliation{Department of Physics, University of Helsinki, Finland}

\date{\today}

\begin{abstract}
In this work, we present a machine-learned interatomic potential for the \aFe-H system based on the tabulated Gaussian Approximation Potential (tabGAP) formalism. Trained on a Density Functional Theory (DFT) dataset of atomic configurations, energies, forces, and virials, the potential is designed for simulations on the mechanisms of hydrogen embrittlement (HE), the issue of H-induced acceleration of mechanical failure of metals. The proposed potential is shown to outperform the widely used classical and machine-learned interatomic potentials in fundamental properties of the \aFe-H system. We show that the tabGAP model reproduces H-point defect properties, H-dislocation interaction, H-H interaction, and elastic constants with nearly DFT-level accuracy at a computational cost that is competitive with the efficient classical Embedded Atom Method (EAM) potentials. As an application of the tabGAP model we simulate the effect of H on the mobility of the $\tfrac{1}{2}\hkl<111>$ screw dislocation, which show that the model predicts an enhancement of the mobility of the screw dislocation via trapped H atoms lowering the energy barrier of kink-pair nucleation, resulting in a decreased critical shear stress of dislocation motion at \SI{300}{K}.
\end{abstract}


\maketitle

\section{Introduction}\label{intro}

Hydrogen embrittlement (HE) is the phenomenon of H-induced increase of mechanical failure probability in various metals such as nickel, iron, and different steels. It is one of the long-standing problems in material science due to its high complexity, arising from processes spanning multiple spatial and temporal scales. Although the effect of HE is mainly observed as premature failure in structural metals at the macroscopic scale, the currently accepted HE mechanisms are triggered at the nanoscale, involving various interactions of hydrogen atoms with different microstructural imperfections. These include vacancies and their complexes, dislocations, grain boundaries, and crack tips \cite{yuHydrogenEmbrittlementConspicuous2024, robertsonHydrogenEmbrittlementUnderstood2015}. Gradually, it has become clear that HE is not driven by a single type of H-defect interaction but rather the result of multiple processes that play key roles in driving the material toward premature mechanical failure \cite{yuHydrogenEmbrittlementConspicuous2024}. Understanding these nanoscale phenomena and their emergent outcomes requires spatial and temporal resolutions that are currently inaccessible even by the most advanced experimental methods. Hence, if a high-accuracy description of interatomic interactions is available, computational modeling can become an irreplaceable technique in increasing the understanding of H effects in metals to enable the development of H-based green energy applications.

Various computational studies have addressed HE in iron and steels using the molecular dynamics (MD) method \cite{haywardInterplayHydrogenVacancies2013, songAtomicMechanismPrediction2013, fuAtomisticInvestigationHydrogen2019, xingQuantificationTemperatureThreshold2021, kumarEffectHydrogenPlasticity2023}. The known roadblock in these simulations is the large size difference between the metal and H atoms. This gives rise to considerable differences in the characteristic time and length scales of the two species, where the light H atoms might move unacceptably far during a single timestep, leading to numerical errors and invalidating the simulation result. In addition, the details on kinetic paths of the light H atoms are lost. Due to this, the satisfactory capture of H-related kinetics requires shortening the simulation timestep. To enable simulations at relevantly long timescales along with a reduced timestep, a computationally efficient interatomic potential (IAP) is needed. 

MD simulations of HE effects have typically used simple and computationally efficient IAPs based on the classical embedded-atom method (EAM) \cite{matsumotoAtomisticSimulationsHydrogen2009, fuAtomisticInvestigationHydrogen2019, xingQuantificationTemperatureThreshold2021, kumarEffectHydrogenPlasticity2023}. Although very fast, classical IAPs suffer from poor accuracy outside the interactions and material properties to which they were specifically fitted. This is partially attributable to their fixed functional forms. Thus, machine-learning IAPs (ML-IAPs) have become increasingly popular because of significant improvements in the areas of interatomic interactions where classical potentials usually fail \cite{mortazaviAtomisticModelingMechanical2023, zhangEfficiencyAccuracyTransferability2024}. The flexible functional forms of ML-IAPs have proved to be capable of reproducing the properties of multiple different types of defects present in metallic and ceramic solids in addition to the excellent description of bulk mechanical properties. This also holds for concentrated alloys with high chemical complexity \cite{byggmastarMultiscaleMachinelearningInteratomic2022,byggmastarSimpleMachinelearnedInteratomic2022, fellmanRadiationDamagePhase2026, zhangNeuralNetworkPotential2024, koutnaMachinelearningPotentialsStructurally2025}.

Although superior in the description of material properties, ML-IAPs generally suffer from the main drawback of a fairly high computational cost compared to that of classical IAPs. While the efficiency of individual IAPs varies drastically, in general, the efficiency of classical IAPs is in the order of $\SI{1e-3}{\text{milliseconds}/(\text{atom} \times \text{step})}$, while the performance of many equally or more accurate ML-IAPs is above $\SI{1e-1}{\text{milliseconds}/(\text{atom} \times \text{step})}$; a hundred times slower \cite{byggmastarMultiscaleMachinelearningInteratomic2022, zuoPerformanceCostAssessment2020}.

In this work, we propose an efficient ML-IAP based on the tabulated Gaussian approximation IAP (tabGAP) formalism \cite{bartokGaussianApproximationPotentials2010, bartokRepresentingChemicalEnvironments2013, byggmastarMultiscaleMachinelearningInteratomic2022, byggmastarSimpleMachinelearnedInteratomic2022} for the \aFe-H system. In this potential, we focus on the body‑centered cubic (bcc) phase of Fe, which forms the backbone of low‑carbon steels, the main materials of interest in HE research. We verify the properties of the H-loaded Fe crystal lattice given by our tabGAP against both DFT results and other IAPs, both classical and ML-based, that are available in the literature for this material. In addition, we set up atomistic calculations of screw dislocation mobility by kink-pair nucleation, which is the rate-limiting process for deformation in bcc metals \cite{seegerBildungUndDiffusion1962,rodneyActivationEnthalpyKinkpair2007}, combining CI-NEB calculations as a function of applied shear stress with finite-temperature MD simulations at \SI{300}{K} under a controlled strain rate.

\section{Methodology}\label{meths}

\subsection{Gaussian approximation potentials}

The developed Fe-H potential presented in this work is a tabulated (see Sec. \ref{sec:tabGAP}) version of a model produced with the Gaussian approximation potential (GAP) framework first introduced by Bartok \emph{et al.} \cite{bartokGaussianApproximationPotentials2010, bartokRepresentingChemicalEnvironments2013}. Here we define the developed Fe-H GAP as the sum of energy contributions from a selection of three low-dimensional descriptors, namely a two-body (2b), a three-body (3b), and an EAM descriptor \cite{byggmastarSimpleMachinelearnedInteratomic2022, byggmastarMultiscaleMachinelearningInteratomic2022}. In addition, the energy contribution $E_{rep.}$ from a fixed analytical Coulomb potential is added to reproduce short-range repulsive interactions between species \cite{byggmastarMachinelearningInteratomicPotential2019}. This way the non-short-range interactions are left to be machine-learned. Accordingly, the total energy expression for the GAP is defined as
\begin{align}\label{eq:gap_E}
    E_{tot} &= E_{rep.} + \sum_d \delta_d^2 \sum_n^{N_d}\sum_m^{M_d} \alpha_s K_{se} (\bm{q}_{d,n}, \bm{q}_{d,m}),
\end{align}
where the sum over $d \in \{ \text{2b}, \text{3b}, \text{EAM} \}$ denotes the sum over the descriptor selection. The contribution of each energy term is calculated by summing the number of descriptors $N_d$ of type $d$, and the number of sparse reference environments $M_d$. In addition, the descriptor contributions are scaled by the hyperparameter $\delta_d^2$ defined in Tab. \ref{tab:hyperparams}. The variables $\alpha_s$ are the regression coefficients obtained from independent Gaussian process regressions. For measuring the similarity of the target ($\bm{q}_{d,n}$) and reference ($\bm{q}_{d,m}$) descriptors, we use the squared exponential kernel ($K_{se}$):
\begin{equation}
    K_{se}(\bm{q},\bm{q}') = \exp \left(-\sum_{\xi}\left(\frac{q_{\xi}-q_{\xi}'}{2\theta}\right)^2 \right),
\end{equation}
where we set the length scale hyperparameter $\theta=1$.

The selection of descriptors yielding the energy expression in Eq. \ref{eq:gap_E} has been previously shown to provide an incredible combination of low computational cost (after tabulation; see Sec. \ref{sec:tabGAP}) and flexibility in metallic systems \cite{byggmastarSimpleMachinelearnedInteratomic2022, byggmastarMultiscaleMachinelearningInteratomic2022}.

The descriptors $\bm{q}$ have been defined as the simple two-body distance $\bm{q}_{2b}=r_{ij}$, and $\bm{q}_{ijk}$ for the 3b descriptor, which is a three-valued vector:
\begin{equation}
    \bm{q}_{ijk} =
    \begin{pmatrix}
        r_{ij} + r_{ik}\\ 
        (r_{ij} - r_{ik})^2\\
        r_{jk}
    \end{pmatrix}
    f_{cut}(r_{ij})f_{cut}(r_{jk}),
\end{equation}
where $f_{cut}(r_{ij})$ and $f_{cut}(r_{jk})$ are cutoff functions for the distances between atoms $i$ and $j$, and $j$ and $k$, respectively \cite{bartokGaussianApproximationPotentials2015}. Finally, the definition of the EAM descriptor is 
\begin{equation}    
\bm{q}_{EAM}=\rho_i=\sum_j^N \varphi_{ij}(r_{ij}), 
\end{equation}
the total pair-wise density contributed by all the atoms in the local environment of atom $i$. The pairwise density $\varphi_{ij}$ is the 3rd order polynomial function from Ref. \cite{byggmastarMultiscaleMachinelearningInteratomic2022}.

Furthermore, the repulsive energy term $E_{rep.}$ in Eq. \ref{eq:gap_E} has the Ziegler-Biersack-Littmark (ZBL) potential form
\begin{equation}\label{eq:ZBL}
    E_{rep.} = \sum^N_{i<j}f_{cut}(r_{ij})
                \phi(\frac{r_{ij}}{a})
                \frac{1}{4\pi\varepsilon_0}
                \frac{Z_i Z_j e^2}{r_{ij}},
\end{equation}
where $n$ is the sum over the atoms in the system, and $a$ is defined as
\begin{equation}
    a=\frac{0.46848}{Z^{0.23}_i + Z^{0.23}_j}.
    \label{eq:ZBLa}
\end{equation}
The difference between the Eq. \ref{eq:ZBL} and the universal ZBL potential \cite{zieglerStoppingRangeIons1985} is that the screening function $\phi(\frac{r_{ij}}{a})$ is fit to repulsive Fe-Fe, Fe-H, and H-H dimer data computed with all-electron DFT \cite{nordlundRepulsiveInteratomicPotentials1997a}. The repulsive contribution of this energy term is restrained by the smooth cutoff function $f_{cut}(r_{ij})$, which was set to $\myqtyrange{0.4}{0.6}{\angs}$ for H-H, $\myqtyrange{1.0}{1.3}{\angs}$ for Fe-H, and $\myqtyrange{1.8}{2.2}{\angs}$ for Fe-Fe interactions.

\subsubsection{Hyperparameter optimization}

The definitions of the descriptors within the GAP framework contain several hyperparameters that can be tuned to maximize the performance of the fitted IAP \cite{klawohnGaussianApproximationPotentials2023}. In Tab. \ref{tab:hyperparams} we summarize the most notable hyperparameters used for the training of the Fe-H tabGAP presented in this work. During the training procedure, we noticed that using different energy scales ($\delta_d$ in Eq. \ref{eq:gap_E}) according to the pertaining species led to a remarkable improvement in the performance of the potential. Since the scales of the energy surfaces for the Fe-Fe interactions and for the H-H interactions differ substantially, this double hyperparameter has a physical justification. In other words, we set the $\delta$ parameter for the 2b-FeFe and 2b-FeH descriptors to be $10.0$, while $\delta$ for the 2b-HH descriptor was set to $5.0$. Similarly, $\delta$ for any 3b descriptor for H-H interactions was set to $0.1$, while $\delta$ for the other triplets was set to $1.0$. Separating the $\delta$-parameters for the EAM descriptors showed no improvement; hence, we left them the same for both species.

For the regularization parameters $\sigma$, which encode the assumed accuracy of the target energies, forces, and virials \cite{klawohnGaussianApproximationPotentials2023}, we adjusted them individually for each Fe-H structure category defined in the Supplemental Tab. S.1.1. For pure Fe structures, we followed the scheme in Ref. \cite{byggmastarMultiscaleMachinelearningInteratomic2022}, with default $\sigma$ set to $\SI{1}{meV}$, $\SI{10}{meV/\angs}$, and $\SI{100}{meV/\angs^3}$ for energies, forces, and virials, respectively. For Fe-H structures, we used a stronger regularization of $2\sigma$ for structures with a higher local H concentration, such as vacancy-H structures with $>6$ H atoms near the vacancy. Structures with a single interstitial H atom used a weaker regularization of $0.5\sigma$. The free H-trimer configurations used a stronger regularization of $10\sigma$. See the Supplemental Tab. S.1.1. for more details.

\begin{table}
\centering
\caption{Hyperparameters used for the descriptors in training a GAP model for \aFe-H: cutoff radius $r_\mathrm{cut}$, width of the cutoff region $r_{\Delta\mathrm{cut}}$, energy scale $\delta$, and the number of sparse environments $M$. For clarity, shared hyperparameters between descriptors of the same type are not repeated.}
    \begin{tabular}{@{}lcccc@{}}
        \toprule
        \textbf{Descriptor} & $\boldsymbol{r_{\text{cut}}}\; (\si{\angs})$ & $\boldsymbol{r_{\Delta\text{cut}}}\; (\si{\angs})$ & $\boldsymbol{\delta}$ & $\boldsymbol{M}$ \\
        \midrule
        \multicolumn{5}{c}{\textbf{EAM}} \\
        \midrule
        Fe   & \multirow{2}{*}{4.5} & \multirow{2}{*}{1.0} & \multirow{2}{*}{1.0} & \multirow{2}{*}{30} \\
        H    &     &     &     &    \\
        \midrule
        \multicolumn{5}{c}{\textbf{Two-body}} \\
        \midrule
        FeFe & \multirow{3}{*}{4.5} & \multirow{3}{*}{1.0} & 10.0 & \multirow{3}{*}{20} \\
        FeH  &     &     & 10.0 &    \\
        HH   &     &     & 5.0  &    \\
        \midrule
        \multicolumn{5}{c}{\textbf{Three-body}} \\
        \midrule
        FeFeFe    & \multirow{6}{*}{3.7} & \multirow{6}{*}{0.6} & 1.0 & \multirow{6}{*}{800} \\
        FeFeH     &     &     & 1.0 &     \\
        HFeFe     &     &     & 1.0 &     \\
        FeHH      &     &     & 0.1 &     \\
        HFeH      &     &     & 0.1 &     \\
        HHH       &     &     & 0.1 &     \\
        \bottomrule
    \end{tabular}
\label{tab:hyperparams}
\end{table}

\subsection{Training and testing dataset}

The dataset for the present Fe-H tabGAP is extended from the previously published tabGAP for the pure \aFe \cite{byggmastarMultiscaleMachinelearningInteratomic2022}. In the present work, the extended dataset includes various configurations of isolated H, and \aFe-H structures calculated by means of Density Functional Theory (DFT), using the version 5.4.4 of the Vienna Ab initio Simulation Package (VASP) \cite{kresseEfficiencyAbinitioTotal1996, kresseEfficientIterativeSchemes1996, kresseInitioMolecularDynamics1993, kresseInitioMoleculardynamicsSimulation1994}. The electronic structure calculations were used in deriving the total energies, atomic forces, and virials for the training and testing datasets. These are the target properties of the Gaussian process regression. 

Since the accuracy of the DFT calculations also rely on the selection of several critical input parameters, for consistency of the dataset extension, in this work we used the parameters adopted by Byggmästar \emph{et al.} \cite{byggmastarMultiscaleMachinelearningInteratomic2022}, since the pure \aFe training data were inherited from their work. Maintaining consistent DFT parameters ensures coherency within the dataset, which is important when using ML methods.

The adopted DFT input parameters include the Perdew–Burke–Ernzerhof generalized gradient approximation exchange-correlation functional (PBE-GGA)\cite{perdewGeneralizedGradientApproximation1996}. Alongside this, the projector-augmented wave (PAW) pseudopotential was employed, with the 16 valence electron variant selected for Fe \cite{kresseUltrasoftPseudopotentialsProjector1999}. For H, the standard PAW pseudopotential with one valence electron was used. Additionally, aspherical contributions inside the PAW spheres were included. The plane-wave energy cutoff was set to 500 eV.

The Brillouin zone integration employed a k-point grid spacing of $\SI{0.15}{\angs^{-1}}$. The self-consistent field convergence criterion was set at $\SI{1e-6}{eV}$. All calculations were performed with collinear spin polarization, which yields the ferromagnetic state of \aFe. The Methfessel–Paxton first-order smearing method was employed with a smearing width of $\SI{0.1}{eV}$.

The Fe-H structures in this work were mainly generated with two methods: by randomly displacing atoms in ideal initial structures obtained from DFT relaxation, and by sampling frames from finite-temperature (\myqtyrange{200}{400}{K}) MD simulations using an earlier tabGAP version. The first method was used in the beginning of the tabGAP model development, and the second method was used later once the model became usable. 

The main contents of the final dataset is covered by the following list:

\begin{enumerate}[itemsep=-3pt, topsep=1pt]
    \item bcc Fe lattices with interstitial solute H atoms
    \begin{itemize}[itemsep=-3pt, topsep=-2pt]
        \item $\myqtyrange{2}{20}{\atp}$ H content
    \end{itemize}
    \item Distorted bcc Fe lattices with interstitial H
    \begin{itemize}[itemsep=-3pt, topsep=-2pt]
        \item Both uniaxial, and fully random cell distortions
        \item Strains in the range of $\myqtyrange{-10}{10}{\%}$
        \item $2$ and $\SI{10}{\atp}$ H content
    \end{itemize}
    \item Fe vacancy -H structures
    \begin{itemize}[itemsep=-3pt, topsep=-2pt]
        \item $1$ to $14$ H atoms near or inside the vacancy
    \end{itemize}
    \item Short-range H-H and H-H-H configurations
    \begin{itemize}[itemsep=-3pt, topsep=-2pt]
        \item H atoms in neighboring T-sites ($d_{H-H} \approx \myqtyrange{1.6}{3}{\angs}$)
        \item H atoms near the same T-site ($d_{H-H} \approx \myqtyrange{0.5}{1.6}{\angs}$)
    \end{itemize}
    \item Fe self-interstitial atoms with one or three nearby H atoms
    \item H atoms in the O-site
    \item H in the T-T migration path saddle point 
    \item Free H-dimer and H-trimer configurations
\end{enumerate}

The content of the structure dataset reflects the main focus of the presented potential: the accurate capture of free H interaction, and both H-Fe and H-H interaction in bulk \aFe. Due to the stark difference in the chemical behavior of H in these two environments, this is not an easy fitting problem for low-dimensional IAPs. It is known from DFT calculations that when H is placed inside the \aFe metal lattice, it mainly loses its ability to form covalent bonds with other H atoms, and instead adopts configurations with separations $d_{H-H}\SI{>1.9}{\angs}$ \cite{haywardInterplayHydrogenVacancies2013}, much larger than the free \ce{H2} bond length of $\SI{\sim0.74}{\angs}$. Special care was taken to ensure that the dataset covered this difference in the H-H interaction by introducing \aFe-H lattices with $d_{H-H}$ separations down to $\SI{0.5}{\angs}$.

In total, the final constructed Fe-H dataset contained $2686$ structures, with $1609$ being additions of this work. From these, $\SI{90}{\%}$ were used for training the Fe-H tabGAP. A testing set was constructed by choosing a random subset of $\SI{10}{\%}$ from each subcategory except free dimers/trimers (see Supplementary Tab. S1.1). Consequently, the root-mean-square errors (RMSEs) reported in this work are calculated excluding the free dimer and trimer structures, and liquid Fe structures from Ref. \cite{byggmastarMultiscaleMachinelearningInteratomic2022}.

\subsection{The tabGAP formalism}\label{sec:tabGAP}
To reduce the high computational cost of the developed GAP model \cite{zuoPerformanceCostAssessment2020, byggmastarMachinelearningInteratomicPotential2019}, the tabGAP method introduced by Byggmästar \emph{et al.} \cite{byggmastarModelingRefractoryHighentropy2021} was used, in which the two-body and three-body energy contributions from the selected low-dimensional descriptors are tabulated onto their respective one-dimensional and three-dimensional grids. After tabulation, energies can be efficiently and accurately evaluated with spline interpolation between the grid points. Given a sufficiently dense grid, the interpolation error becomes inconsequential, while the computational cost is reduced by approximately two orders of magnitude. This makes the tabGAP model competitive with the cost-efficient classical EAM potentials while maintaining a nearly quantum-mechanical level of accuracy \cite{byggmastarModelingRefractoryHighentropy2021}. For the Fe-H tabGAP, the interpolation grids of $5000$ and $120\times120\times120$ were adopted for the 2b and 3b descriptors, respectively.
For other previously published tabGAP models, see Refs. \cite{byggmastarMultiscaleMachinelearningInteratomic2022, luoInteratomicForceFields2024,zhaoComplexGa2O3Polymorphs2023,byggmastarSimpleMachinelearnedInteratomic2022,fellmanFastAccurateMachinelearned2025, fellmanRadiationDamagePhase2026}.
\subsection{Molecular statics and dynamics}\label{sec:MD-methods}

For all molecular static (MS) and molecular dynamics (MD) simulations, we used the open-source software LAMMPS, version 29 Aug 2024 \cite{thompsonLAMMPSFlexibleSimulation2022}. In MS calculations, the energy and pressure of the structures were minimized with the Polak-Ribiere conjugate gradient (CG) algorithm \cite{polakNoteConvergenceMethodes1969} until the relative energy change or the residual force norm reached $10^{-14}$. The H migration pathways in bulk \aFe were calculated with the climbing image nudged elastic band method (CI-NEB) with $13$ images and a spring constant of $\SI{1.0}{eV/\angs^{2}}$ for the nudging forces \cite{henkelmanClimbingImageNudged2000}. All MD simulations were started from CG-relaxed structures and run with an integration timestep of $\SI{0.5}{fs}$ for accurately capturing H kinetics. Here we note that at the MD-level our treatment of H is fully classical, and the presented Fe-H tabGAP in this work doesn't account for any nuclear quantum effects (NQEs) such as zero-point motion or tunneling related to H. For example, it's known that accounting for zero-point motion is required to get quantitative agreement between experimental and computational results of H diffusion on \aFe \cite{kimizukaEffectTemperatureFast2011}. Methods such as the path-integral MD \cite{kimizukaEffectTemperatureFast2011} or the quantum thermal bath \cite{brieucQuantumThermalBath2016} could be used as treatments for these effects within MD simulations, but they were left for future work. We expect such corrections to have mostly quantitative and not qualitative effects on the findings presented here.

For information on the LAMMPS-implementation of the tabGAP formalism, see Ref. \cite{JesperByggmastarTabgap2025}.

\subsubsection{Screw dislocation simulations}\label{sec:screw-methods}

Calculations for $1/2\hkl<111>\hkl{110}$ screw dislocations were performed using $30 \times 40 \times n_b$ simulation cells with \hkl[121], \hkl[-101], and \hkl[1-11] aligned with the $x$, $y$, and $z$ axes, respectively. The dislocation line was parallel to the $z$ direction, with the variable $n_b$ defining the length of the dislocation in the number of Burgers vectors, which was varied according to the simulation condition. The screw dislocation was inserted at the lateral center ($(x,y)$-plane) of the simulation box using Atomsk's isotropic elasticity solver \cite{hirelAtomskToolManipulating2015}. For calculations involving the movement of the dislocation, the periodicity of the cell was recovered in $x$ (glide-direction) by adding a tilt-factor of $b/2$ to the $xz$-component of the box. The resulting cell is non-periodic in $y$, with free surfaces being introduced in that direction. In static calculations of the core structure, atomic slabs with a thickness of $\SI{5}{\angs}$ at the non-periodic $x$ and $y$ surfaces were held fixed. We checked that the calculation results were converged with respect to the $x$ and $y$ dimensions.

To simulate the nucleation barrier of the kink-pair screw dislocation with the CI-NEB method, two $n_b=40$ cells with screw dislocations in neighboring Peierls valleys in the \hkl(-101) glide plane were constructed and relaxed under applied shear stress. The stress was applied to the cell according to the scheme specified in Ref. \cite{rodneyActivationEnthalpyKinkpair2007}, i.e. the external shear forces were added to two groups of atoms on the free top and bottom surfaces in $y$-direction.

The initial CI-NEB path was constructed with a spatially progressing interpolation scheme in which the displacement field between the endpoint structures was introduced smoothly with a $\tanh$-based weight function centered at the midpoint of the dislocation line, and then progressively extended in the $\pm z$ directions. This gives an initial reaction path close to the nucleation of the kink-pair and the migration pathway (see Supplemental Fig. S.5.6). It also allows us to predefine the kink-pair nucleation site at the center of the dislocation line. We attempted to use ordinary linear interpolation for the initial path, but this led to the dislocation migrating as a straight line when no H was introduced to break the symmetry of the lattice. 

In the final barrier calculations we used $25$ images. The applied shear stress $\tau_{yz}$ was varied between \SI{0.0}{GPa} and \SI{1.0}{GPa}. In the calculations of the barriers in the presence of H atoms, one H atom was introduced into a E1 site (see Fig. \ref{fig:H_Eb_screw_pos}) adjacent to the core in the glide direction, corresponding to line density of $\sim\SI{0.1}{nm^{-1}}$ for the $n_b=40$-long dislocation. We looked at the effect that the position of the H atom relative to the kink-pair nucleation site had on the barrier by introducing the H atom at three different positions along the dislocation line (see inset images in Fig. \ref{fig:screw_kink_H_effect}). 

The kink-pair nucleation enthalpy was calculated from each CI-NEB energy profile as the maximum energy along the path relative to the initial image,
\begin{equation}
    \Delta H_{\mathrm{kp}}(\tau) = \max_i \left[ E_i(\tau) - E_0(\tau) \right],
    \label{eq:neb_barrier}
\end{equation}
where $E_i(\tau)$ is the energy of CI-NEB image $i$ under applied shear stress $\tau$, and $E_0(\tau)$ is the energy of the initial image.

Finite-temperature screw-dislocation simulations were carried out at \SI{300}{K} for a $n_b=40$ screw dislocation as well with the strain rate controlled scheme from Ref. \cite{rodneyActivationEnthalpyKinkpair2007}. Ten independent runs for pure Fe and H-loaded conditions were sampled with different initial velocity seeds and H atom positions. Four H atoms ($\sim\SI{0.4}{nm^{-1}}$) were randomly introduced within $\SI{2}{\angs}$ of the dislocation core. Each run started with a $\SI{10}{ps}$ long NPT ensemble equilibration, after which a strain rate of $\dot{\gamma}=\SI{2e-5}{ps^{-1}}$ was imposed by assigning an initial velocity $\pm v_z=\pm \dot{\gamma} \frac{l_y}{2}$ to the y-top and y-bottom atom groups, and conserving the average z-component of the force of these groups. The simulations were continued up to a total shear strain of $\gamma=0.02$. The resolved shear stress was obtained from the reaction forces acting on the driven boundary atom groups, and the first dislocation slip event was identified from the dislocation position data extracted with the Dislocation Extraction Algorithm (DXA) implemented in version 3.14.1 of the program OVITO \cite{stukowskiVisualizationAnalysisAtomistic2009, stukowskiExtractingDislocationsNondislocation2010, stukowskiAutomatedIdentificationIndexing2012}. We applied DXA to the Fe sublattice only, excluding the interstitial H atoms from the DXA mesh construction.

\section{Results}

\subsection{Training and testing errors}
Tab. \ref{tab:train_test_errs} reports the energy and force RMSEs for the Fe-H GAP and tabGAP potentials calculated for the training and testing sets. In general, both the energy and force errors are low, only a few $\si{meV/atom}$ and $\sim \SI{.1}{\eV/\angs}$, respectively. The H-specific force RMSEs are approximately double that of the total. A large fraction of this error originates from structures with high local concentrations of H atoms, for example, structures with a Fe vacancy that contain $14$ H atoms. If such structures are ignored in the error calculation, the force RMSE drops to $\sim \SI{0.11}{\eV/\angs}$ for both training and testing sets with the tabulated potential. However, having the dataset encompass the high local H concentration structures in the training of the Fe-H tabGAP was important for improving the performance of the potential in said conditions, such as in the maximum H capacity of the Fe vacancy.

\begin{table}
\centering
\caption{Root-mean-square errors (RMSE) for Fe–H GAP and tabGAP models on the training and testing sets. Energy errors ($E_{\text{total}}$, $E_{\text{FeH}}$) are reported in $\si{meV/atom}$ and force errors ($F_{\text{total}}$, $F_{H}$) in $\si{eV/\angs}$. $E_{\text{total}}$ and $F_{\text{total}}$ are computed over all structures, $E_{\text{FeH}}$ is computed over structures that contain both Fe and H, and $F_H$ is computed over all H atom forces.}
    \begin{threeparttable}
        \begin{tabular}{@{}ccccc@{}}
            \toprule
             & \multicolumn{2}{c}{$\si{meV/atom}$} & \multicolumn{2}{c}{$\si{eV/\angs}$} \\
             \cmidrule{2-3} \cmidrule{4-5}
             & $E_{\text{total}}$ & $E_{\text{FeH}}$ & $F_{\text{total}}$ & $F_{\text{H}}$ \\
            \midrule
            \multicolumn{5}{c}{\textbf{Training}} \\
            \midrule
              GAP    & 2.20  & 1.18  & 0.073 & 0.135 \\
              tabGAP & 2.23  & 1.29  & 0.077 & 0.154 \\
            \midrule
            \multicolumn{5}{c}{\textbf{Testing}} \\
            \midrule
              GAP    & 1.87  & 1.76  & 0.069 & 0.150 \\
              tabGAP & 1.88  & 1.75  & 0.069 & 0.157 \\
            \bottomrule
        \end{tabular}
    \end{threeparttable}
\label{tab:train_test_errs}
\end{table}

\subsection{Computational efficiency}
We evaluated the computational efficiency of the different IAPs by performing a $10000$ time step NVT MD simulation for a ($6\times6\times6$) \aFe supercell with three different interstitial H concentrations: $\SI{0.25}{\atp}$, $\SI{1}{\atp}$, and $\SI{10}{\atp}$. The simulations were run on a single core of a Xeon Gold 6230 CPU.

The efficiency of the potentials is compared in terms of the computational time required to perform the simulation per atom and per time step. The comparison is presented in Tab. \ref{tab:comp_eff} in units of milliseconds/(atom $\times$ step). The lower this value, the more efficient is the used IAP. The presented Fe-H tabGAP appeared to be somewhat less efficient than the classical EAM IAPs (less than one order of magnitude difference), while its efficiency is two orders of magnitude higher than that of Meng\_NNIP, which is another ML-IAP available in the literature. Computational efficiency is a complex function of many parameters, but it is roughly proportional to the average number of neighboring atoms per an atom. This is reflected in the table as a decrease in efficiency for all potentials when the concentration of H increases in the \aFe lattice. The differences in scaling between the IAPs as a function of the H concentration are roughly explained by differences in interaction cutoff distances, which resulted in slightly worse scaling of the Fe-H tabGAP as compared to the other IAPs.

\begin{table*}
\centering
\caption{The computational efficiency (milliseconds/(atom $\times$ step)) of an MD simulations of a ($6\times6\times6$) \aFe structure at three different H concentrations. All simulations were run with a single CPU core.}
    \begin{threeparttable}
        \begin{tabular}{@{}cccccc@{}}
            \toprule
            H concentration ($\si{\atp}$) & tabGAP\tnote{a} & Ram\_EAM\tnote{b} & Wen\_EAM\tnote{c} & Kumar\_EAM\tnote{d} & Meng\_NNIP\tnote{e} \\
            \midrule
              $0.25$ & 0.0096 & 0.0018 & 0.0017 & 0.0022 & 2.13 \\
              $1.0$  & 0.0105 & 0.0019 & 0.0019 & 0.0024 & 2.09 \\
              $10.0$ & 0.0172 & 0.0023 & 0.0023 & 0.0032 & 2.41 \\
            \midrule
        \end{tabular}
    
    \begin{tablenotes}[para, flushleft]
        \item[a] This work
        \item[b]  \cite{ramasubramaniamInteratomicPotentialsHydrogen2009} 
        \item[c]  \cite{wenNewInteratomicPotential2021}
        \item[d]  \cite{kumarEffectHydrogenPlasticity2023}
        \item[e]  \cite{mengGeneralpurposeNeuralNetwork2021}
    \end{tablenotes}
    \end{threeparttable}
\label{tab:comp_eff}
\end{table*}

\subsection{Potential validation}\label{sec:validation}

\subsubsection{H as a point defect in \aFe}

To validate the performance of the proposed potential, in Tab. \ref{tab:H_energetics} we compare the most relevant characteristic energies describing H in the \aFe lattice: solution energies for H in tetrahedral (T) and octahedral (O) interstitial sites ($E_{s,T}^{H}$ and $E_{s,O}^{H}$), the solution energy for H in a vacancy ($E_{s,V}^{H}$), and migration energy barriers from a direct transition between two T-sites ($E_{m,T-T}^{H}$) or through an O-site between two T-sites ($E_{m,T-O-T}^{H}$). For the properties related to the more stable T-site, we also report the energy values modulated by hydrostatic compressive/tensile strains ($\varepsilon=\SI{\pm2}{\%}$).

We calculate the solution energies as
\begin{equation}
    E_{s,X}^{H}=E(X_H) - E(X) - \frac{1}{2}E(H_2),
\end{equation}\label{eq:solution_energy}
where $E(X_H)$ is the energy of a structure with H in the site $X$, $E(X)$ is the energy of the corresponding H-free structure, and $E(H_2)$ is the energy of an isolated \ce{H2} molecule. A positive value of $E_{s,X}^{H}$ indicates the endothermic solution of H, and a negative value indicates the exothermic solution of H.

The T-site is known to be more stable than the O-site in \aFe for a H atom, with solution energies ranging between $\myqtyrange{0.20}{0.34}{eV}$ according to DFT estimates from the literature \cite{tateyamaStabilityClusterizationHydrogenvacancy2003, ramasubramaniamInteratomicPotentialsHydrogen2009, haywardInterplayHydrogenVacancies2013}. All IAPs in Tab. \ref{tab:H_energetics} capture this property sufficiently well. Because zero-point energy (ZPE), which has been shown to increase the solution energy of H in the T-site \cite{haywardInterplayHydrogenVacancies2013}, is not taken into account in either of the listed ML-IAPs, the present tabGAP and Meng\_NNIP \cite{mengGeneralpurposeNeuralNetwork2021} give somewhat lower $E_{s,T}^{H}$-values than the ZPE-corrected EAM potentials.

The stability of the T-site persists also at moderate hydrostatic strains in the bcc lattice. In our calculations we see that the solution energies at tensile and compressive strains of $\SI{2}{\%}$ are consistent with the non-ZPE corrected DFT estimates \cite{ramasubramaniamInteratomicPotentialsHydrogen2009, mengGeneralpurposeNeuralNetwork2021}. All potentials show $E_{s,T}^{H}$ increasing with the density of the lattice, indicating that H would segregate to less dense regions of \aFe.

\begin{figure}
\centering
    \subcaptionsetup[figure]{margin={0.12\textwidth, 0pt}}
    \begin{subfigure}{.45\textwidth}
        \includegraphics[width=\linewidth]{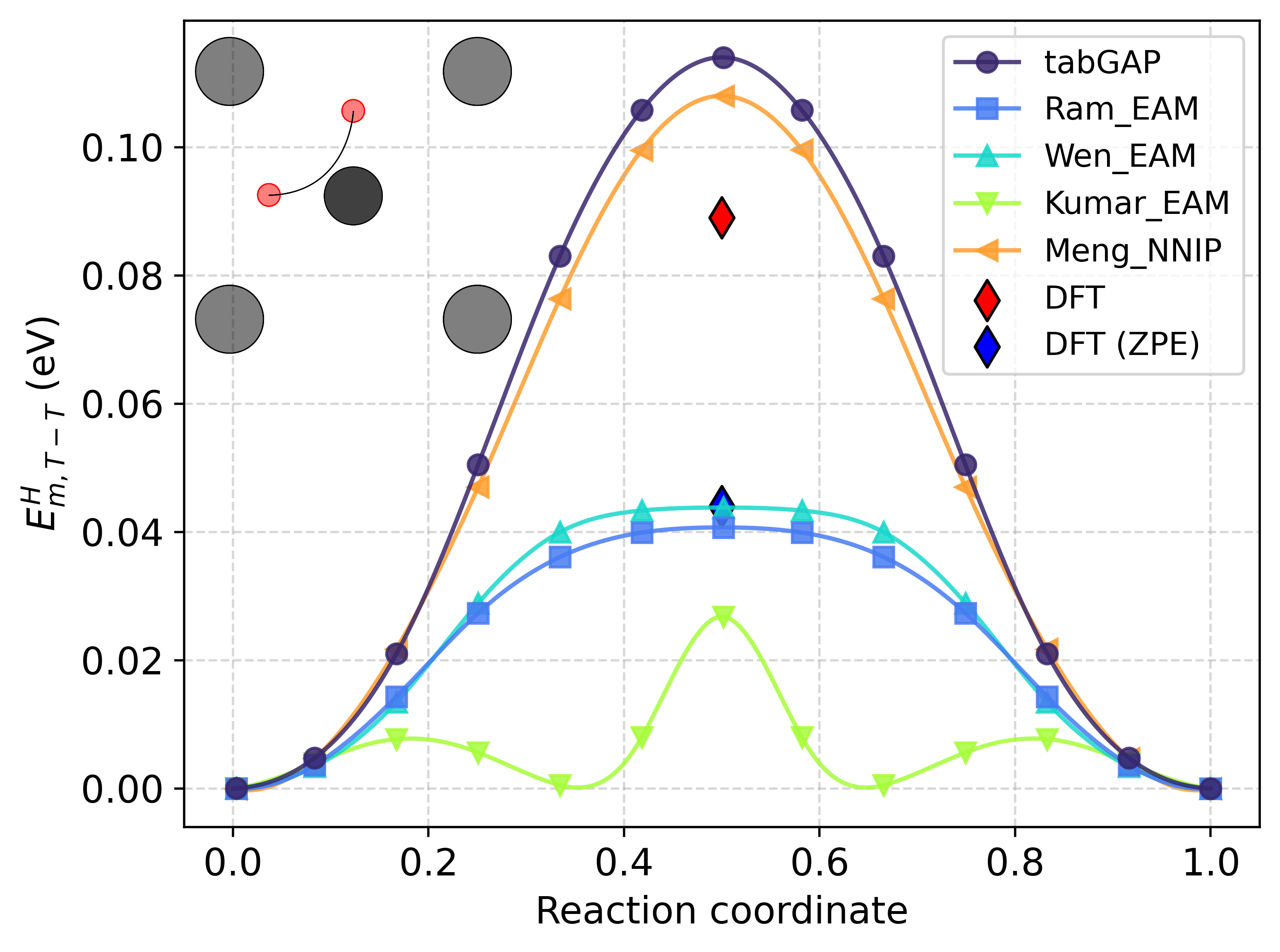}
        \caption{}
        \label{fig:TT_paths}
    \end{subfigure}
    \begin{subfigure}{.45\textwidth}
        \includegraphics[width=\linewidth]{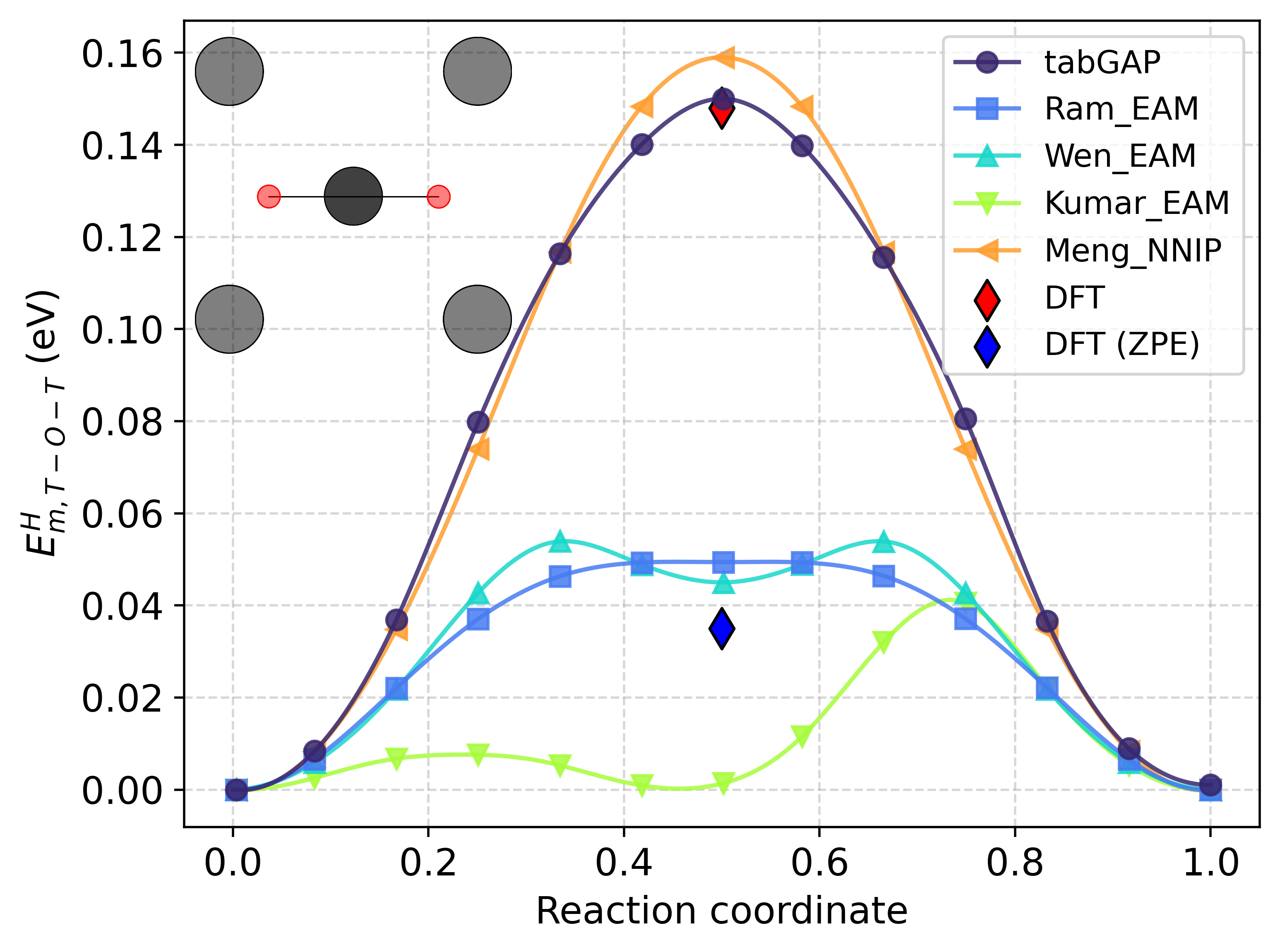}
        \caption{}
        \label{fig:TOT_paths}
    \end{subfigure}
\caption{The migration energy barriers of the T-T (\subref{fig:TT_paths}) and T-O-T transitions (\subref{fig:TOT_paths}) for a H atom in bulk \aFe, with sketches provided for both. The sketches show a \hkl[100]-view of the bcc unit cell. The H atoms are colored red, and Fe atoms are gray. DFT data is taken from Ref. \cite{haywardInterplayHydrogenVacancies2013}.}
\label{fig:mig_paths}
\end{figure}

Given the geometry of interstitial sites in the bcc lattice, two diffusion paths connect neighboring T-sites: a direct T-T hop (Fig. \ref{fig:TT_paths}) and a T-O-T hop via the intervening O-site (Fig. \ref{fig:TOT_paths}). The Fe-H tabGAP reproduces both barriers well (Tab. \ref{tab:H_energetics}); in particular, $E_{m,T-O-T}^{H}$ closely matches non-ZPE-corrected DFT \cite{haywardInterplayHydrogenVacancies2013}, while the T-T barrier is $\sim\SI{0.02}{eV}$ higher than commonly cited values but remains within the experimental range $\myqtyrange{0.035}{0.142}{eV}$ \cite{jiangDiffusionInterstitialHydrogen2004}. According to DFT calculations \cite{jiangDiffusionInterstitialHydrogen2004, haywardInterplayHydrogenVacancies2013}, both the T-T and T-O-T paths contain a single saddle point, which is reproduced by the Fe-H tabGAP. In addition, the T-T path is not straight but has a slight curvature toward the neighboring O-site. DFT calculations have predicted the distance between the saddle point of this T-T transition and the O-site to be $\SI{0.407}{\angs}$ \cite{jiangDiffusionInterstitialHydrogen2004}, and the Fe-H tabGAP predicts a very similar distance of $\SI{0.372}{\angs}$.

The H migration profiles for the other IAPs are also reported in Fig. \ref{fig:mig_paths}. The classical EAM IAPs, which are used in this study for comparison, are parameterized to include the ZPE-correction. As has been previously shown, the ZPE correction reduces the migration energy barrier for H diffusion in \aFe \cite{jiangDiffusionInterstitialHydrogen2004, haywardInterplayHydrogenVacancies2013}. Hence, both Ram\_EAM and Wen\_EAM predict lowered energy barriers. However, Wen\_EAM IAP reports the O-site to be a metastable minimum and not a saddle point (see cyan curve in Fig. \ref{fig:TOT_paths}), in contrast to DFT \cite{jiangDiffusionInterstitialHydrogen2004, bodaAdsorptionAbsorptionDiffusion2019}. Kumar\_EAM produces multi-saddle energy profiles for both paths (green curves in Fig. \ref{fig:mig_paths}). The other ML-IAP, Meng\_NNIP, gives barriers close to those obtained with DFT and tabGAP. For the detailed dependence of the migration energy barriers on hydrostatic strain, we guide the reader to the Supplementary Material.

\begin{table*}
\centering
\caption{Summary of the point-defect energetics of H in \aFe given by DFT and the tested IAPs. $E_{s,X}^{H}$ is the solution energy of H in the three sites X=T, X=O, X=vac, or octahedral and tetrahedral interstitial sites, and in a Fe-vacancy, respectively. The subscript $\varepsilon\%$ refers to a hydrostatic engineering strain applied to the Fe lattice. The terms $E_{m,T-T}^{H}$ and $E_{m,T-O-T}^{H}$ are the energy barrier heights of the T-T and T-O-T transition paths calculated with CI-NEB.}
    \begin{threeparttable}
    \begin{adjustbox}{width=0.9\textwidth}
    \begin{tabular}{@{}lccccccc@{}}
        \toprule
                 &  DFT  & DFT (ZPE) & tabGAP\tnote{a} & Ram\_EAM\tnote{b} & Wen\_EAM\tnote{c} & Kumar\_EAM\tnote{d} & Meng\_NNIP\tnote{e} \\
        \midrule
            $E_{s,T}^{H}$ (\si{eV})
            & 0.228\tnote{a} \, 0.20\tnote{g} & 0.30\tnote{g} \, 0.32\tnote{h} & 0.227 & 0.291 & 0.326 & 0.317 & 0.232 \\
            $E_{s,T,-2\varepsilon\%}^{H}$ (\si{eV})
            & 0.56\tnote{e} & 0.67\tnote{b} & 0.559 & 0.564 & 0.590 & 0.629 & 0.601 \\
            $E_{s,T,+2\varepsilon\%}^{H}$ (\si{eV})
            & -0.036\tnote{e} & 0.040\tnote{b} & -0.033 & 0.055 & 0.096 & 0.089 & -0.076 \\
            $E_{s,O}^{H}$ (\si{eV})
            & 0.33\tnote{g} & 0.35\tnote{i} & 0.375 & 0.340 & 0.370 & 0.350 & 0.391 \\
            $E_{s,V}^{H}$ (\si{eV})
            & -0.383\tnote{j} & -0.50\tnote{j} & -0.350 & -0.312 & -0.284 & -0.290 & -0.367 \\
            \midrule
            $E_{m,T-T}^{H}$ (\si{eV})
            & 0.088\tnote{k} \, 0.090\tnote{j} & 0.035\tnote{k} \, 0.044\tnote{j} & 0.114 & 0.041 & 0.044 & 0.027\tnote{*} & 0.107 \\
            $E_{m,T-T,-2\varepsilon\%}^{H}$ (\si{eV})
            & 0.095\tnote{e} & 0.044\tnote{b} & 0.127 & 0.018 & 0.019 & 0.049\tnote{*} & 0.112 \\
            $E_{m,T-T,+2\varepsilon\%}^{H}$ (\si{eV})
            & 0.072\tnote{e} & 0.049\tnote{b} & 0.093 & 0.040 & 0.044 & 0.044\tnote{*} & 0.092 \\
            $E_{m,T-O-T}^{H}$ (\si{eV})
            & 0.148\tnote{j} & 0.035\tnote{j} & 0.150 & 0.048 & 0.054\tnote{*} & 0.041\tnote{*} & 0.159 \\
        \bottomrule
    \end{tabular}
    \end{adjustbox}

    \begin{tablenotes}[para, flushleft]
        \item[a] This work
        \item[b] \cite{ramasubramaniamInteratomicPotentialsHydrogen2009}
        \item[c] \cite{wenNewInteratomicPotential2021}
        \item[d] \cite{kumarEffectHydrogenPlasticity2023}
        \item[e] \cite{mengGeneralpurposeNeuralNetwork2021}
        \item[f] \cite{haywardInterplayHydrogenVacancies2013}
        \item[g] \cite{jiangDiffusionInterstitialHydrogen2004}
        \item[h] \cite{tateyamaStabilityClusterizationHydrogenvacancy2003}
        \item[i] \cite{chenTemperaturedependentDissolutionDiffusion2017}
        \item[j] \cite{haywardInterplayHydrogenVacancies2013}
        \item[k] \cite{hirataFirstPrinciplesStudyHydrogen2018}
        \item[*] Shape of the migration path does not agree with DFT.
    \end{tablenotes}
    \end{threeparttable}
\label{tab:H_energetics}
\end{table*}

\subsubsection{H-vacancy interaction}

The interaction between H and vacancies plays a crucial role in some mechanisms proposed to explain HE. Among these are the H-enhanced formation of strain-induced vacancies (HESIV) \cite{nagumoHydrogenRelatedFailure2004a, yuHydrogenEmbrittlementConspicuous2024}, and a possible synergistic fracture mechanism with grain boundaries \cite{momidaHydrogenenhancedVacancyEmbrittlement2013}. DFT studies show that Fe vacancies are able to trap up to six H atoms \cite{haywardInterplayHydrogenVacancies2013}, which preferentially occupy sites close to the O-sites around the vacancy. The detailed energetics of the H-vacancy interaction are complex; DFT calculations predict that the binding energy of the second H atom inside a single \aFe vacancy is higher than that of the first H atom \cite{haywardInterplayHydrogenVacancies2013}. These findings make the H-vacancy interaction a challenging but important benchmark for \aFe-H IAP performance.

The binding energy of subsequently added H within a vacancy can be defined as
\begin{equation}
    E_b^{inc} = \left[E(H_{T})-E(bulk)\right] - \left[E(H_{n})-E(H_{n-1})\right],
\label{eq:vac_nH_Eb}
\end{equation}
where $H_{T}$ is a bulk \aFe structure with a H atom in a T-site, $bulk$ refers to a pure \aFe bulk structure, and $H_n$ is a structure with $n$ H atoms inserted into the vacancy. Using Eq. \ref{eq:vac_nH_Eb} we calculated the H binding energies for the tabGAP and the tested IAPs by inserting one to six H atoms into the O-sites surrounding a Fe-vacancy in a $10\times10\times10$ supercell. The H insertion sites and the insertion order were adopted from Ref. \cite{mengGeneralpurposeNeuralNetwork2021}.

The results are summarized in Tab. \ref{tab:vac_nH_Eb}. Most notably, our tabGAP correctly reproduces the increase in binding energy from one to two H atoms in the vacancy from DFT results, and the strongly reduced binding energy of the sixth H atom. In contrast, this behavior is not captured by any of the classical EAMs, while the Meng\_NNIP shows a trend similar to that of DFT and tabGAP.
\begin{table*}
\centering
\caption{The binding energy of incrementally added H in a single Fe vacancy. The H insertion sites and order were adopted from Ref. \cite{mengGeneralpurposeNeuralNetwork2021}.}
    \begin{threeparttable}
    \begin{tabular}{@{}ccccccccc@{}}
        \toprule
        $E_b^{inc}$ (eV)& n & DFT\tnote{f} & DFT (ZPE)\tnote{f} & tabGAP\tnote{a} & Ram\_EAM\tnote{b} & Wen\_EAM\tnote{c} & Kumar\_EAM\tnote{d} & Meng\_NNIP\tnote{e} \\
        \midrule
          &  1       & 0.498 & 0.616 & 0.577 & 0.603 & 0.610 & 0.607 & 0.598 \\
          &  2       & 0.543 & 0.651 & 0.608 & 0.603 & 0.572 & 0.578 & 0.616 \\
          &  3       & 0.337 & 0.381 & 0.449 & 0.425 & 0.393 & 0.454 & 0.364 \\
          &  4       & 0.304 & 0.351 & 0.451 & 0.361 & 0.327 & 0.348 & 0.338 \\
          &  5       & 0.269 & 0.269 & 0.322 & 0.242 & 0.244 & 0.285 & 0.318 \\
          &  6       &-0.043 & 0.045 & 0.047 & 0.315 & 0.149 & 0.178 & 0.042 \\
        \cmidrule{2-9}
    \end{tabular}
    \begin{tablenotes}[para, flushleft]
        \item[a] This work
        \item[b] \cite{ramasubramaniamInteratomicPotentialsHydrogen2009}
        \item[c] \cite{wenNewInteratomicPotential2021}
        \item[d] \cite{kumarEffectHydrogenPlasticity2023}
        \item[e] \cite{mengGeneralpurposeNeuralNetwork2021}
        \item[f] \cite{haywardInterplayHydrogenVacancies2013}
    \end{tablenotes}
    \end{threeparttable}
\label{tab:vac_nH_Eb}
\end{table*}

\subsubsection{Interaction of H atoms with dislocations}

We further verified the interactions of H atoms with dislocations using the Fe-H tabGAP. Multiple different proposed HE mechanisms are closely linked to the interaction of H atoms with dislocations, namely hydrogen enhanced localized plasticity (HELP) \cite{birnbaumHydrogenenhancedLocalizedPlasticity1994}, adsorption-induced dislocation emission (AIDE) \cite{lynchFractographicStudyHydrogenassisted1986}, and the Defactant concept \cite{kirchheimReducingGrainBoundary2007}. In bcc metals, the deformation behavior is strongly controlled by the movement of the $1/2\hkl<111>$ screw dislocation \cite{duesberyPlasticAnisotropyBcc1998}. Therefore, assessing the effect of solute H on this defect is of great importance. Before this, we assessed the pure Fe screw dislocation properties given by the tabGAP using the differential displacement map (DDM) \cite{vitekCoreStructure1/21111970}. The tabGAP gives the correct nondegenerate core structure with the atomic displacements spreading in the three \hkl{110} planes of the \hkl[111] zone, as shown in Fig. S.4.5a \cite{vitekAtomicLevelComputer2011}. Alongside this, our model reproduces the single-humped Peierls barrier between two neighboring easy-core configurations in the \hkl(-101) glide plane (Fig. S.4.4) \cite{ventelonInitioInvestigationPeierls2013, itakuraEffectHydrogenAtoms2013a}. It is a known problem of the Ram\_EAM and Wen\_EAM IAPs that they produce a double-humped Peierls barrier in contradiction to the DFT results \cite{kumarEffectHydrogenPlasticity2023}.

\begin{figure*}
\centering
    \subcaptionsetup[figure]{margin={-0.19\textwidth, 0pt}}
    \begin{subfigure}{.48\textwidth}
        \includegraphics[width=\linewidth]{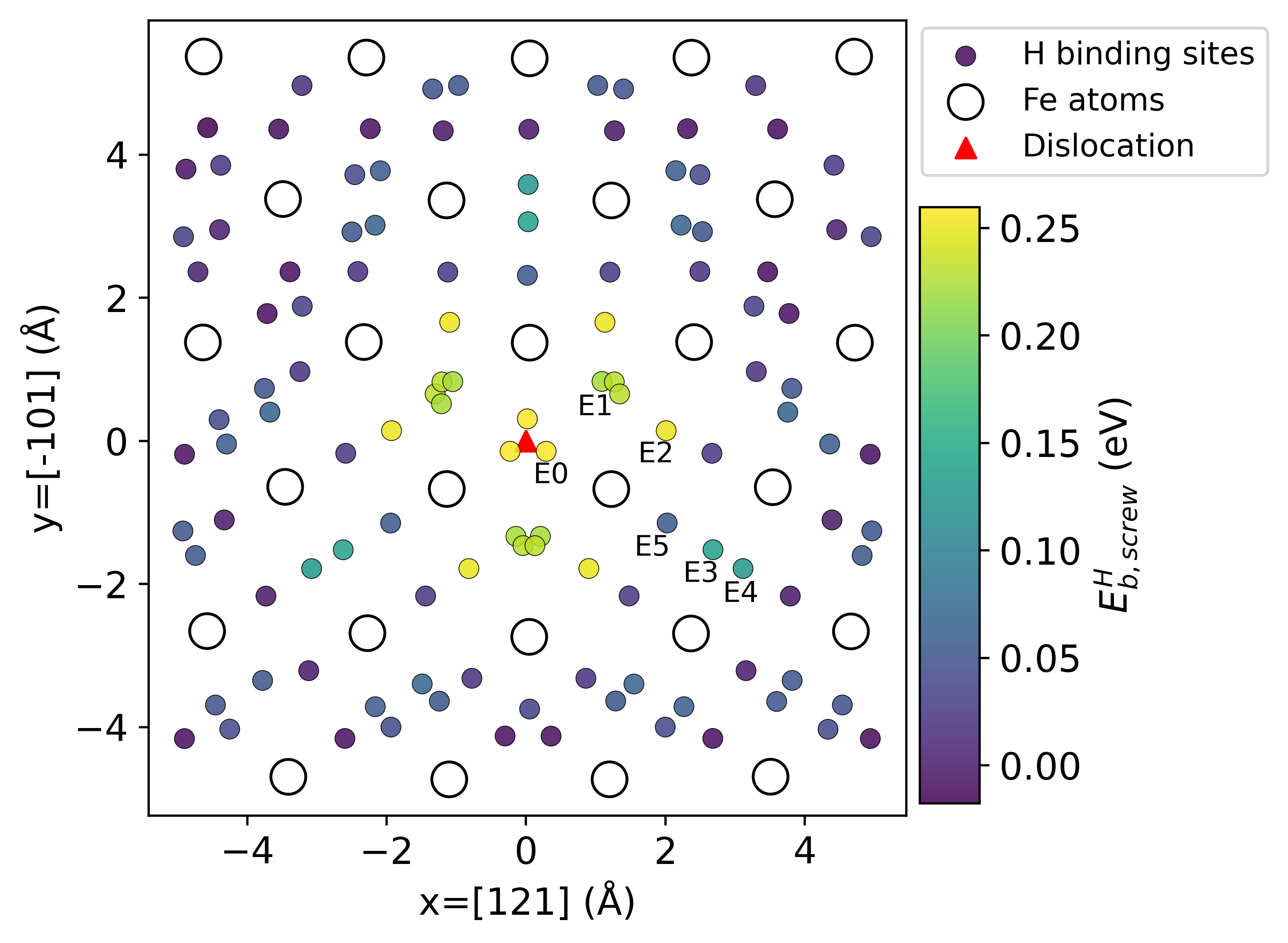}
        \caption{}
        \label{fig:H_Eb_screw_pos}
    \end{subfigure}
    \subcaptionsetup[figure]{margin={0.13\textwidth, 0pt}}
    \begin{subfigure}{.48\textwidth}
        \includegraphics[width=\linewidth]{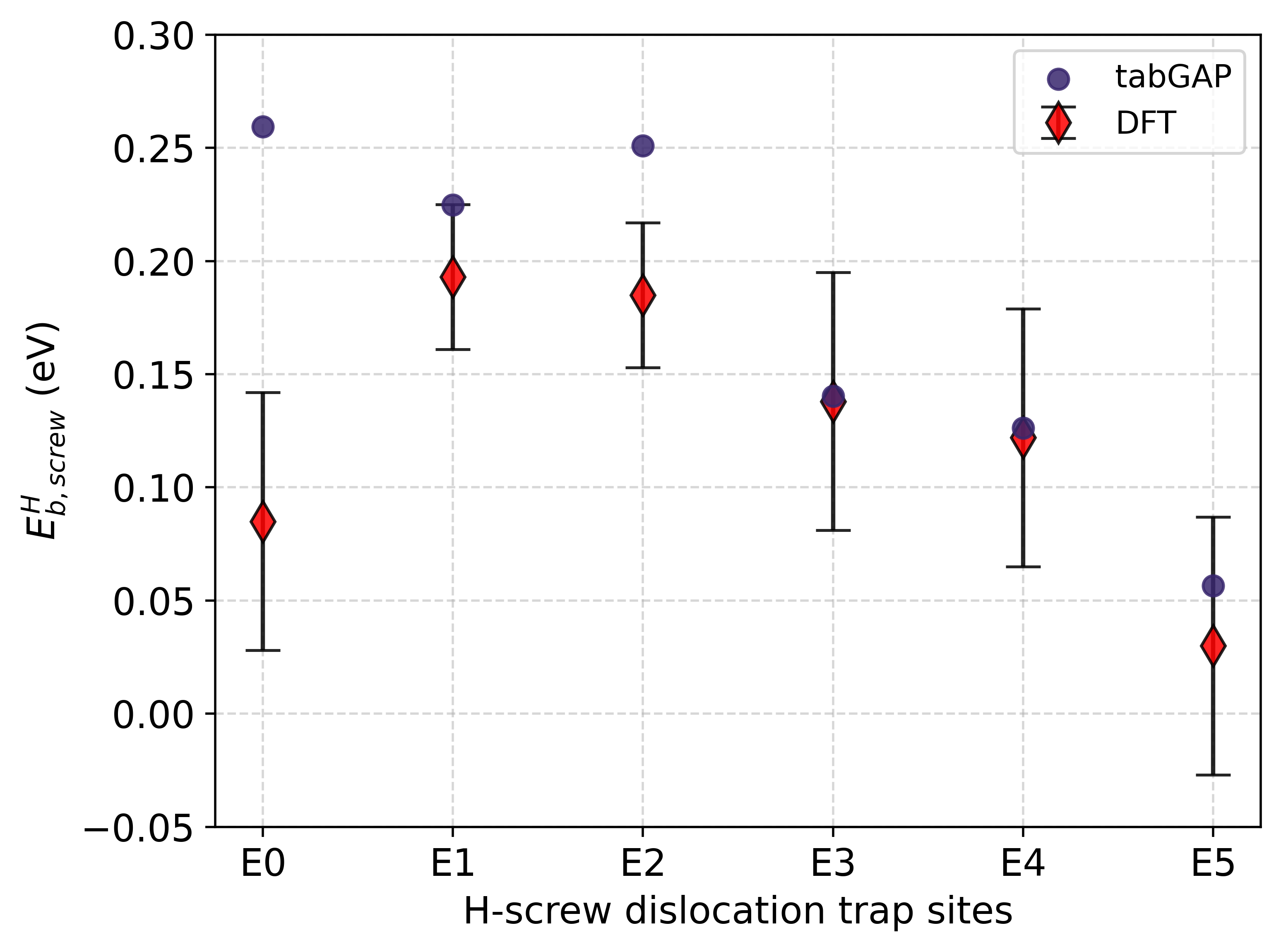}
        \caption{}
        \label{fig:H_Eb_screw_dft}
    \end{subfigure}
    \begin{subfigure}{.57\textwidth}
        \includegraphics[width=\linewidth]{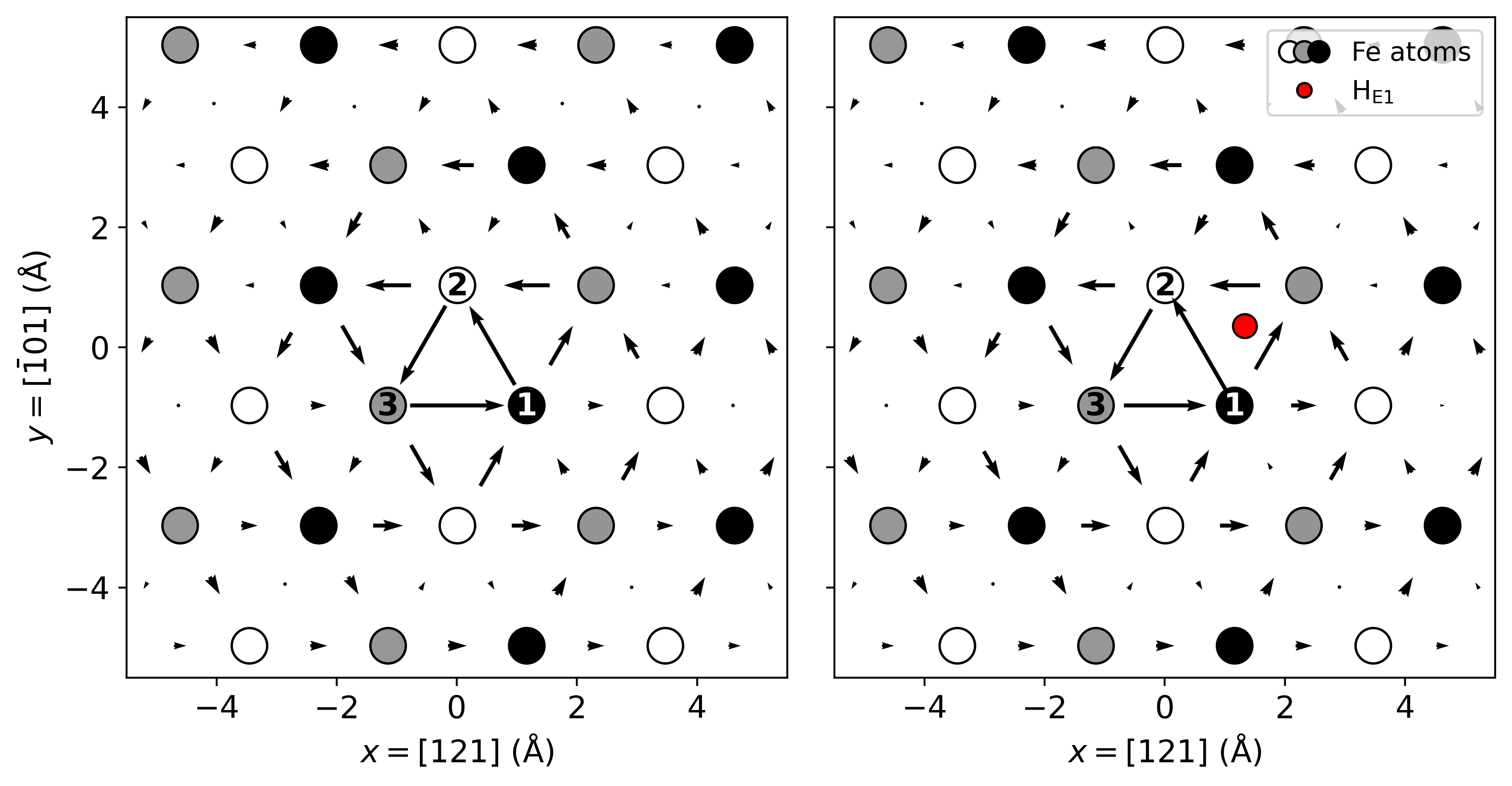}
        \caption{}
        \label{fig:screw_H_ddm}
    \end{subfigure}
    \begin{subfigure}{.39\textwidth}
        \includegraphics[width=\linewidth]{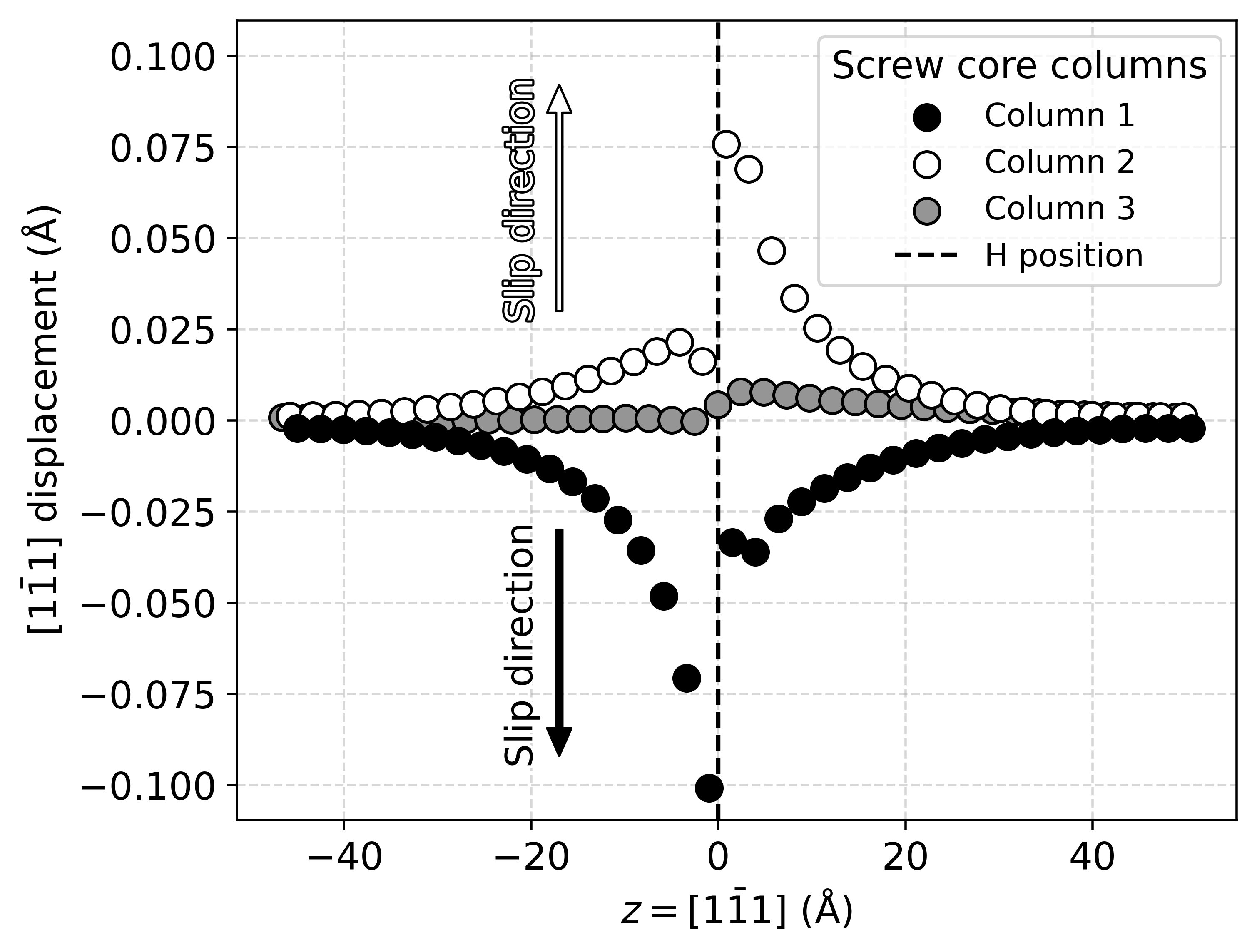}
        \caption{}
        \label{fig:screw_H_core_displacements}
    \end{subfigure}
\caption{(\subref{fig:H_Eb_screw_pos}) The binding sites and energies for a H atom near a $1/2\hkl<111>\hkl{110}$ easy-core screw dislocation calculated with the tabGAP. (\subref{fig:H_Eb_screw_dft}) The binding energies of a few annotated cites compared to DFT calculations from Ref. \cite{itakuraEffectHydrogenAtoms2013a}. (\subref{fig:screw_H_ddm}) The differential displacement map of a $40b$ long screw dislocation visualizing the core structure near a H atom in the E1-site. Only the arrows within a $b/2$ distance of the H atom z-coordinate are shown for clarity. The arrows lengths are scaled by $b/2$. The Fe atoms are colored according to their position in the \hkl[1-11] direction in the bulk configuration, and the Fe atoms comprising the screw core are labeled 1,2 and 3 for reference. (\subref{fig:screw_H_core_displacements}) The screw core Fe atom \hkl[1-11] displacements caused by the presence of a H atom in the E1-site. For columns 1 (black) and 2 (white), the direction of slip that would make the dislocation move towards the H atom is annotated.}
\label{fig:H_Eb_screw}
\end{figure*}

Firstly, we identified the unique binding sites for H atoms around a $1/2\hkl<111>\hkl{110}$ easy-core screw dislocation. For this, we randomly inserted singular H atoms within a rectangular region of $5\times5\times3 \; \si{\angs}$ around the dislocation core and relaxed the atomic positions. The relaxed H binding sites are illustrated in Fig. \ref{fig:H_Eb_screw_pos} and compared to the non-ZPE corrected DFT data from Ref. \cite{itakuraEffectHydrogenAtoms2013a} in Fig. \ref{fig:H_Eb_screw_dft}. The annotation of the binding sites was adopted from Ref. \cite{itakuraEffectHydrogenAtoms2013a}. The detailed comparison between the tabGAP and DFT binding sites, as well as those predicted by the classical IAP Ram\_EAM, can be seen in Fig. \ref{appfig:H_Eb_screw} in the Appendix. From this figure we see that the binding site locations with the tabGAP are in excellent agreement with the DFT predictions. Most notably, the three strongly binding basins around the dislocation core, those that include the E1 and E2 sites in the configuration E2-E1-E1-E2, are predicted by the tabGAP similarly to the DFT from Ref. \cite{itakuraEffectHydrogenAtoms2013a}. In comparison, there are three binding site locations predicted by the Ram\_EAM near the dislocation core that are not in agreement with the locations identified in the DFT calculations.

Moreover, Fig. \ref{fig:H_Eb_screw_dft} shows a good quantitative agreement for the binding energies between tabGAP and DFT for all the sites E1 to E5. The only substantial disagreement between tabGAP and DFT is the binding energy of the E0-site, where tabGAP reports a value of $\SI{0.26}{eV}$, and DFT gives $\SI{0.085}{eV}$. Here we note that our tabGAP training dataset does not include structures with dislocations and H atoms simultaneously. These configurations were left out due to being bound to exaggerate the concentration of H atoms near the dislocation because of the limited size of the atomic configurations used for the training of the potential. Despite this, our potential is able to reproduce the H-screw dislocation binding energetics remarkably well. We attribute this to the strongly strained bulk \aFe-H structure data containing local environments that are similar to dislocation cores.

In Fig. \ref{fig:screw_H_ddm} we show the DDM of a $40b$ long screw dislocation containing a H atom relaxed into an E1-site. The H atom produces a weak and localized split-like transformation of the core toward the H-occupied site, while the rest of the core structure remains close to the pure-Fe configuration. The corresponding pure-Fe and H-induced difference DDMs are shown in Fig. S.4.5. The effect of H on the core structure remained localized to the immediate region of the H atom when the length of the dislocation was varied between \myqtyrange{2}{40}{b}. In the extreme case of a $1b$-long dislocation, or when there's one H atom per Burgers vector period, a stronger split-core transformation of the dislocation was observed. This is shown in Fig. S.4.5d.

The \hkl[1-11] displacements in Fig. \ref{fig:screw_H_core_displacements} show the H atom effect on the core Fe atoms along the screw dislocation line. The most notable outcome is that the displacements caused by the H atom are not volumetric but shear-like, displacing column 1 and column 2 Fe atoms uniformly in their respective slip directions. This essentially creates an initial kink-pair to the dislocation that is toward the H atom.

\subsubsection{H-H interaction and clustering}

Due to the limitations in spatial and temporal scales accessible within MD simulations, it is challenging to directly observe H-induced effects that may contribute to HE. To circumvent this, in simulations it's common to use H concentrations far above experimental levels, reaching several \unit{\atp} \cite{songAtomicMechanismPrediction2013,starikovAngulardependentInteratomicPotential2021, lvHydrogenDiffusionVacancy2018, zhuSoluteHydrogenEffects2018, wangMolecularDynamicsStudy2024a}, whereas experimentally measured H concentrations in bulk \aFe are on the order of $\SI{0.001}{\atp}$ even under charging conditions \cite{kiuchiSolubilityDiffusivityHydrogen1983a,matsuiEffectHydrogenMechanical1979}. Therefore, when performing simulations, special care must be taken to ensure that the selected IAP describes the behavior of the H atoms under such artificially high H concentrations. Song and Curtin \cite{songAtomicMechanismPrediction2013} reported that the Ram\_EAM IAP \cite{ramasubramaniamInteratomicPotentialsHydrogen2009} leads to spontaneous formation of interstitial H clusters at room temperature already at $\SI{1}{\atp}$. They attributed this clustering to an unphysical attractive interaction between H atoms in neighboring T-sites, identified via calculations of the pairwise binding energy (Eq. \ref{eq:Eb_HH}). To remedy this, they modified the IAP by incorporating an additional repulsive H–H term (see the supplementary information of Ref. \cite{songAtomicMechanismPrediction2013}). Nonetheless, Starikov \emph{et al.} \cite{starikovAngulardependentInteratomicPotential2021} still detected H clustering at concentrations above $\SI{2}{\atp}$ when using the modified Ram\_EAM potential.

To evaluate the physical relevance of these simulation results one can compare them with experimental data on iron hydride formation, where results indicate a threshold H concentration of $\myqtyrange{4}{13}{\atp}$ \cite{hiroiPhaseDiagramSuperabundant2005, baddingHighPressureChemistryHydrogen1991, fukaiHydrogenIronIts1982}, that is calculated from the approximate volume expansion caused by solute H. Therefore, based on the experimental evidence, H clustering in simulations with the concentration below the threshold of $\SI{4}{\atp}$ could be interpreted as an artifact of the selected IAP.

\begin{figure}
\centering
\includegraphics[width=\linewidth]{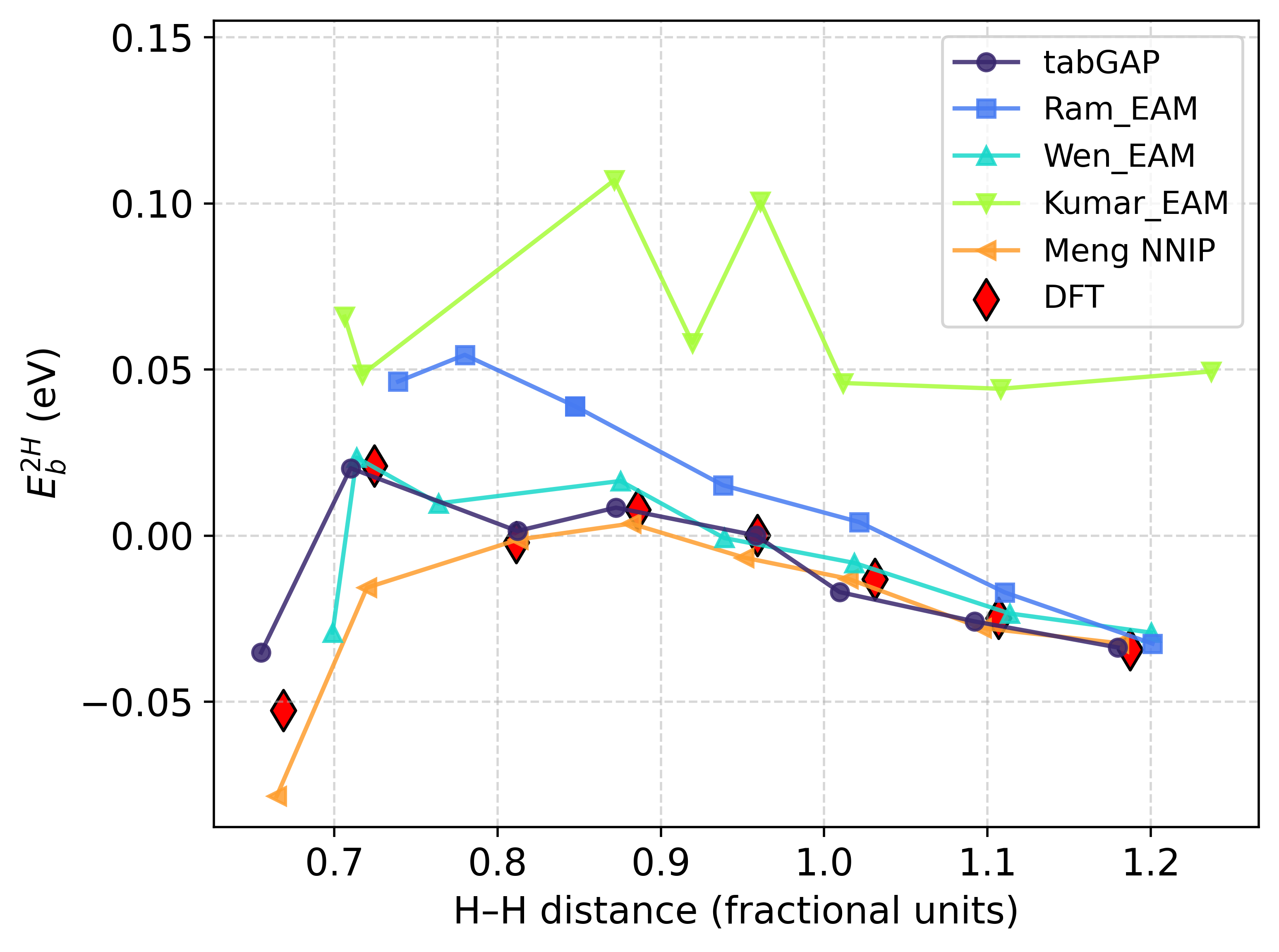}
\caption{The binding energy of interstitial H-pairs within the \aFe matrix calculated with the Eq. \ref{eq:Eb_HH}. The x-axis is the distance (fractional units) between the pair of H atoms after energy minimization. DFT results are from Ref. \cite{haywardInterplayHydrogenVacancies2013}.}
\label{fig:Eb_HH}
\end{figure}

To quantitatively assess the prediction of H clustering by an IAP, we examine the binding energy of two interstitial H atoms using the following equation:
\begin{equation}
    E_b^{2H} = 2E(H_{T}) - E(2H) + E(bulk),
\label{eq:Eb_HH}
\end{equation}
where $E(2H)$ is the energy of a structure containing two neighboring interstitial H atoms. A positive $E_b^{2H}$ indicates a preferential occupation of closely located H sites relative to isolated H positions ($H_{T}$). To assess the accuracy of the Fe–H tabGAP, the values of $E_b^{2H}$ were calculated by placing H atoms into eight unique T-site pairs found within $\SI{\sim3}{\angs}$ of each other in the \aFe lattice (see Fig. \ref{fig:Eb_HH}), and relaxing the structure. The tabGAP accurately reproduces these binding energies obtained from DFT \cite{haywardInterplayHydrogenVacancies2013}.

\begin{figure}
\centering
    \includegraphics[width=\linewidth]{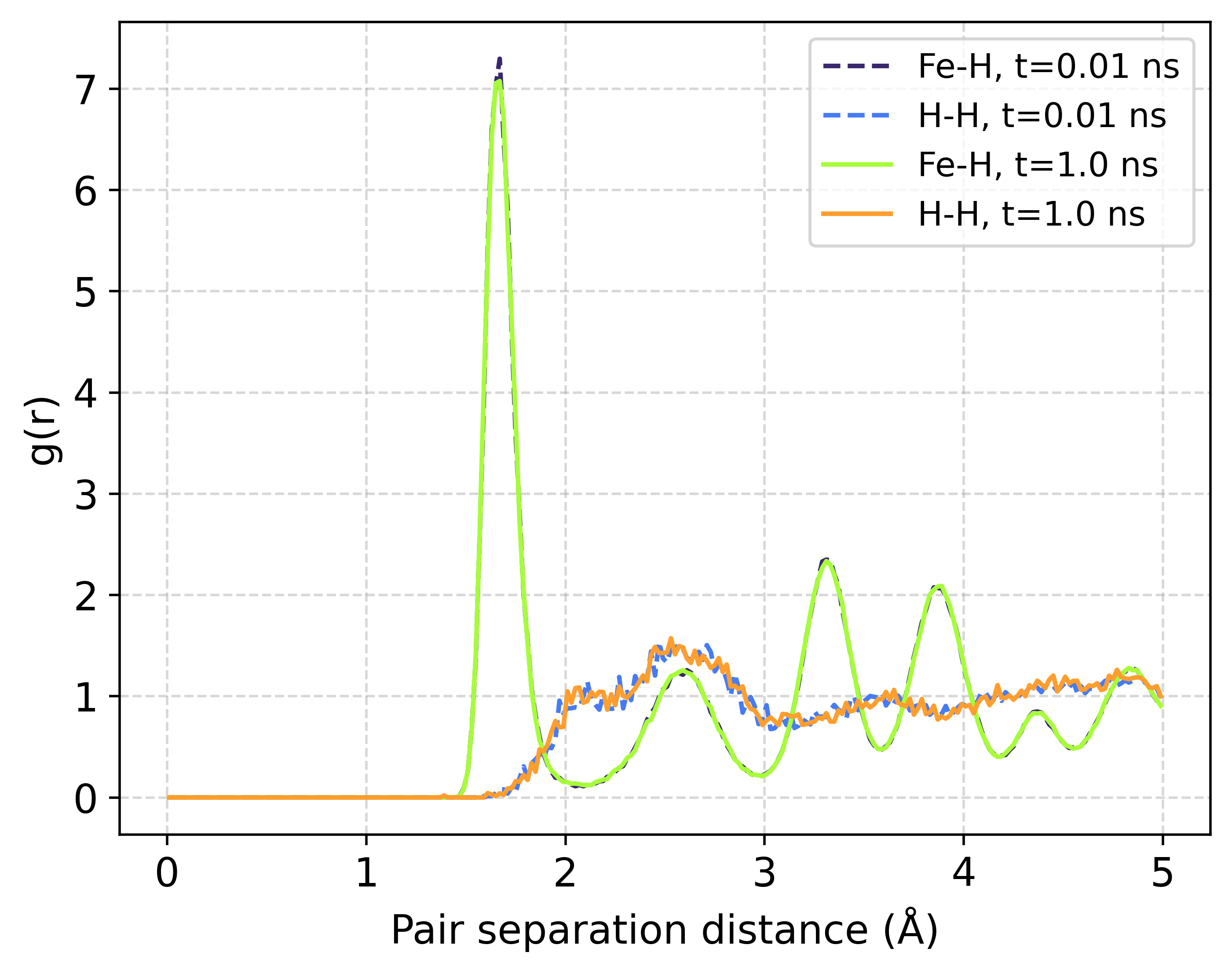}
\caption{The radial distribution functions of Fe-H and H-H pair separation distances in a ($37\times37\times37$)-sized \aFe $\SI{10}{\atp}$-H-structure before and after equilibrating for $\SI{1}{ns}$ at $\SI{300}{K}$ with the NPT ensemble.}
\label{fig:H_clust_rdf}
\end{figure}

Taking into consideration only the $E_b^{2H}$ values is insufficient for the proof of absence of clustering at low H concentrations because it only describes two neighboring H atoms in a zero-temperature environment. Therefore, we performed further test simulations at finite temperature with an interstitial H concentration range of $\myqtyrange{0.07}{30}{\atp}$. In these simulations the H atoms were randomly distributed within ($10\times10\times10$)-sized \aFe cells, and the structure was equilibrated at $\SI{300}{K}$ for $\SI{1}{ns}$, controlling the temperature and pressure via the NPT ensemble. The evolution of the H distribution was monitored by following the radial distribution function (RDF) of H-pair separation distances. No clustering of H atoms was observed in these simulations, and the initial shape of the RDF did not change. To validate the accuracy of this result, we repeated the simulation for a larger ($37\times37\times37$) cell and the H concentration of $\SI{10}{\atp}$. The result is shown in Fig. \ref{fig:H_clust_rdf}. The mean displacement of the H atoms during this simulation was $\SI{33.13}{\angs}$, showing that the H atoms are mobile. 

These simulations demonstrate the Fe-H tabGAP's applicability in modeling the solute H-H interactions in \aFe at finite temperature. The Fe-H tabGAP accurately reproduces the interstitial H pair binding energies from DFT calculations, and doesn't produce clustering distributions of interstitial H in room temperature MD simulations. Both of these results support the conclusion of a low probability of H clustering within the defect-free Fe matrix at realistic solute concentrations. However, there are some computational results indicating H-H interaction-driven clustering around dislocations \cite{vonpezoldHydrogenenhancedLocalPlasticity2011,houHydrogenClusteringBcc2020}. These considerations are, however, out of the scope of the present study.

\subsubsection{Zero-temperature elastic constants}\label{sec:0K_Econsts}

The elastic constants of \aFe as a function of the concentration of solute interstitial H were calculated at $\SI{0}{K}$ by averaging the results of simulations with \aFe lattices containing different random distributions of H atoms. For each distribution, we randomly selected T-sites to be occupied by H atoms up to the target concentration. After the structure was loaded with the H atoms, it was relaxed using the conjugate gradient algorithm. For each H concentration, $20$ different H distributions were sampled in the concentration range of $\myqtyrange{0}{5}{\atp}$. We want to re-emphasize here that the experimentally measured concentration of H in \aFe is $<\SI{0.001}{\atp}$ \cite{kiuchiSolubilityDiffusivityHydrogen1983a,matsuiEffectHydrogenMechanical1979}, and the H concentration range used in our MD simulations was selected to extract a visible trend in the elastic constants alongside being able to draw comparisons between available DFT results \cite{psiachosInitioStudyModification2011}.

The prepared structures were then deformed by applying the strain of $\varepsilon=\SI{5e-3}{}$, and the elastic stiffness tensor was evaluated by measuring the change in the stress tensor. With the tabGAP, the lattice parameters were pre-calculated using the following linear equation: 
\begin{equation}\label{eq:alat_T_cH}
    a = \num{3.428e-5}T + \num{3.639e-3}cH + 2.82944 \;\si{\angs},
\end{equation}
where $T$ is the temperature in Kelvin and cH is the concentration of H in $\si{\atp}$. Eq. \ref{eq:alat_T_cH} was derived from a least-squares fit to lattice parameter data calculated with NPT simulations. The explanatory power of the fit was $R^2=0.9999$, indicating a very linear relationship. Similar equations were determined for all the other tested IAPs.
\begin{figure*}
\centering
    \setkeys{Gin}{keepaspectratio}
    \subcaptionsetup[figure]{margin={0.10\textwidth, 0pt}}
    \begin{subfigure}[c]{.45\textwidth}
      \centering
      \includegraphics[width=\linewidth]{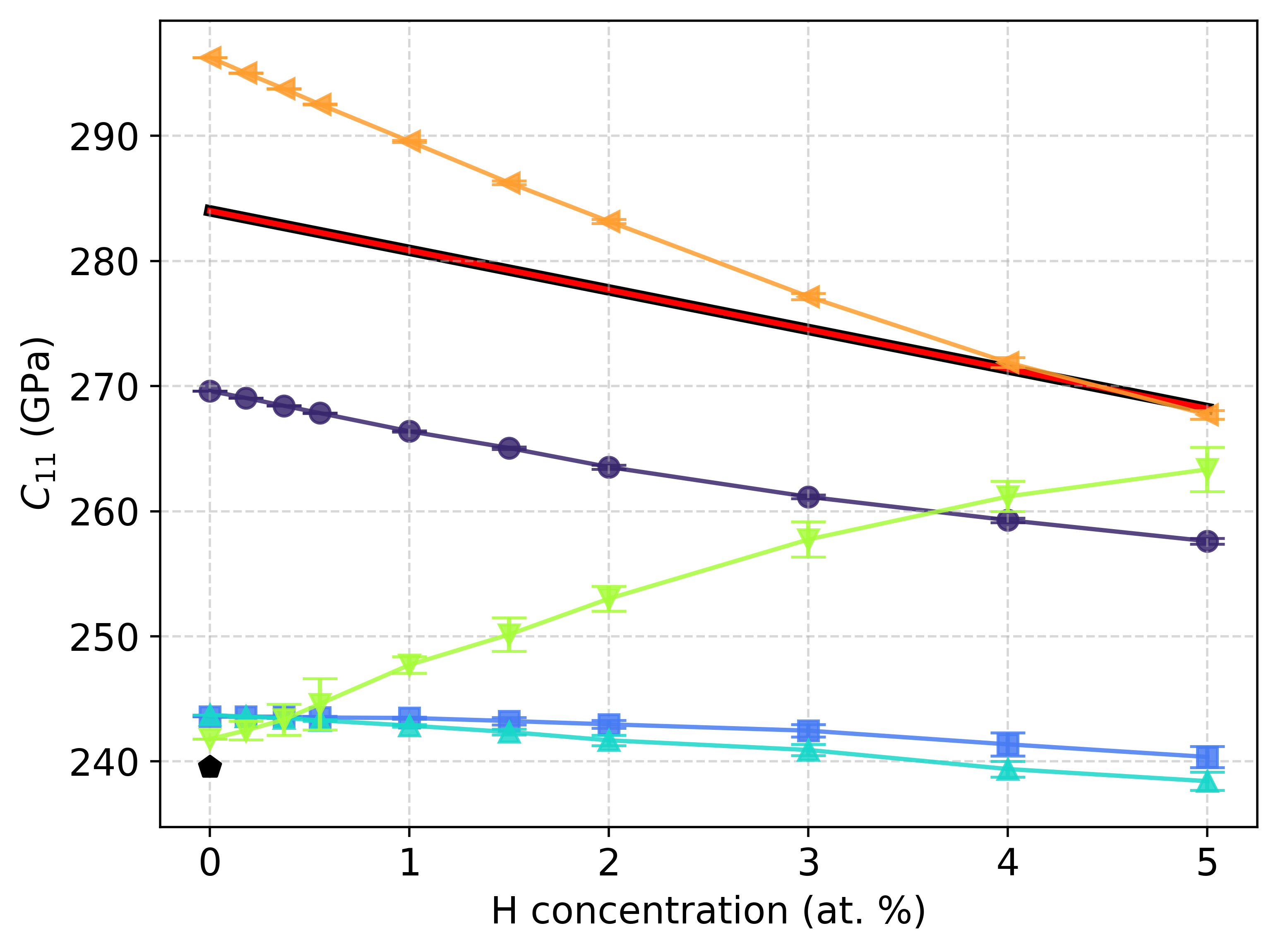}
      \caption{$C_{11}$}
      \label{fig:C11_0K}
    \end{subfigure}%
    \subcaptionsetup[figure]{margin={0.10\textwidth, 0pt}}
    \begin{subfigure}[c]{.45\textwidth}
      \centering
      \includegraphics[width=\linewidth]{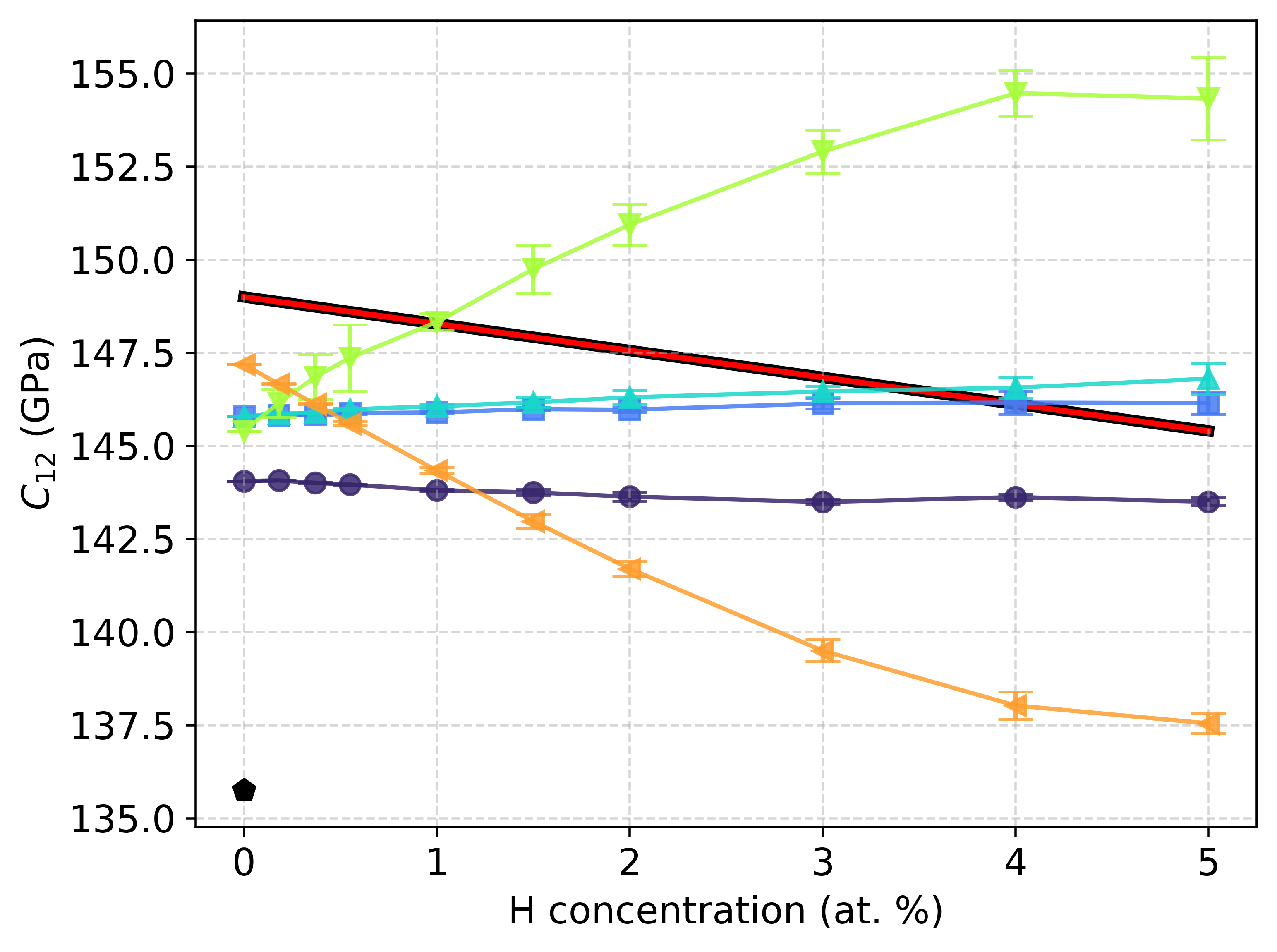}
      \caption{$C_{12}$}
      \label{fig:C12_0K}
    \end{subfigure}

    \hspace*{0.13\textwidth}
    \begin{subfigure}[c]{.45\textwidth}
      \centering
      \includegraphics[width=\linewidth]{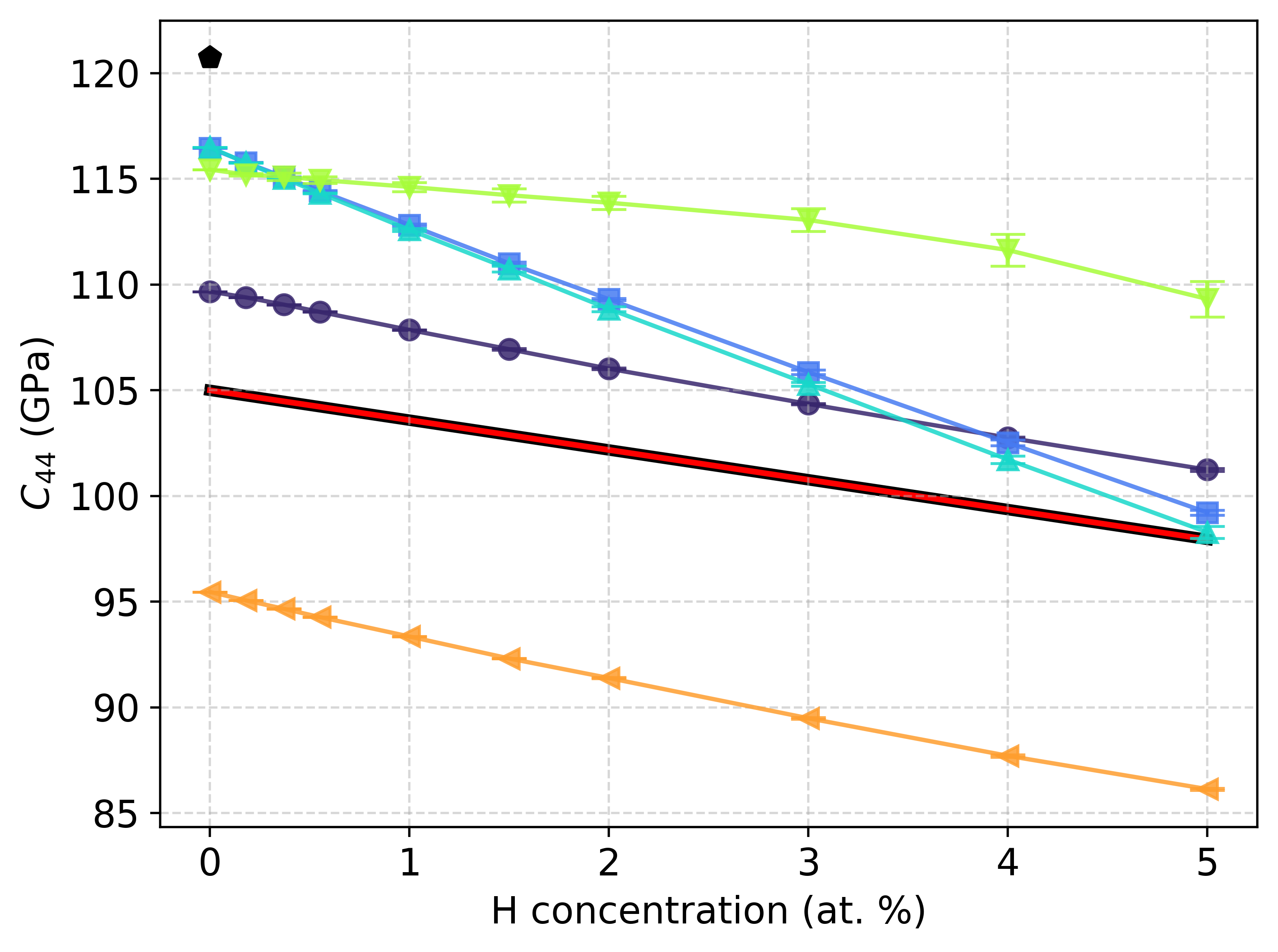}
      \caption{$C_{44}$}
      \label{fig:C44_0K}
    \end{subfigure}%
    \begin{subfigure}[c]{0.14\textwidth}
        \centering
        \includegraphics[width=\linewidth]{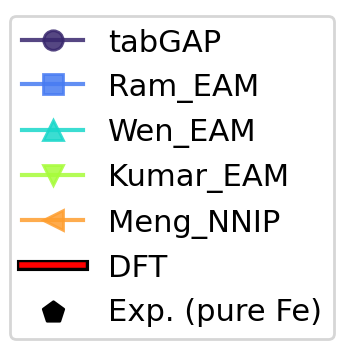}
        \vspace{.8cm}
    \end{subfigure}
\caption{The \aFe-H elastic constants of the tested IAPs at $\SI{0}{K}$ as a function of H concentration. The DFT results are linear best fits from Ref. \cite{psiachosInitioStudyModification2011}, and the experimental results are from Ref. \cite{adamsElasticConstantsMonocrystal2006}.}
\label{fig:Econsts_0K}
\end{figure*}

Fig. \ref{fig:Econsts_0K} shows the trends of the $C_{11}$, $C_{12}$, and $C_{44}$ dependencies on H concentration in comparison to DFT results \cite{psiachosInitioStudyModification2011}. There is a strong disagreement in the absolute values of $C_{11}$ and $C_{44}$ between the different models, while $C_{12}$ agrees better between the different models. This is mainly explained by the fact that the EAM models are fit to the experimental values of the elastic constants, which the DFT calculations disagree with by $\myqtyrange{10}{18}{\%}$ \cite{adamsElasticConstantsMonocrystal2006}. Overall, we see that the tabGAP results agree much better with those calculated by DFT in comparison to all the tested IAPs. For example, the linear dependencies of $C_{11}$ and $C_{44}$ on H concentration with the tabGAP have almost identical slopes to the DFT data. For $C_{12}$, tabGAP reproduces the DFT trend of a negative slope, though the absolute slope is smaller.

\subsubsection{Finite-temperature elastic constants}\label{sec:finT_Econsts}

The finite-temperature dependencies of the elastic constants of \aFe with H concentration were obtained for the structures with lattice parameters calculated using Eq. \ref{eq:alat_T_cH} similarly to those constructed for the $\SI{0}{K}$ simulations. In this case, because of thermal fluctuations, the average change in the stress tensor of a deformed structure was sampled from an NVT simulation. To examine the chemical and volumetric effects of H on the elastic constants separately, we additionally performed volume-equivalent calculations. In these calculations pure \aFe lattices were constructed with the lattice expansion defined by Eq. \ref{eq:alat_T_cH} according to the studied H concentration, but no H atoms were inserted into these lattices. We supply the numerical data from these calculations in the Supplemental Tab. S.3.2.

Fig. \ref{fig:Econsts_T} shows the temperature dependence of the elastic constants at various H concentrations. For pure \aFe (dark-purple curves in Fig. \ref{fig:Econsts_T}), the tabGAP model predicts an approximately linear decrease for all three constants. The effect of H is negligible up to concentrations of $\sim\SI{1}{\atp}$. At higher concentrations, we see a decrease in both $C_{11}$ and $C_{44}$, with the reduction becoming more pronounced as the H content increases. By contrast, $C_{12}$ shows the opposite trend, with higher solute H concentrations causing an increase in $C_{12}$, with the increase becoming more prominent at higher temperatures.

At room temperature, the H content decreases both $C_{11}$ and $C_{44}$ by $\SI{1.5}{\%}$, and increases $C_{12}$ by $\SI{0.5}{\%}$ per $\SI{1}{\atp}$ increase in H concentration. Moreover, the volume-equivalent calculations for pure Fe indicated a more significant reduction for all three constants per $\SI{1}{\atp}$ increase in H concentration; $\SI{2.3}{\%}$, $\SI{2.5}{\%}$, and $\SI{1.5}{\%}$ for $C_{11}$, $C_{12}$, and $C_{44}$, respectively. These results imply that the chemical contribution of H generally increases all three constants, consistent with the DFT calculations by Psiachos \emph{et al.} \cite{psiachosInitioStudyModification2011}. However, the decrease due to volumetric expansion is more significant than the chemical contribution by H for $C_{11}$ and $C_{44}$. 

While experimental results on single-crystal Fe samples are lacking, a decrease of $\SI{8}{\%}$ of the shear modulus with $\SI{1}{\atp}$ H-content was deduced by Lunarska \emph{et al.} \cite{lunarskaEffectHydrogenShear1977} for polycrystal \aFe at $\SI{100}{K}$. This result can be compared to the tabGAP-derived elastic constants via the Voigt-Reuss-Hill average shear modulus $G_H$ \cite{hillElasticBehaviourCrystalline1952}, defined by the equations:
\begin{align}
    G_H&= \frac{G_R+G_V}{2}, \\
    G_R&= \frac{5(C_{11}-C_{12})C_{44}}{4C_{44}+3(C_{11}-C_{12})}, \\
    G_V&= \frac{C_{11}-C_{12}+3C_{44}}{5}.
\end{align}

The tabGAP model predicts a $G_H(T=\SI{100}{K})$ of $85.9$ and $\SI{84.3}{GPa}$ at the H concentrations of $0$ and $\SI{1}{\atp}$, respectively. This is a decrease of $\SI{1.9}{\%}$.

\begin{figure*}
\centering
    \setkeys{Gin}{keepaspectratio}
    \subcaptionsetup[figure]{margin={0.10\textwidth, 0pt}}
    \begin{subfigure}[c]{.45\textwidth}
      \centering
      \includegraphics[width=\linewidth]{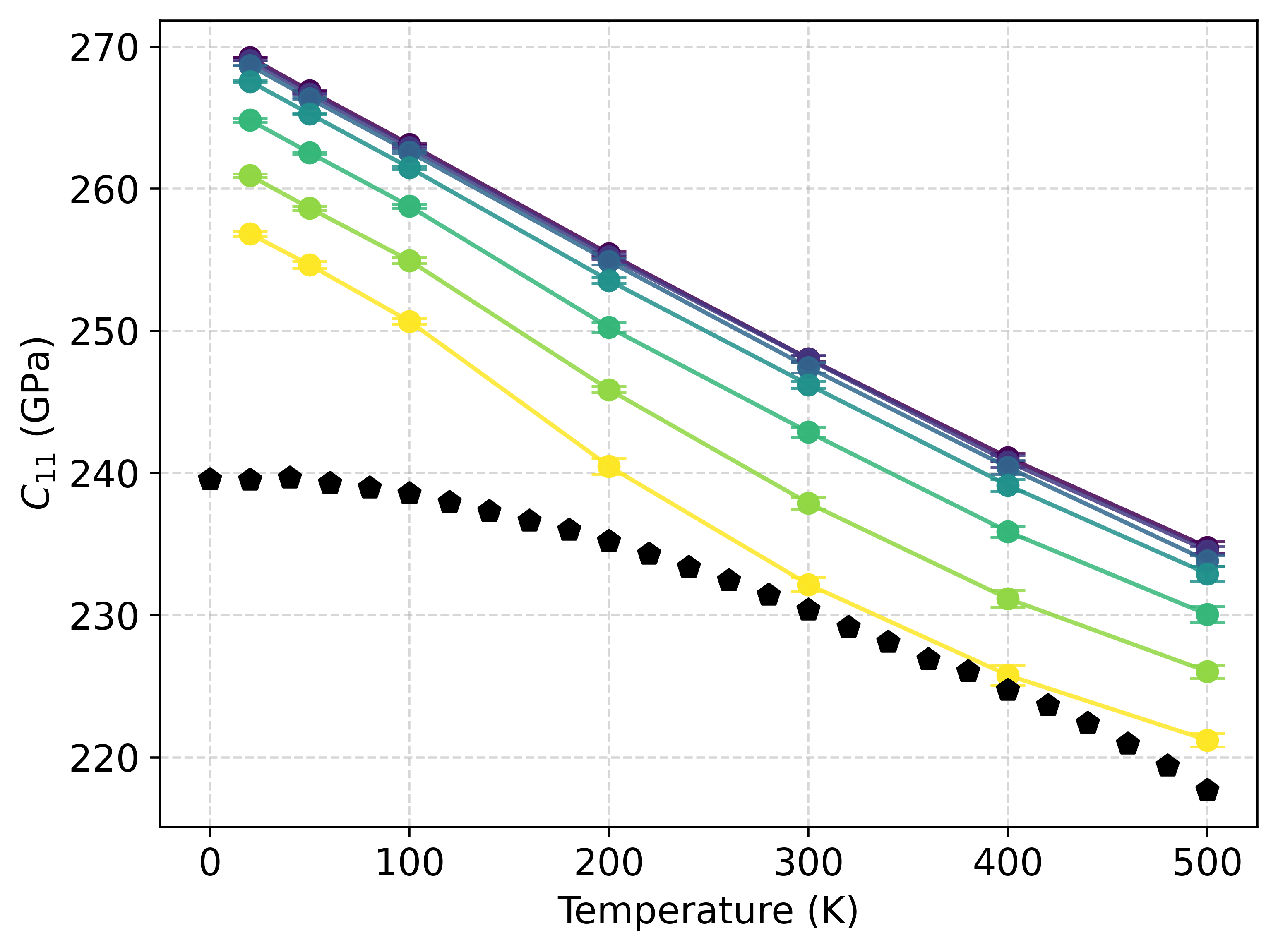}
      \caption{$C_{11}$}
      \label{fig:C11_T}
    \end{subfigure}%
    \subcaptionsetup[figure]{margin={0.10\textwidth, 0pt}}
    \begin{subfigure}[c]{.45\textwidth}
      \centering
      \includegraphics[width=\linewidth]{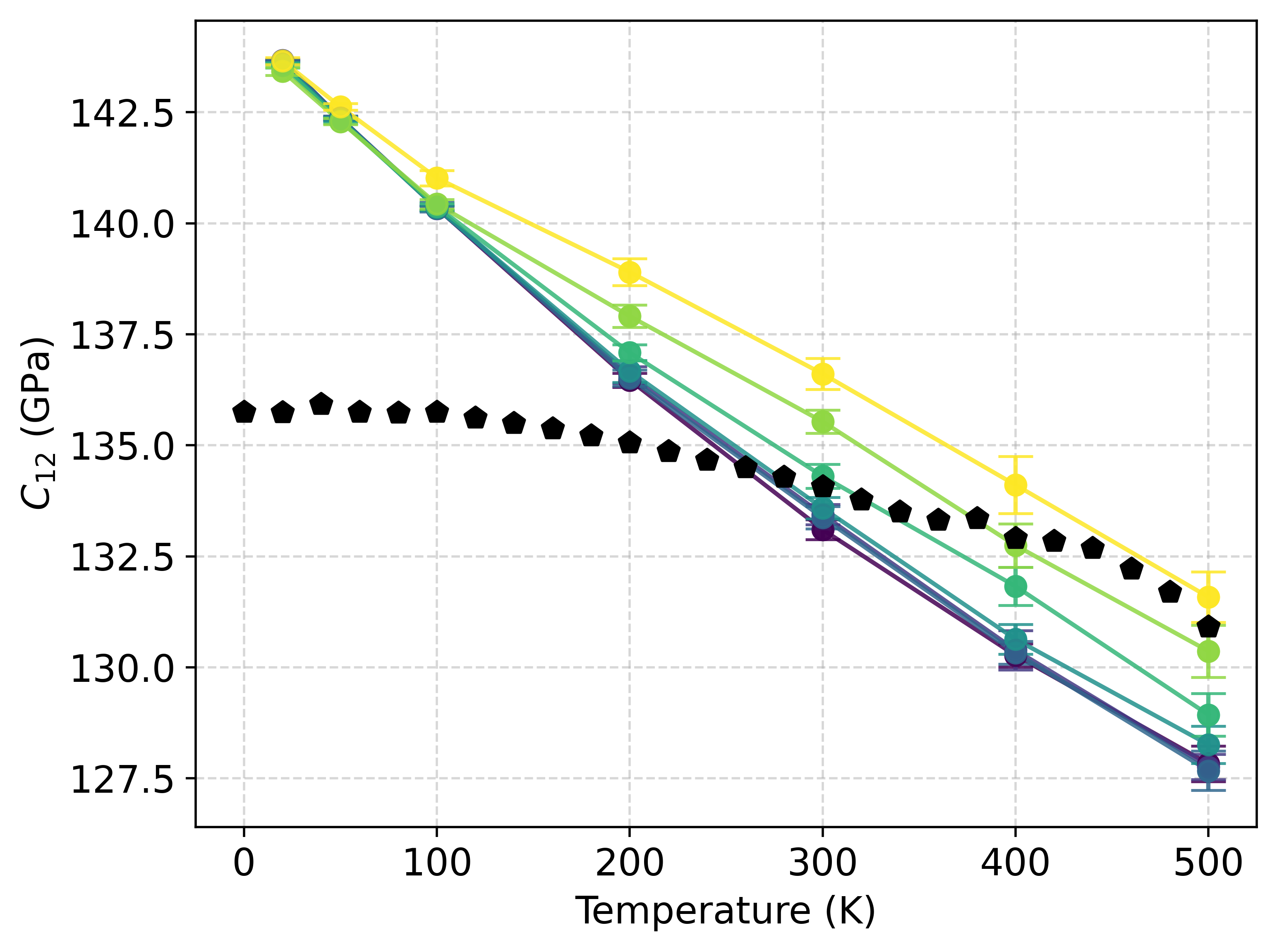}
      \caption{$C_{12}$}
      \label{fig:C12_T}
    \end{subfigure}

    \hspace*{0.13\textwidth}
    \begin{subfigure}[c]{.45\textwidth}
      \centering
      \includegraphics[width=\linewidth]{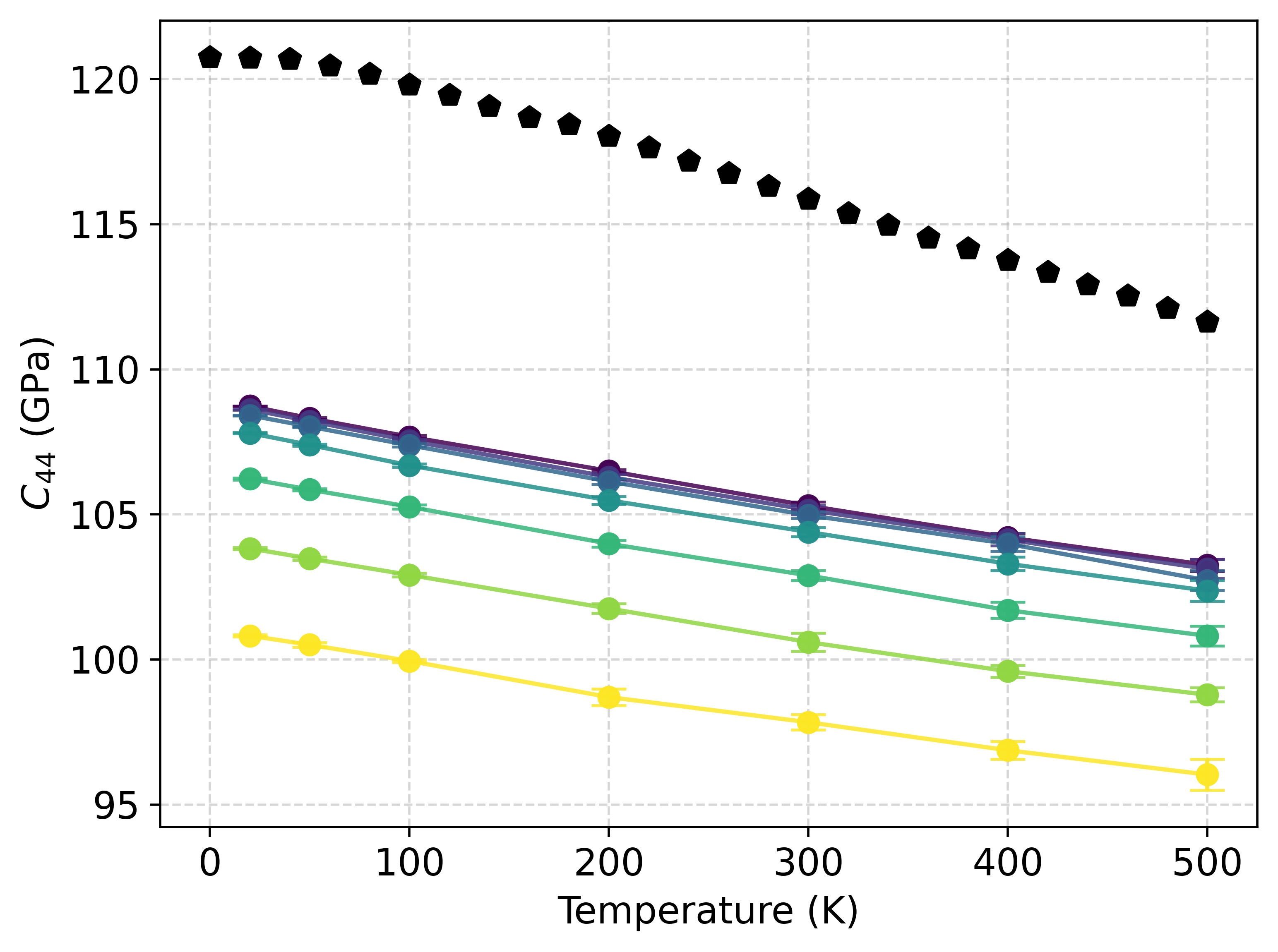}
      \caption{$C_{44}$}
      \label{fig:C44_T}
    \end{subfigure}%
    \begin{subfigure}[c]{0.14\textwidth}
        \centering
        \includegraphics[width=\linewidth]{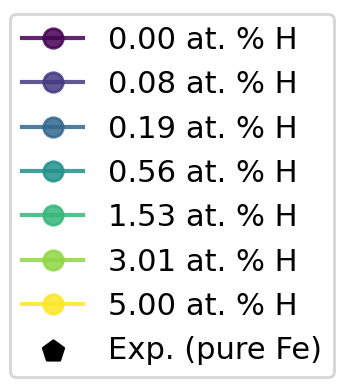}
        \vspace{.8cm}
    \end{subfigure}
\caption{The \aFe-H elastic constants for the tabGAP model as a function of both temperature and H concentration. Experimental results for pure \aFe are from Ref. \cite{adamsElasticConstantsMonocrystal2006}.}
\label{fig:Econsts_T}
\end{figure*}

\subsection{Influence of H on screw dislocation mobility}\label{sec:screw_mobility_results}

\subsubsection{CI-NEB kink-pair nucleation enthalpy}

\begin{figure*}
    \centering
    \subcaptionsetup[figure]{margin={0.10\textwidth, 0pt}}
    \begin{subfigure}[c]{.45\textwidth}
        \includegraphics[width=\linewidth]{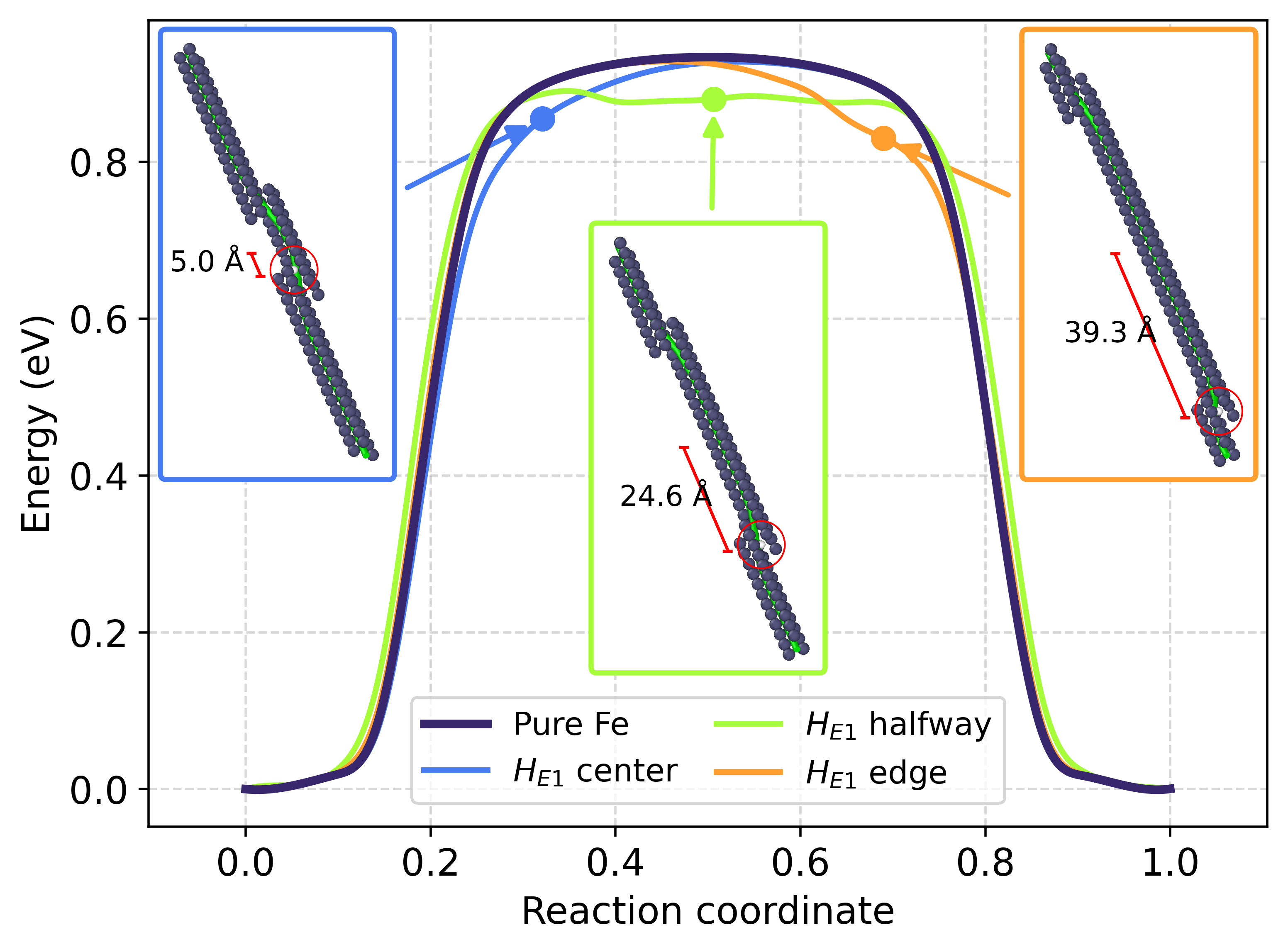}
        \caption{}
        \label{fig:screw_kink_H_effect}
    \end{subfigure}
    \subcaptionsetup[figure]{margin={0.10\textwidth, 0pt}}
    \begin{subfigure}[c]{.45\textwidth}
        \includegraphics[width=\linewidth]{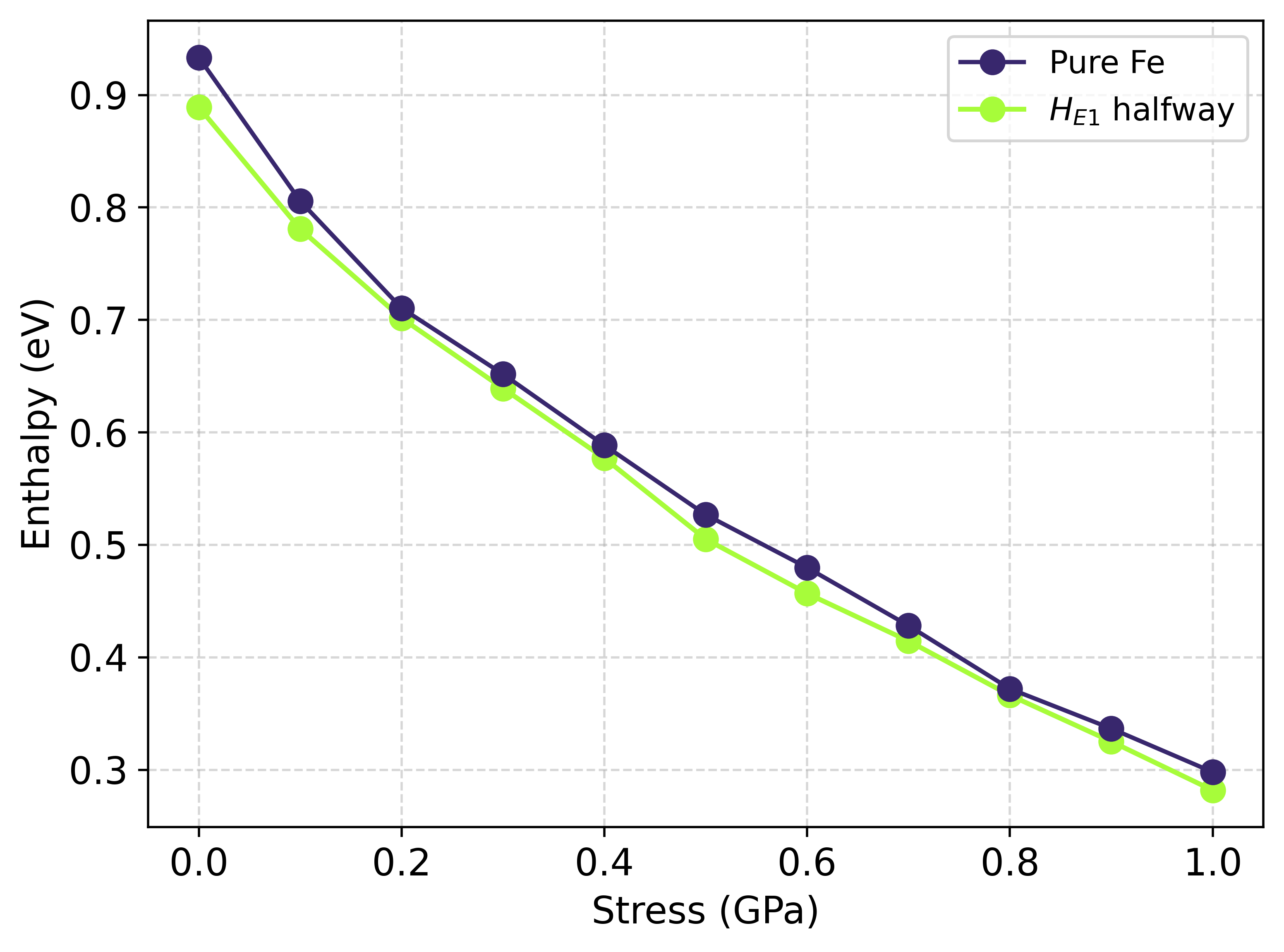}
        \caption{}
        \label{fig:screw_kink_stress}
    \end{subfigure}
    \caption{(\subref{fig:screw_kink_H_effect}) The zero-stress kink-pair nucleation energy with a $\text{H}_{E1}$ atom at three different distances from the kink-pair nucleation site. The dislocation and its core atoms extracted with DXA \cite{stukowskiVisualizationAnalysisAtomistic2009, stukowskiExtractingDislocationsNondislocation2010, stukowskiAutomatedIdentificationIndexing2012} alongside the solute hydrogen atoms (circled in red) are visualized in the inset images at three different images along the nucleation path.  (\subref{fig:screw_kink_stress}) The dependence of the kink-pair nucleation enthalpy on the applied shear stress with a $\text{H}_{E1}$ atom at the halfway point of the kink-pair nucleation path.}
    \label{fig:screw_kink_barriers}
\end{figure*}

The effect of H on the kink-pair nucleation enthalpy ($\Delta H_{kp}$) is shown in Fig. \ref{fig:screw_kink_barriers}. Overall, the presence of a H atom in the E1-site causes a reduction in the energy barrier, but only when the distance between the H atom and the kink is small. This is evident from Fig. \ref{fig:screw_kink_H_effect}, where the zero-stress kink-pair nucleation NEB profile is shown for three different initial kink-pair-H distances, $5.0$, $24.6$, and $\SI{39.3}{\angs}$. When the H atom is located close to the kink-pair nucleation site, the energy barrier is reduced in the first half of the nucleation path. However, after the kink-pair has passed the H atom, the energy barrier increases to match the H-free case. When the H atom is located at the halfway point of the nucleation path, a flatter reduction of $\sim\SI{50}{meV}$ is observed at the high plateau of the energy barrier. The magnitude of the reduction roughly matches between the three cases when comparing to the pure Fe profile at the reaction coordinates where the H atom is close to the kink. Previous works report reductions of $90$ \cite{mengGeneralpurposeNeuralNetwork2021} and $110 \si{meV}$ \cite{itakuraEffectHydrogenAtoms2013a}.

In Fig. \ref{fig:screw_kink_stress}, we show the dependence of the kink-pair nucleation enthalpy, calculated with Eq. \ref{eq:neb_barrier} on the applied shear stress for the case where the H atom is located at the halfway point of the nucleation path. A slight reduction in the enthalpy, between $\SI{6}{meV}$ and $\SI{44}{meV}$, is observed across the entire range of applied shear stresses, with no obvious dependence on the applied stress. This treatment using Eq. \ref{eq:neb_barrier} is, in a certain way, an underestimate of the true effect. This is because of the H atom effect being localized to the CI-NEB images where the kink is close to the H atom, as shown in Fig. \ref{fig:screw_kink_H_effect}. This means that the calculated enthalpy is only affected if the H atom is close to the kink at the image that is used in Eq. \ref{eq:neb_barrier} to calculate the enthalpy. If we instead define the H-related reduction in the kink-pair nucleation enthalpy as the lowering of the CI-NEB profile at the image where the H atom meets the kink, we get a larger reduction of $\myqtyrange{37}{115}{meV}$ across the applied shear stresses. We relate this reduction in $\Delta H_{kp}$ back to Fig. \ref{fig:screw_H_core_displacements} which shows that the interstitial H atom creates an initial kink-pair-like displacement of the core Fe atoms toward the H atom.

\subsubsection{Mobility at finite temperature}

\begin{figure}
\centering
    \includegraphics[width=\linewidth]{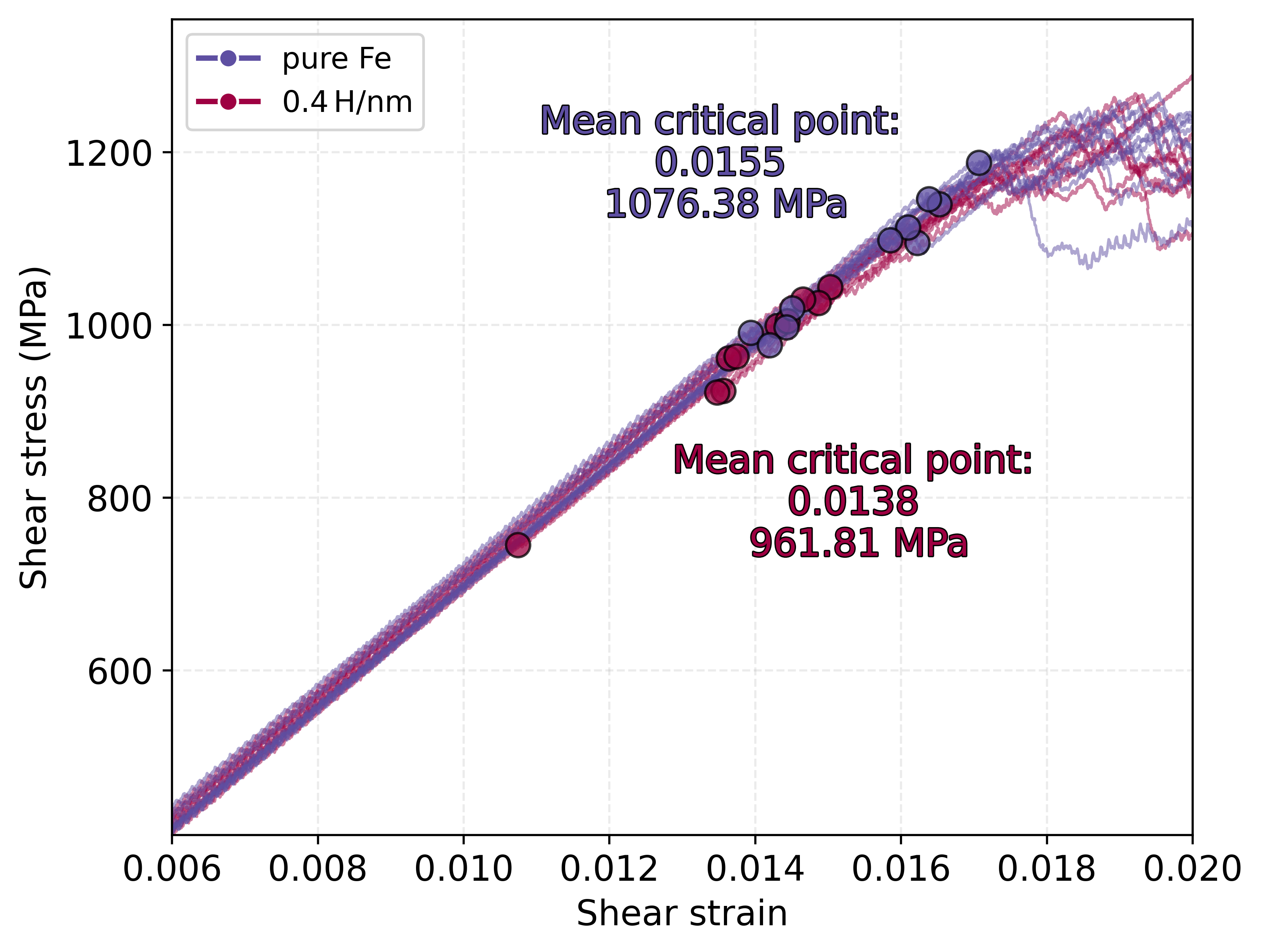}
    \caption{Stress-strain curves for constant strain rate simulations of a $40b$-long screw dislocation with (red) and without (blue) $\SI{0.4}{nm^{-1}}$ of H atoms at $\SI{300}{K}$. The first observed dislocation slip event is annotated for each of the $20$ independent trajectories simulations, $10$ for pure Fe and $10$ with H.}
    \label{fig:screw_stress_strain}
\end{figure}

We also simulated the H-effect on the kink-pair nucleation at \SI{300}{K} by employing the constant strain-rate scheme mentioned in Sec. \ref{sec:screw-methods}. In Fig. \ref{fig:screw_stress_strain} we show the stress-strain curves for the $20$ independent trajectories we calculated, with $10$ being H-free and $10$ with $\sim\SI{0.4}{nm^{-1}}$ of H inserted into the dislocation core. We observe a meaningful reduction in the critical shear stress ($\tau_c$) of dislocation motion in the H-containing simulations, with the mean $\tau_c$ being $\SI{962}{MPa}$ and $\SI{1076}{MPa}$ with and without solute H, respectively. Using the two-sided Welch's unequal variances t-test \cite{ruxtonUnequalVarianceTtest2006} on the two groups of $N=10$ $\tau_c$ values, we get a p-value of $0.0056$, implying a statistically significant result.

One of the H-containing simulations showed a markedly reduced $\tau_c$ of $\SI{745}{MPa}$, for which we provide snapshots in the Appendix \ref{app:screw_mobility_finT} to illustrate the observed dislocation slip event. The kink-pair is observed to nucleate nearby three H atoms (Fig. \ref{fig:screw_finT_ss2}), which are at the time located in the E1/E2 basin toward the \hkl[121] slip direction. The kinks then rapidly propagate along the dislocation line in opposite directions, and the slip event is completed within $\SI{20}{ps}$ (Figs. \ref{fig:screw_finT_ss3}-\ref{fig:screw_finT_ss6}). This slip event is compatible with the reduction in the kink-pair nucleation enthalpy caused by the presence of H in the \hkl[121]-E1-site shown in Fig. \ref{fig:screw_kink_stress}. As in the finite-temperature simulations the interstitial H atoms are free to move around in the dislocation structure, they don't occupy only the \hkl[121]-E1-site, and in our simulations they frequently explored all three triaxially symmetric E1/E2 basins around the dislocation core. Due to the direction of the applied shear strain, only the H-atoms in the \hkl[121]-E1/E2 basin would result in more frequent dislocation slip events, which is the effect we see in Fig. \ref{fig:screw_stress_strain}.

\section{Discussion}
In this section we further discuss the results we obtained in the screw dislocation mobility simulations using the developed Fe-H tabGAP.

As we saw in Sec. \ref{sec:screw_mobility_results}, the presented tabGAP model predicts the presence of H atoms near a $1/2\hkl<111>\hkl{110}$ screw dislocation core to significantly increase the mobility of the dislocation, both in \SI{0}{K} CI-NEB calculations of the kink-pair nucleation enthalpy, and in $\SI{300}{K}$ constant strain rate simulations with $\SI{0.4}{nm^{-1}}$ of hydrogen. These findings support the HELP mechanism of HE, which also has strong experimental evidence in the form of \textit{in situ} transmission electron microscopy (TEM) studies, among other methods \cite{birnbaumHydrogenenhancedLocalizedPlasticity1994}. Some previous MD studies have also reported results that align with the HELP mechanism \cite{wenAtomisticSimulationsEffect2003, mengGeneralpurposeNeuralNetwork2021, kumarEffectHydrogenPlasticity2023, huangQuantitativeTestsRevealing2023a, kimHydrogenCanBoth2025}, mostly based on $\SI{0}{K}$ NEB or consecutive energy minimization calculations. Reports of finite-temperature MD results demonstrating the HELP mechanism are rare, with the only study, to the best of our knowledge, being the recent work by Huang \emph{et al.} \cite{huangQuantitativeTestsRevealing2023a} showing a reduction in the critical shear stress of screw dislocation glide from $\SI{32}{MPa}$ to $\SI{22}{MPa}$ also with $\SI{0.4}{nm^{-1}}$ of H at $\SI{300}{K}$. The reported critical shear stress is an order of magnitude lower than the stresses observed in our simulations, and it is reported to qualitatively match with experimental results \cite{huangQuantitativeTestsRevealing2023a}. The large discrepancy between our stresses and those of Huang \emph{et al.} is likely explained by differences between the used IAPs, with empirical EAM IAPs being fit to reproduce experimentally found \aFe properties. For relating the results of our simulations to experimental conditions, differences between properties such as the short dislocation length and high strain rate would need to be taken into account, but the full treatment of these effects is outside the scope of the present study. We refer the reader to Refs. \cite{domainSimulationScrewDislocation2005a,alleraCarboninducedStrengtheningBcc2022} for further information on these treatments. Furthermore, in the CI-NEB calculations of the kink-pair nucleation enthalpy we only explored the effect of H-atoms in the E1-site in the direction of dislocation slip, and to fully understand phenomena such the softening-hardening transition \cite{itakuraEffectHydrogenAtoms2013a}, all H-atom binding sites around the screw dislocation should be considered in future studies. In the finite-temperature simulations we only tracked the first observed dislocation slip event, which approximates the condition where dislocation velocity is much slower than H-diffusion. These are both highly dependent on temperature, and as such future studies should simulate multiple temperatures and track multiple consecutive slip events to get quantitative results on the softening/hardening regimes of the H-effect on screw dislocation mobility.

\section{Conclusions}\label{sec:concs}

In this work, we present a new machine-learned interatomic potential (tabGAP) for the \aFe-H system based on a Density Functional Theory (DFT) generated dataset of atomic structures, forces, and energies. We have quantified the accuracy of the tabGAP model via benchmarking it against DFT, and many of the other currently published \aFe-H potentials in H-point defect energetics, H-vacancy interaction, H-dislocation interaction, H-H interaction in the \aFe matrix, and the effect of H on the \aFe elastic constants. The presented tabGAP model strikes a new balance between accuracy and performance, offering near-DFT accuracy in the benchmarked properties, while having a computational cost close to that of the classical EAM potentials.

Furthermore, we applied the developed tabGAP model in molecular dynamics simulations of $1/2\hkl<111>\hkl{110}$ screw dislocation mobility using both $\SI{0}{K}$ CI-NEB calculations and $\SI{300}{K}$ constant strain rate dynamics. In these simulations, we observed the following:
\begin{itemize}
    \item The Fe-H tabGAP model gives the kink-pair nucleation as the screw dislocation glide mechanism at finite temperature, with good agreement with previous models on the nucleation barrier at $\SI{0}{K}$.
    \item H trapped near the screw dislocation core was found to enhance the mobility of the screw dislocation by introducing an energetically preferential site for kink-pair nucleation.
\end{itemize}
The above prediction made with the tabGAP model is in line with previous computational studies, and it supports the hydrogen-enhanced localized plasticity (HELP) mechanism of hydrogen embrittlement in the case of screw dislocations. We attribute this to the trapped hydrogen's effect on the core structure of the screw dislocation, where the H atom creates an initial kink-pair-like displacement of the core Fe atoms toward the H atom.

As the presented tabGAP model and its training dataset have been demonstrated to offer excellent performance and accuracy for the \aFe-H system, it lays the groundwork for future augmentations, with plans to extend the model's capabilities to H effects in grain boundaries and free surfaces.



\section*{Acknowledgments}

This work was funded by the European Union under Grant Agreement n° 101135374. Views and opinions expressed are however those of the author(s) only and do not necessarily reflect those of the European Union or the European Health and Digital Executive Agency (HADEA). Neither the European Union nor the granting authority can be held responsible for them. A.L.C. acknowledges the AI4S fellowship within the “Generaci{ó}n D” initiative by Red.es, Ministerio para la Transformaci{ó}n Digital y de la Funci{ó}n P{ú}blica, for talent attraction (C005/24-ED CV1), funded by NextGenerationEU through the Recovery, Transformation and Resilience Plans (PRTR). The authors would like to thank the CSC-IT Center for Science and the Finnish Computing Competence Infrastructure (FCCI) for providing the computational and data storage resources for this work.



\renewcommand\thefigure{\thesection.\arabic{figure}}
\setcounter{figure}{0}

\appendix
\section*{Appendices}
\section{Binding sites of H to a screw dislocation}\label{app:Eb_H_screw}%

\begin{figure}
    \centering
    \setkeys{Gin}{keepaspectratio}
    \subcaptionsetup[figure]{margin={-0.18\textwidth, 0pt}}
    \begin{subfigure}{.45\textwidth}
      \centering
      \includegraphics[width=\linewidth]{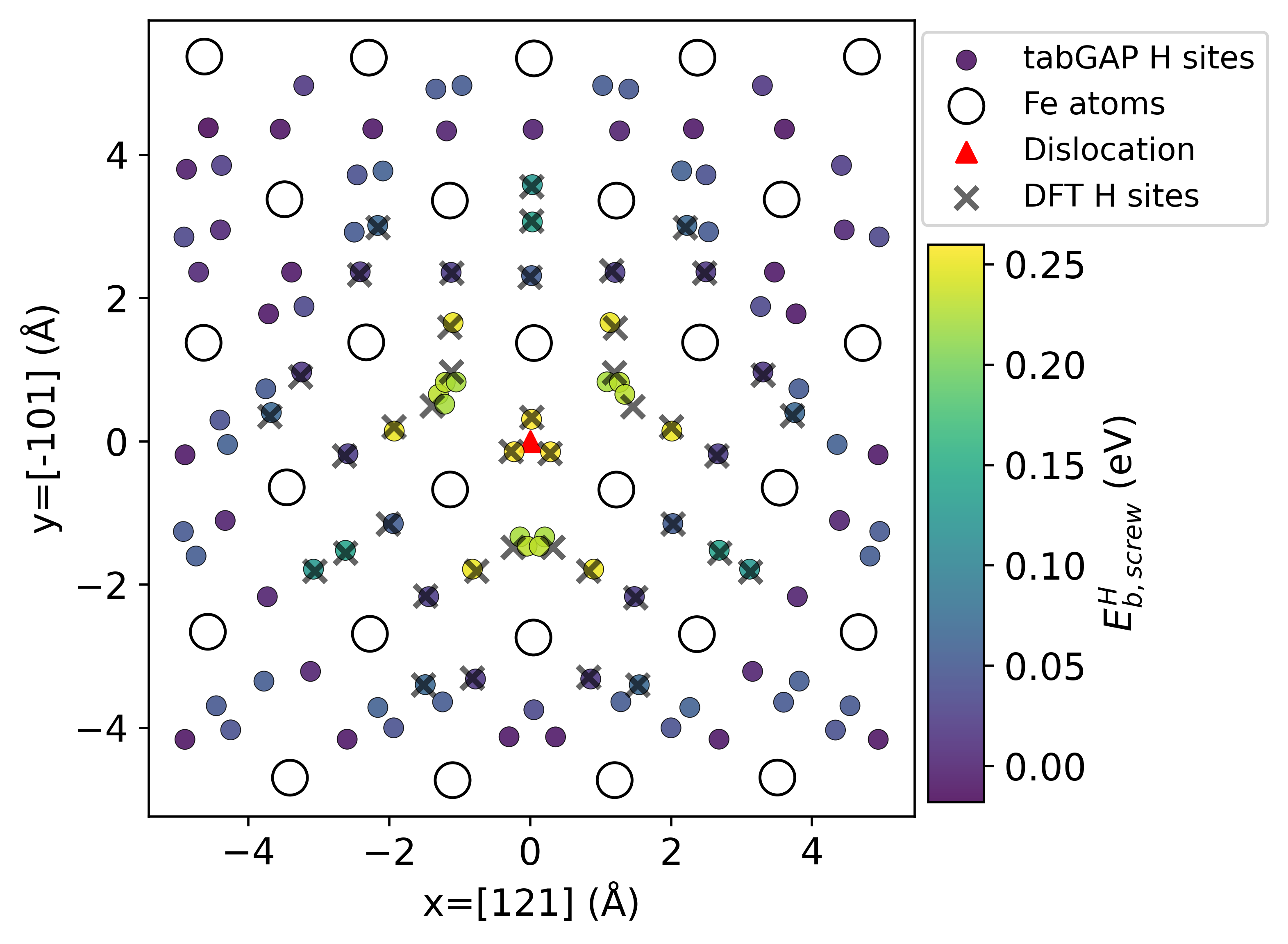}
      \caption{}
      \label{appfig:H_Eb_screw_tabGAP}
    \end{subfigure}%
    \subcaptionsetup[figure]{margin={-0.21\textwidth, 0pt}}
    \begin{subfigure}{.45\textwidth}
      \centering
      \includegraphics[width=\linewidth]{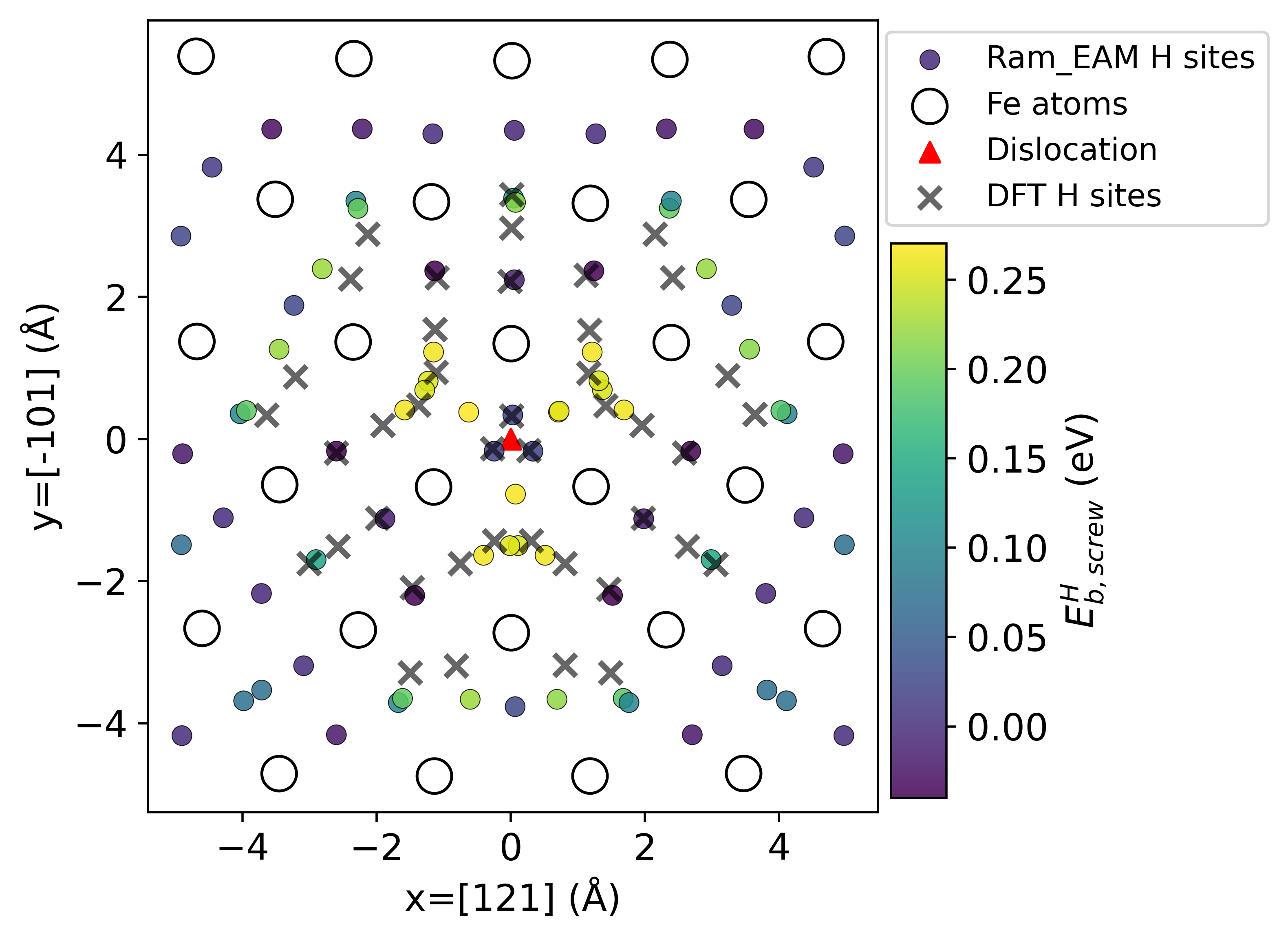}
      \caption{}
      \label{appfig:H_Eb_screw_ram}
    \end{subfigure}
\caption{The binding sites and energies for a H atom near a $1/2\hkl<111>\hkl{110}$ soft-core screw dislocation. (\subref{appfig:H_Eb_screw_tabGAP}) The tabGAP model results, and in (\subref{appfig:H_Eb_screw_ram}) the Ram\_EAM results are shown. The binding sites given by DFT (reproduced from Fig. 6a in Ref. \cite{itakuraEffectHydrogenAtoms2013a}) are marked on top with gray crosses.}
\label{appfig:H_Eb_screw}
\end{figure}

The H binding site positions around a $1/2\hkl<111>\hkl{110}$ easy-core screw dislocation given by the presented Fe-H tabGAP, and the Ram\_EAM models are compared to the sites reported by Itakura \emph{et al.} \cite{itakuraEffectHydrogenAtoms2013a}. As the \aFe lattice parameters of these models differ from each other, the H binding site positions were scaled by matching the positions of the three central dislocation core Fe atoms. From Fig. \ref{appfig:H_Eb_screw} we can see excellent agreement in the binding site positions between the tabGAP and DFT, while there are larger differences between DFT and Ram\_EAM. The Ram\_EAM also reports three additional high-$E_b$ sites located next to the screw dislocation core which don't exist in the DFT and tabGAP results.

\section{Screw dislocation slip at finite temperature}\label{app:screw_mobility_finT}

\begin{figure*}
    \centering
    \setlength{\fboxsep}{0pt}
    \setlength{\fboxrule}{0.4pt}
    \setkeys{Gin}{keepaspectratio}
    \begin{subfigure}[c]{0.70\textwidth}
        \centering
        \begin{subfigure}{.46\linewidth}
            \centering
            \fbox{\includegraphics[width=\dimexpr\linewidth-2\fboxrule\relax]{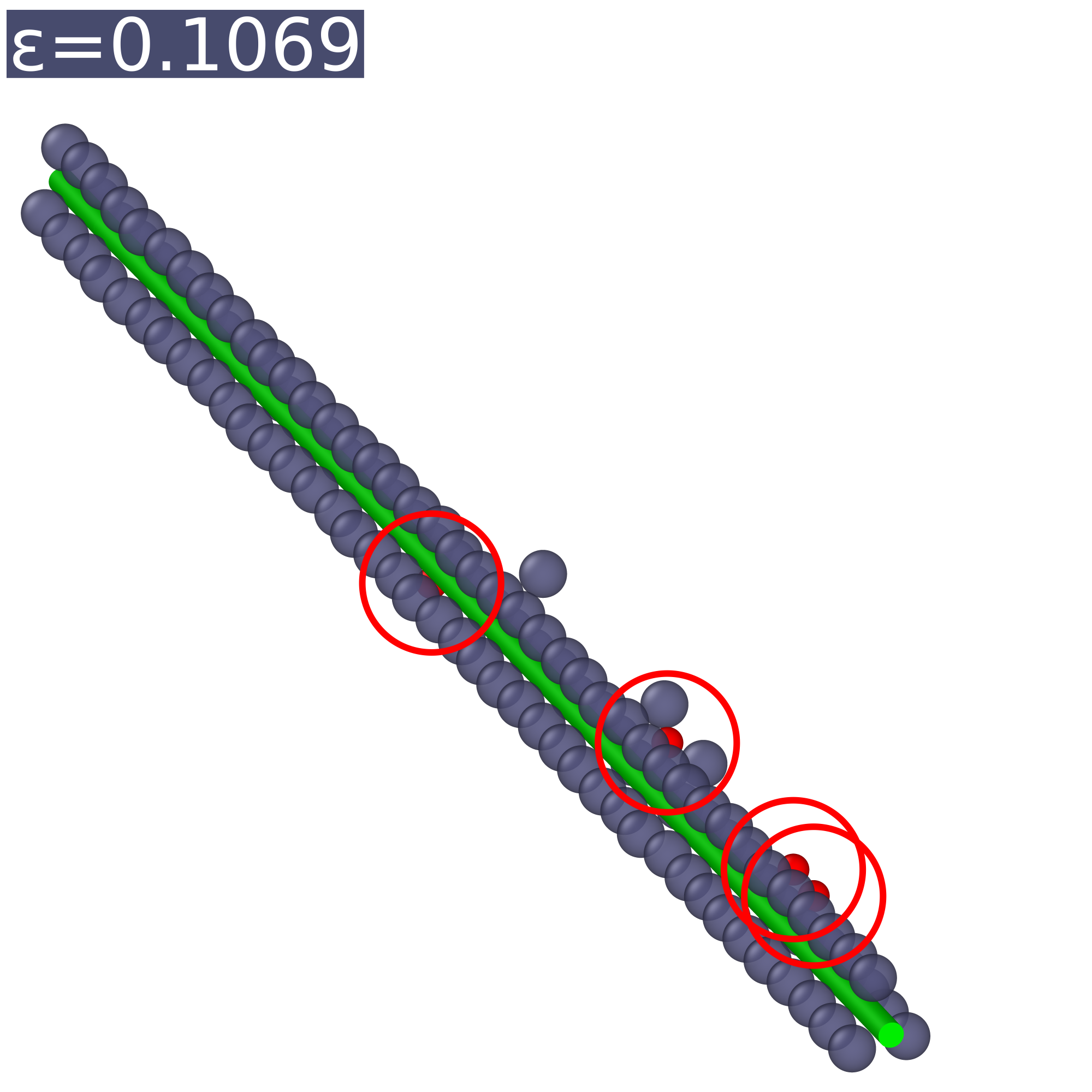}}
            \caption{}
            \label{fig:screw_finT_ss1}
        \end{subfigure}%
        \begin{subfigure}{.46\linewidth}
            \centering
            \fbox{\includegraphics[width=\dimexpr\linewidth-2\fboxrule\relax]{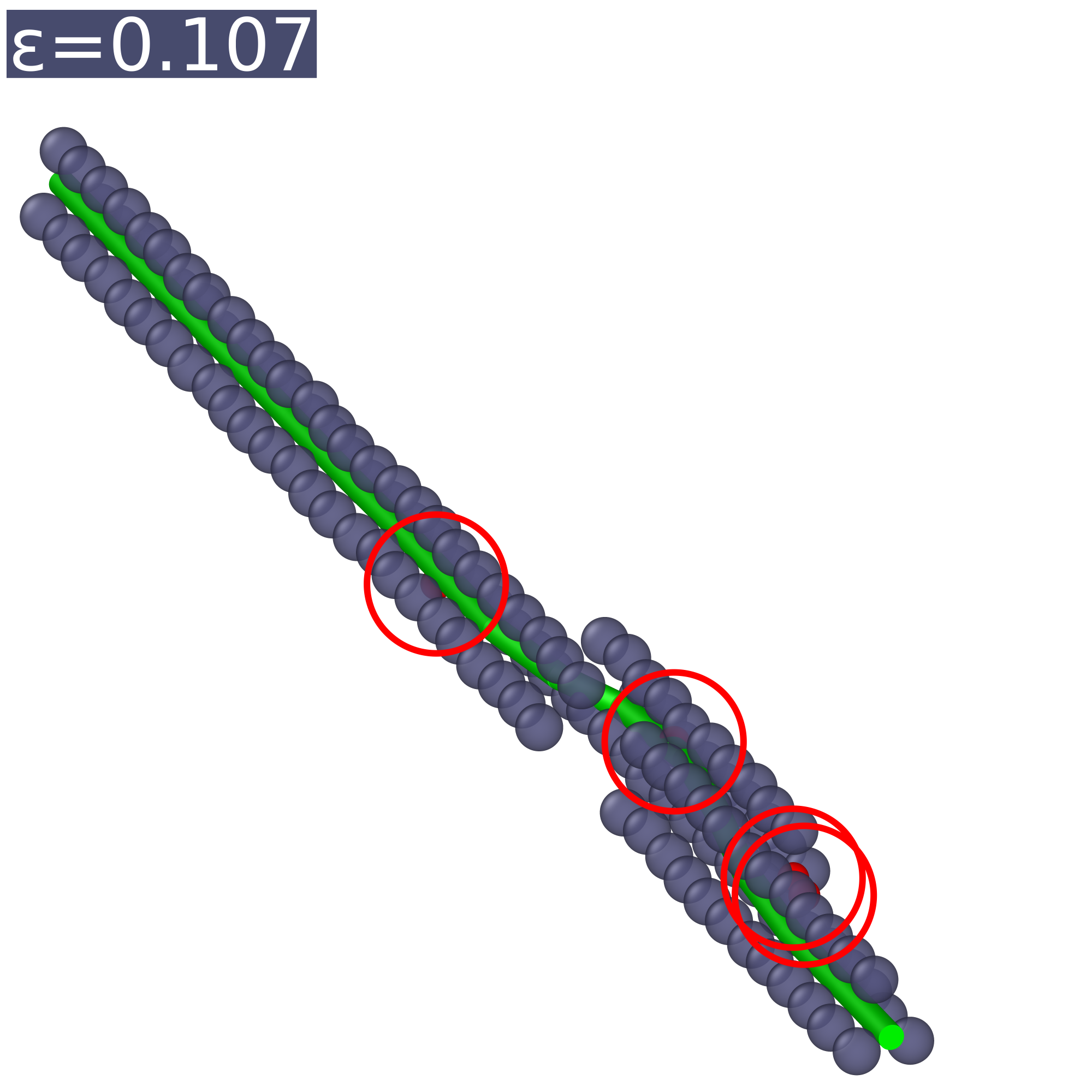}}
            \caption{}
            \label{fig:screw_finT_ss2}
        \end{subfigure}

        \begin{subfigure}{.46\linewidth}
            \centering
            \fbox{\includegraphics[width=\dimexpr\linewidth-2\fboxrule\relax]{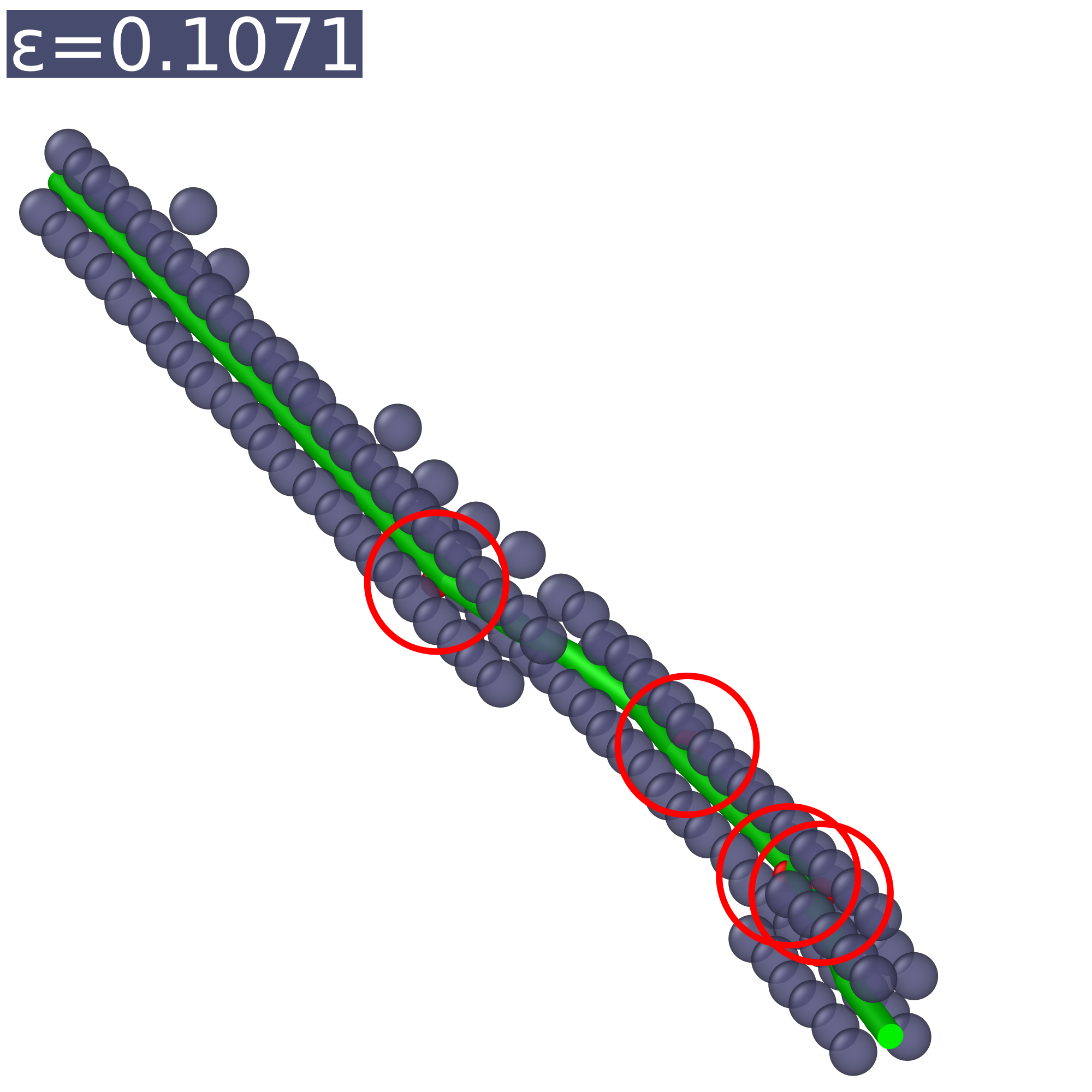}}
            \caption{}
            \label{fig:screw_finT_ss3}
        \end{subfigure}%
        \begin{subfigure}{.46\linewidth}
            \centering
            \fbox{\includegraphics[width=\dimexpr\linewidth-2\fboxrule\relax]{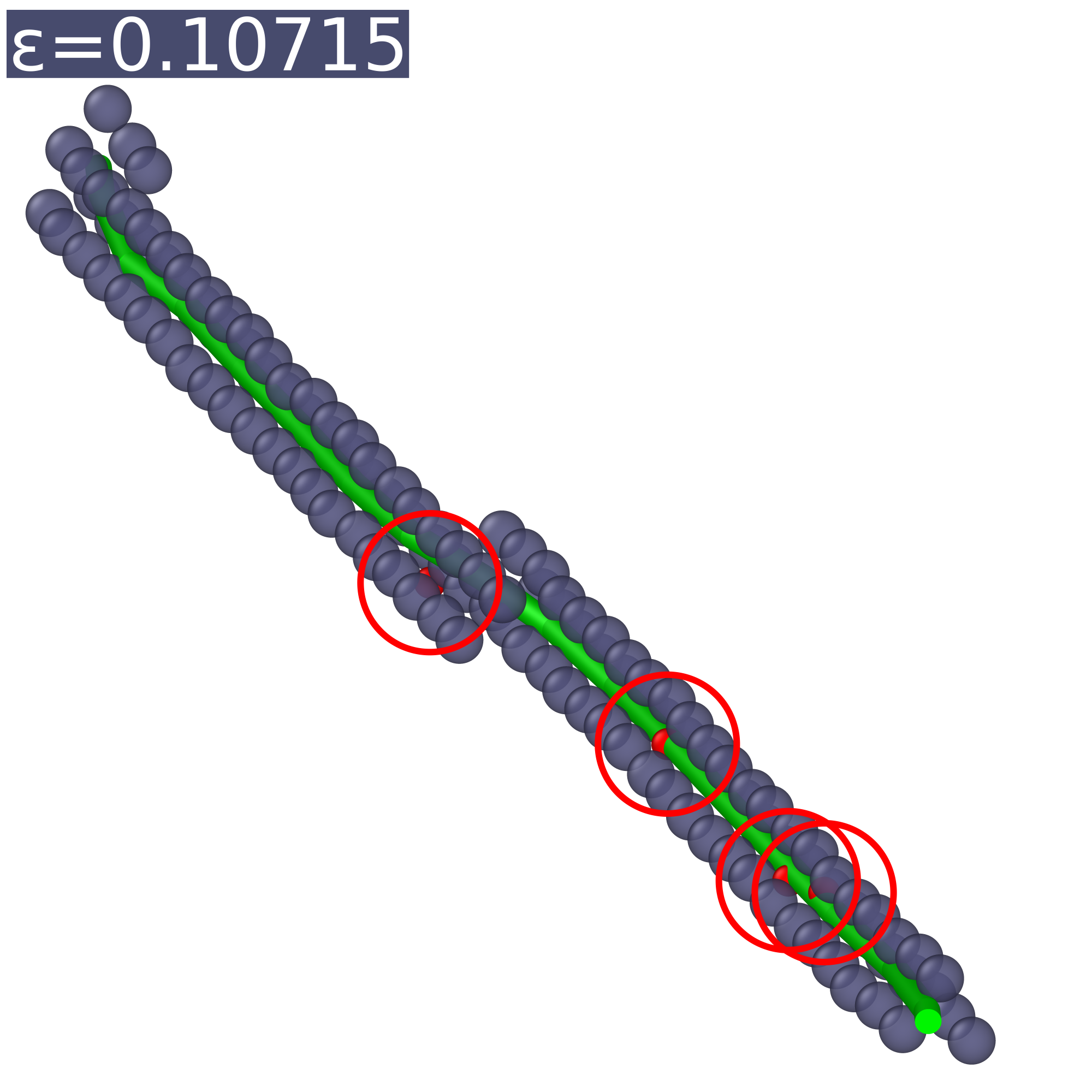}}
            \caption{}
            \label{fig:screw_finT_ss4}
        \end{subfigure}

        \begin{subfigure}{.46\linewidth}
            \centering
            \fbox{\includegraphics[width=\dimexpr\linewidth-2\fboxrule\relax]{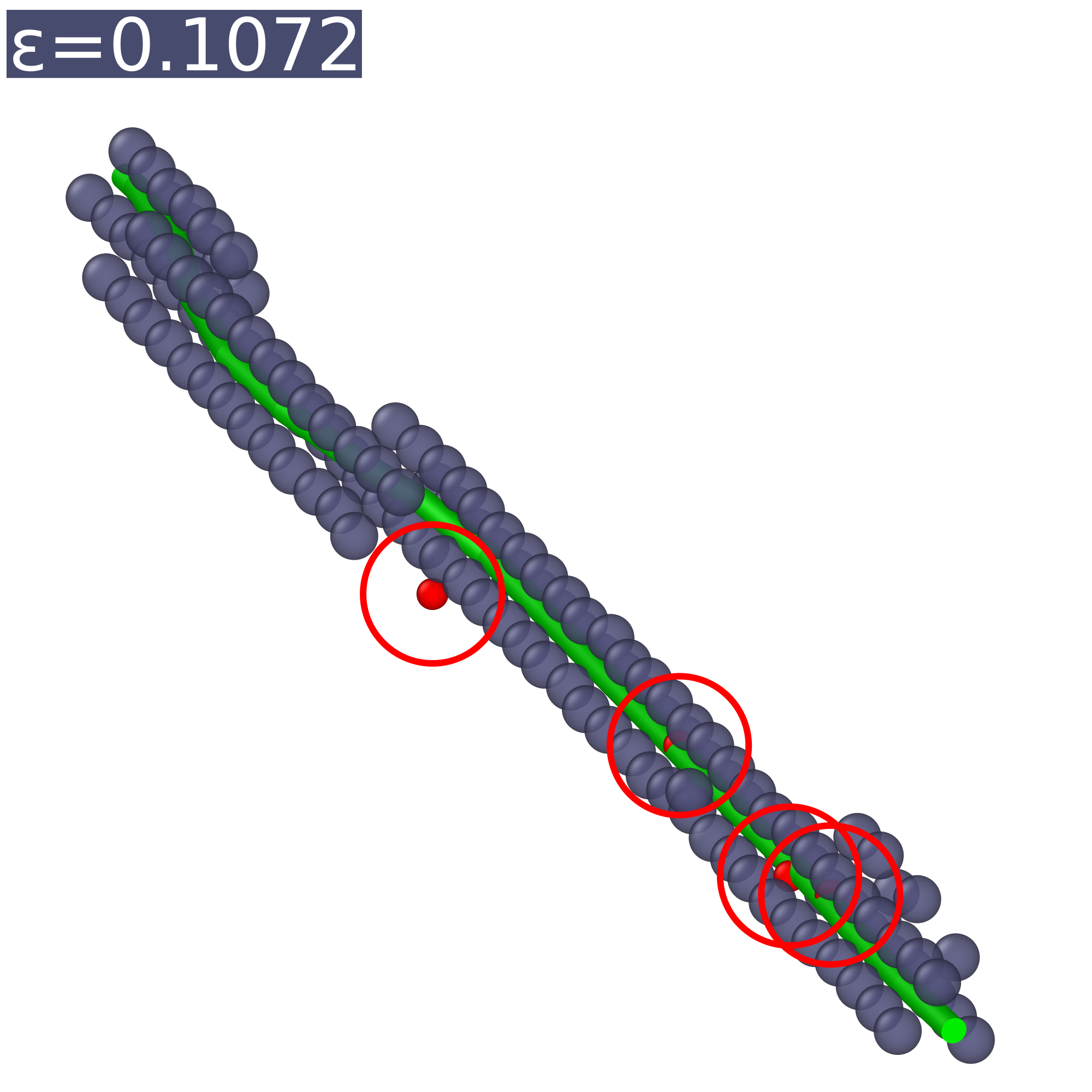}}
            \caption{}
            \label{fig:screw_finT_ss5}
        \end{subfigure}%
        \begin{subfigure}{.46\linewidth}
            \centering
            \fbox{\includegraphics[width=\dimexpr\linewidth-2\fboxrule\relax]{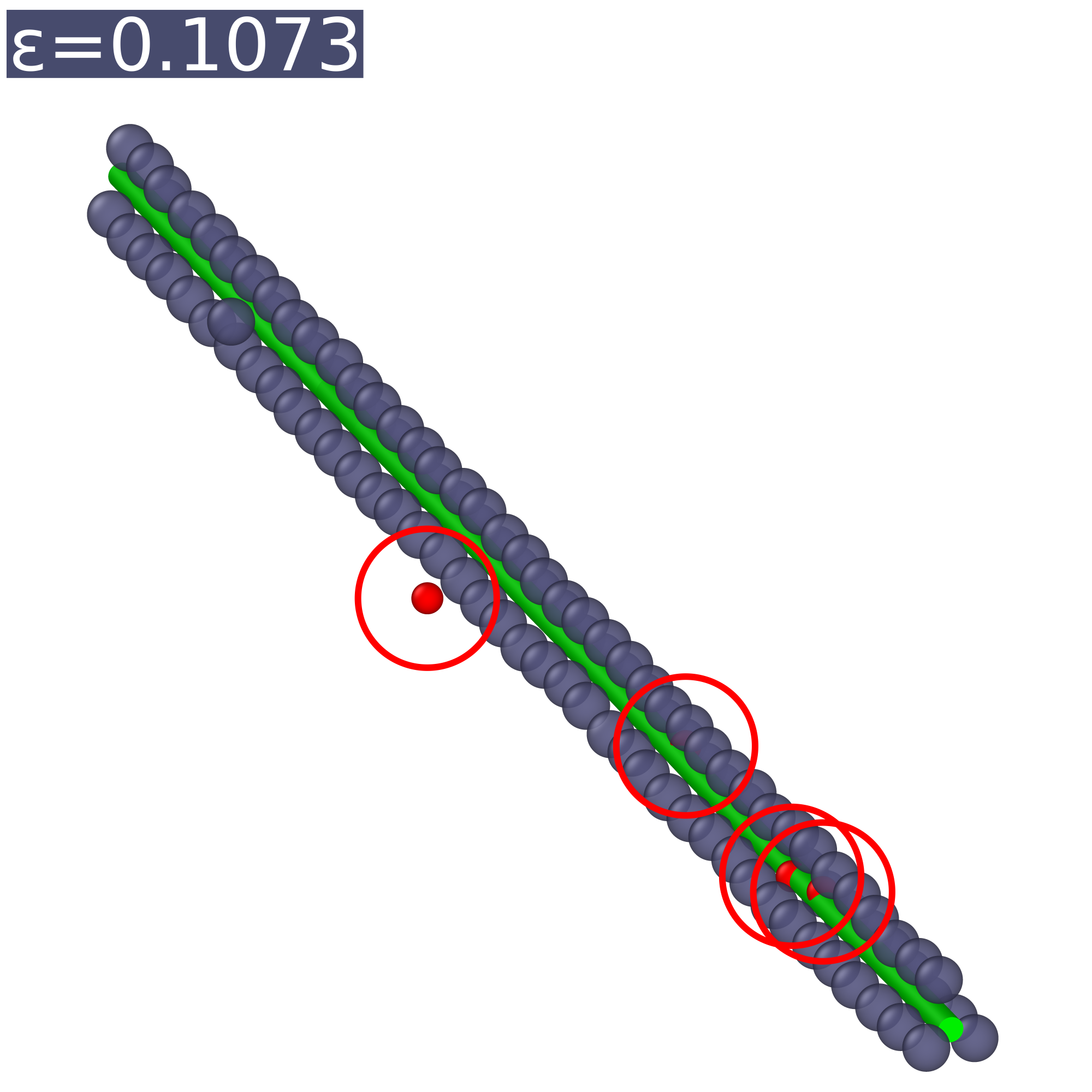}}
            \caption{}
            \label{fig:screw_finT_ss6}
        \end{subfigure}

    \end{subfigure}\hspace{-0cm}%
    \begin{subfigure}[c]{0.20\textwidth}
        \centering
        \includegraphics[width=\linewidth]{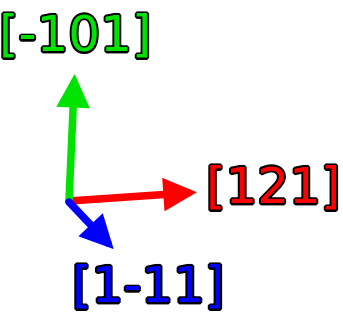}
    \end{subfigure}
\caption{Snapshots of the dislocation slip event in the H-containing simulation with the reduced critical shear stress of $\SI{745}{MPa}$. The Fe and H atoms are indicated by dark blue and red spheres, respectively, and the H atoms are circled for further clarity. The $1/2\hkl<111>\hkl{110}$ screw dislocation is indicated by the green line.  (a) The initial configuration just before the slip event, with the screw dislocation laying straight. (b) The kink-pair nucleates near three of the four H atoms. (c-f) The propagation of the kinks along the dislocation line.}
\label{fig:screw_finT_snapshots}
\end{figure*}

\FloatBarrier

\bibliography{bibliography}


\end{document}


\title{Supplemental information for \\
Fast machine learned \aFe-H interatomic potential for hydrogen embrittlement}

\author{Eetu Makkonen}
    \email[Contact author: ]{eetu.makkonen@helsinki.fi}
    \affiliation{Department of Physics, University of Helsinki, Finland}

\author{Alvaro Lopez-Cazalilla}
    \affiliation{Barcelona Supercomputing Center, Pla{ç}a Eusebi G{ü}ell 1-3, 08034 Barcelona, Spain}

\author{Flyura Djurabekova}
    \affiliation{Department of Physics, University of Helsinki, Finland}

\maketitle

\renewcommand\thesection{S.\arabic{section}}
\renewcommand\thefigure{\thesection.\arabic{figure}}
\renewcommand\thetable{\thesection.\arabic{table}}
\setcounter{figure}{0}

\section{Details on the tabGAP development}

In the Fe-H tabGAP development process, a detailed regularization parameter optimization was performed to improve the properties of the resulting IAP \cite{bartokMachineLearningGeneralPurpose2018a}. The default regularization $\sigma$ was set to $\SI{1}{meV}$, $\SI{10}{meV/\angs}$, and $\SI{100}{meV/\angs^3}$, for the energy, forces, and virials, respectively. The Tab. \ref{supp_tab:reg_params} lists all the structure categories, their associated regularization parameters, and the training and testing RMSEs for each category. The regularization parameters not specifically mentioned in this document were inherited from Ref. \cite{byggmastarMultiscaleMachinelearningInteratomic2022}. Also note that only the virial data in the pure Fe structure category \texttt{bcc\_distorted} was used in fitting, following the methodology in Ref. \cite{byggmastarMultiscaleMachinelearningInteratomic2022}.

The results of the regularization optimization resulted in the following details: lower regularization of $0.5\sigma$ for the relatively simple structure categories of one to three neighboring interstitial H atoms (\texttt{TT\_saddle\_H, tet\_1H, tet\_2H, tet\_3H, oct\_1H, oct\_2H, vac\_1H}) improved H-H interaction accuracy, H solution energy, and migration barriers of H. Higher regularization of $2\sigma$ for the high local H-concentration structures \texttt{vac\_14H} and \texttt{vac\_H\_clusters} improved training and testing errors for said structures. A strong regularization of $10\sigma$ was used for the \texttt{Fe-H\_dimer} category for customary reasons. 

\begin{figure}
\centering
    \includegraphics[width=0.48\textwidth]{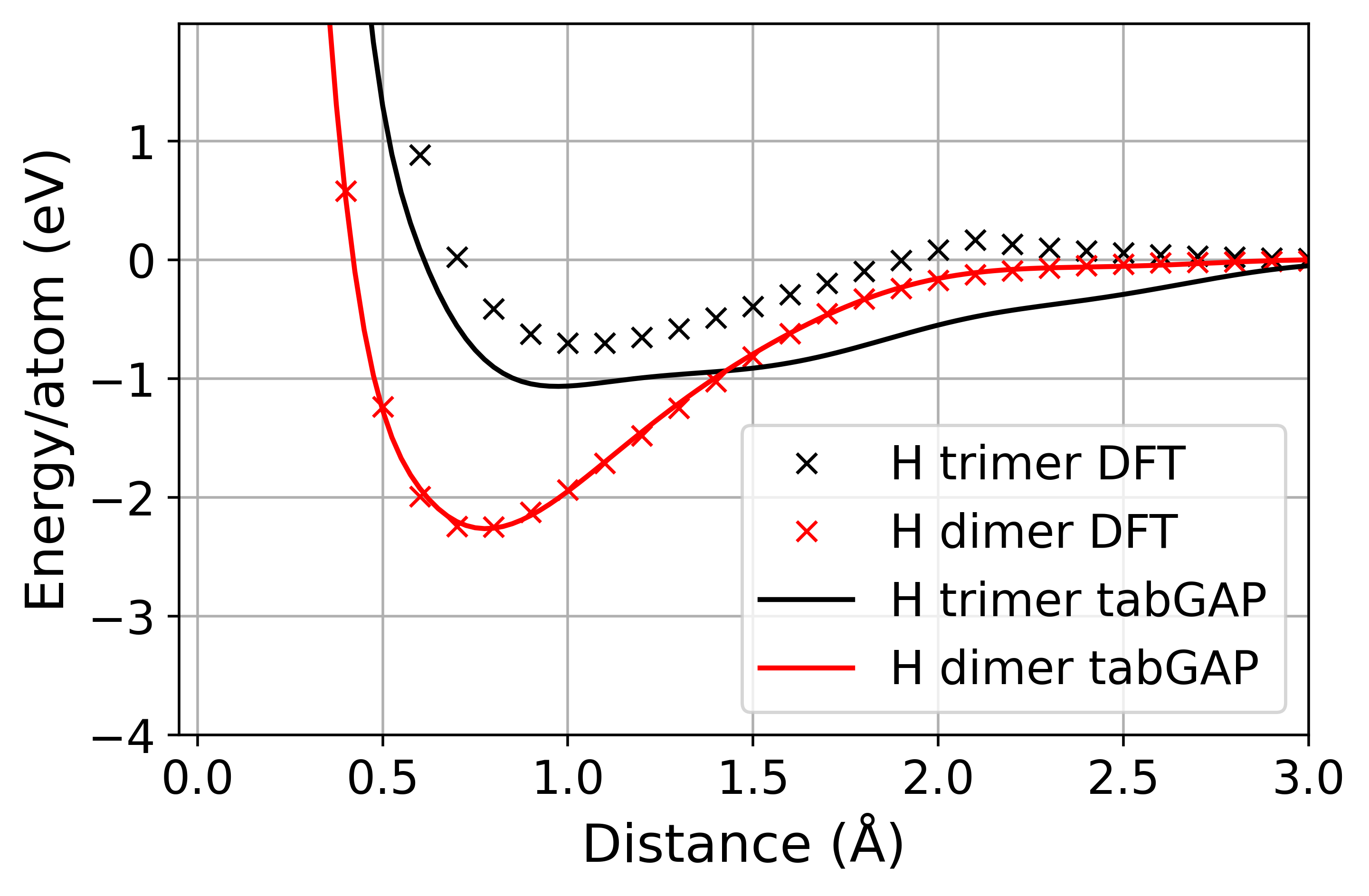}
\caption{The energy curves of the H-dimer and the symmetric H-trimer structures as a function of the H-H distance calculated with the Fe-H tabGAP and DFT.}
\label{supp_fig:H_trimer}
\end{figure}

To accurately capture the difference in H-H interaction between the vacuum and \aFe environments, both free H-dimer and H-trimer structures were included in the training structure dataset of the Fe-H tabGAP. For the tabGAP to predict accurate solution energies of H in \aFe, the energy of the free \ce{H2} molecule was sought to be replicated accurately. This was achieved by using the default regularization. The same regularization was found to be unsuitable for the H-trimer structures, as both training and testing errors increased substantially for even unrelated structures in this case. Instead, using a regularization of $10\sigma$ for the H-trimers resulted in a smooth energy surface of the symmetric H-trimer configuration. The energy of this is plotted as a function of the H-H distance in Fig. \ref{supp_fig:H_trimer}. The tabGAP successfully predicts the less stable nature of the H-trimer configurations compared to the H-dimer.

\begin{table*}
\centering
\caption{Summary of the Fe-H structure dataset, and Fe-H tabGAP training properties. The regularization parameters $\sigma$ are given as multiples of the default value, and the number of structures contained in the training and testing structure sets. The training and testing energy, force, and virial RMSEs are given for the Fe-H tabGAP. The list index refers to the entry number for the structure categories listed in Sec. II B of the main text.}
    \begin{threeparttable}
    \begin{tabular}{@{}ccccccccccc@{}}
    \toprule
    \textbf{Structure category name} & List index & $N_{train}$ & $N_{test}$ & $\sigma$ & $E_{train}$ & $E_{test}$ & $F_{train}$ & $F_{test}$ & $S_{train}$ & $S_{test}$\\
    \midrule
    \centering\textbf{Fe-H categories} & & & & & \multicolumn{2}{c}{$\si{meV/atom}$} & \multicolumn{2}{c}{$\si{eV/\angs}$} & \multicolumn{2}{c}{$\si{eV}$} \\
    \midrule
    	isolated\_atom & - & 2 & 0 & $0.1\sigma$ & 1.004 & - & 0.000 & - & - & - \\
    	TT\_saddle\_H & 7. & 37 & 4 & $0.5\sigma$ & 0.299 & 0.365 & 0.033 & 0.036 & 4.732 & 4.639 \\
    	tet\_1H & 1. & 57 & 6 & $0.5\sigma$ & 0.184 & 0.225 & 0.019 & 0.019 & 5.205 & 5.281 \\
    	tet\_2H & 1. & 78 & 8 & $0.5\sigma$ & 0.545 & 0.307 & 0.047 & 0.051 & 5.054 & 5.042 \\
    	tet\_3H & 1. & 33 & 3 & $0.5\sigma$ & 0.438 & 0.600 & 0.031 & 0.036 & 4.987 & 5.083 \\
    	oct\_1H & 6. & 19 & 2 & $0.5\sigma$ & 0.204 & 0.219 & 0.018 & 0.018 & 4.882 & 4.843 \\
    	oct\_2H & 6. & 46 & 4 & $0.5\sigma$ & 0.746 & 1.024 & 0.067 & 0.072 & 5.077 & 5.144 \\
    	vac\_1H & 4. & 82 & 8 & $0.5\sigma$ & 0.637 & 0.648 & 0.050 & 0.043 & 5.434 & 5.374 \\
    	H\_dimer & 8. & 38 & 0 & $1\sigma$ & 23.760 & - & 0.722 & - & - & - \\
    	SIA\_with\_1H & 5. & 11 & 1 & $1\sigma$ & 0.787 & 0.937 & 0.060 & 0.043 & 11.992 & 12.289 \\
    	SIA\_with\_3H & 5. & 11 & 1 & $1\sigma$ & 0.532 & 0.319 & 0.059 & 0.050 & 12.690 & 13.571 \\
    	bulk\_p10\_H & 1. & 96 & 10 & $1\sigma$ & 0.488 & 0.375 & 0.042 & 0.045 & 5.348 & 5.006 \\
    	bulk\_p15\_H & 1. & 46 & 4 & $1\sigma$ & 1.181 & 0.831 & 0.093 & 0.100 & 3.543 & 3.363 \\
    	bulk\_p20\_H & 1. & 96 & 10 & $1\sigma$ & 0.701 & 0.962 & 0.066 & 0.086 & 2.484 & 2.374 \\
    	distorted\_1H & 2. & 128 & 14 & $1\sigma$ & 1.152 & 1.394 & 0.037 & 0.041 & 5.280 & 5.643 \\
    	distorted\_6H & 2. & 130 & 14 & $1\sigma$ & 1.340 & 0.754 & 0.062 & 0.045 & 5.486 & 5.870 \\
    	int\_H & 1. & 10 & 1 & $1\sigma$ & 0.355 & 0.060 & 0.030 & 0.027 & - & - \\
    	short\_range\_2H & 4. & 12 & 1 & $1\sigma$ & 2.325 & 5.313 & 0.167 & 0.111 & - & - \\
    	short\_range\_3H & 4. & 28 & 2 & $1\sigma$ & 3.891 & 2.610 & 0.286 & 0.298 & - & - \\
    	vac\_2H & 3. & 82 & 8 & $1\sigma$ & 0.716 & 0.430 & 0.053 & 0.048 & 5.685 & 5.394 \\
    	vac\_3H & 3. & 19 & 2 & $1\sigma$ & 0.191 & 0.169 & 0.027 & 0.032 & 6.181 & 6.324 \\
    	vac\_6H & 3. & 75 & 8 & $1\sigma$ & 1.014 & 0.706 & 0.066 & 0.064 & 6.899 & 5.120 \\
    	vac\_14H & 3. & 63 & 6 & $2\sigma$ & 2.721 & 1.506 & 0.195 & 0.080 & 5.776 & 3.156 \\
    	vac\_H\_clusters & 3. & 55 & 5 & $2\sigma$ & 2.363 & 7.126 & 0.211 & 0.285 & 4.755 & 4.925 \\
    	Fe-H\_dimer & - & 35 & 0 & $10\sigma$ & 173.016 & - & 2.634 & - & - & - \\
    	H\_trimer & 8. & 198 & 0 & $10\sigma$ & 206.564 & - & 2.118 & - & - & - \\
    \midrule
    \centering\textbf{pure Fe categories} & & & & & & & & & \\
    \midrule
    	bcc\_distorted\tnote{*} & - & 487 & 0 & $1\sigma$ & 2.125 & - & 0.000 & - & 0.230 & - \\
    	di-sia & - & 19 & 0 & $1\sigma$ & 0.840 & - & 0.048 & - & 25.172 & - \\
    	di-vacancy & - & 35 & 0 & $1\sigma$ & 0.975 & - & 0.058 & - & 11.033 & - \\
    	edge & - & 9 & 0 & $1\sigma$ & 1.065 & - & 0.084 & - & 27.863 & - \\
    	multi-sia & - & 12 & 0 & $1\sigma$ & 1.192 & - & 0.085 & - & 24.518 & - \\
    	phonon & - & 50 & 0 & $1\sigma$ & 1.002 & - & 0.078 & - & 4.298 & - \\
    	screw & - & 10 & 0 & $1\sigma$ & 0.411 & - & 0.047 & - & 13.147 & - \\
    	short\_range & - & 48 & 0 & $1\sigma$ & 0.692 & - & 0.080 & - & 4.458 & - \\
    	sia & - & 25 & 0 & $1\sigma$ & 1.045 & - & 0.061 & - & 10.921 & - \\
    	tri-vacancy & - & 15 & 0 & $1\sigma$ & 1.105 & - & 0.064 & - & 10.265 & - \\
    	vac\_saddle & - & 11 & 0 & $1\sigma$ & 0.638 & - & 0.040 & - & 11.448 & - \\
    	vacancy & - & 28 & 0 & $1\sigma$ & 1.784 & - & 0.064 & - & 4.113 & - \\
    	gamma\_surface & - & 178 & 0 & $2\sigma$ & 3.133 & - & 0.123 & - & - & - \\
    	surf100 & - & 17 & 0 & $2\sigma$ & 2.728 & - & 0.134 & - & - & - \\
    	surf110 & - & 17 & 0 & $2\sigma$ & 14.966 & - & 0.059 & - & - & - \\
    	surf111 & - & 20 & 0 & $2\sigma$ & 5.105 & - & 0.120 & - & - & - \\
    	surf211 & - & 20 & 0 & $2\sigma$ & 4.187 & - & 0.120 & - & - & - \\
    	Fe\_dimer & - & 32 & 0 & $10\sigma$ & 93.041 & - & 1.676 & - & - & - \\
    	liquid & - & 32 & 0 & $10\sigma$ & 12.453 & - & 0.329 & - & 15.127 & - \\
    	liquid\_high & - & 12 & 0 & $10\sigma$ & 30.868 & - & 0.753 & - & 28.109 & - \\
    	bcc\_rattled & - & 0 & 20 & - & - & 0.545 & - & 0.023 & - & 5.207 \\
    	bcc\_rattled\_high & - & 0 & 20 & - & - & 3.198 & - & 0.095 & - & 3.708 \\
    	random-FPs & - & 0 & 5 & - & - & 0.902 & - & 0.092 & - & 17.842 \\
    \bottomrule
    \end{tabular}
    
    \begin{tablenotes}[para,flushleft]
        \item[*] Virial data used for fitting
    \end{tablenotes}

    \end{threeparttable}
\label{supp_tab:reg_params}
\end{table*}

\section{H migration barriers with hydrostatic strain}\label{supp_app:H_mig_strain}

\begin{figure*}
\centering
    \begin{subfigure}{.48\textwidth}
        \includegraphics[width=\linewidth]{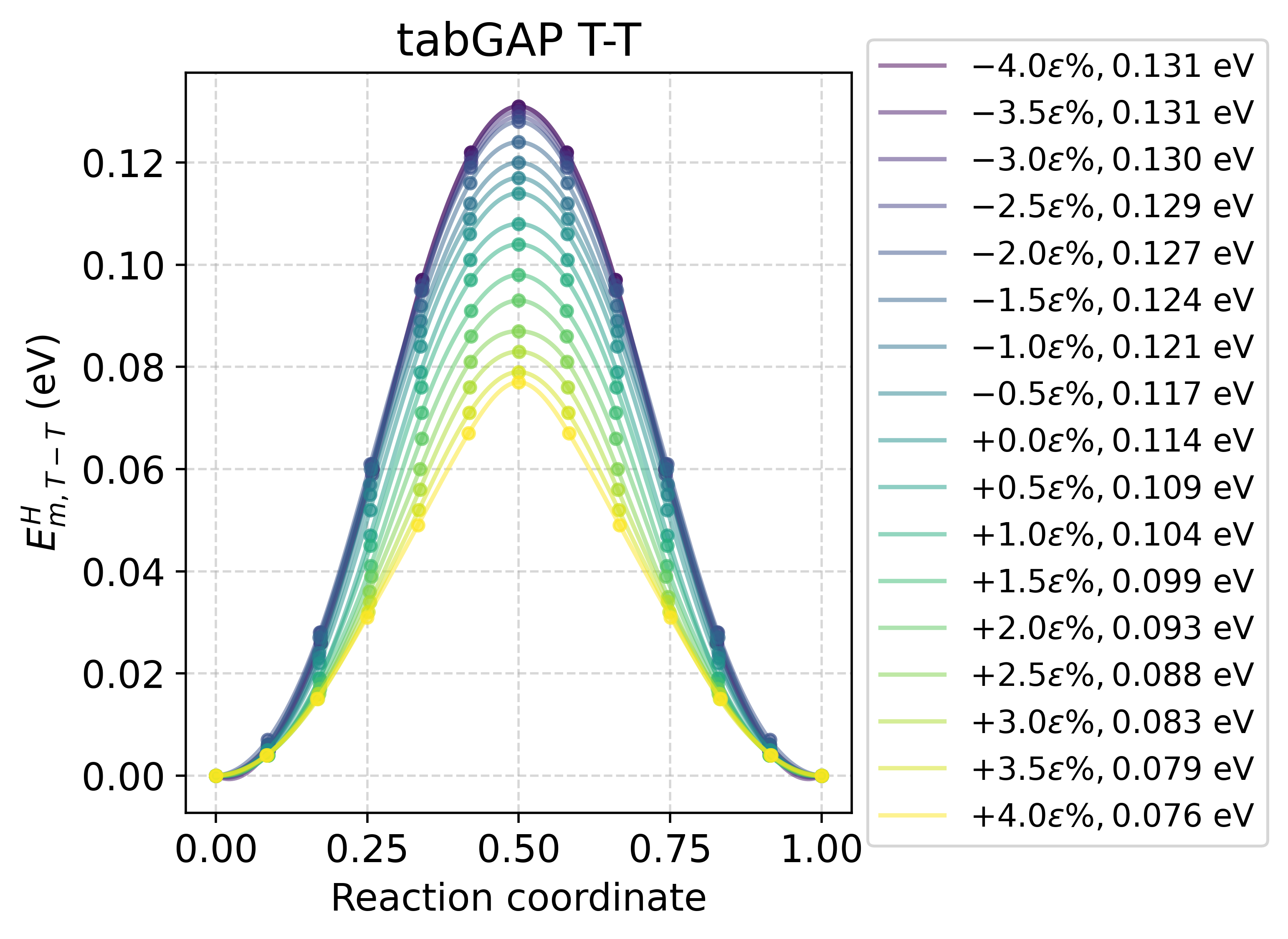}
        \caption{}
        \label{supp_fig:tabGAP_neb_H_TT_strain}
    \end{subfigure}%
    \begin{subfigure}{.48\textwidth}
        \includegraphics[width=\linewidth]{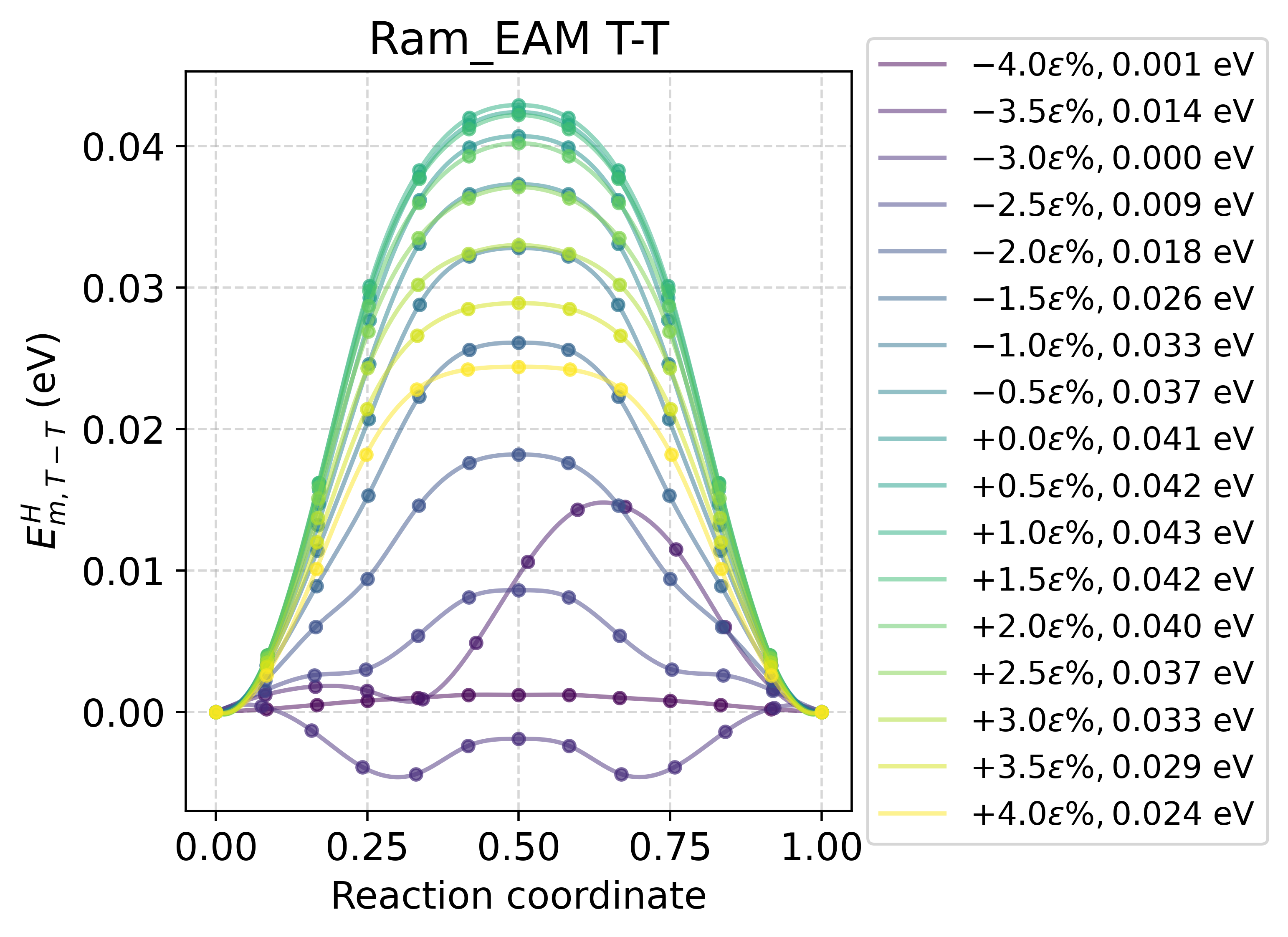}
        \caption{}
        \label{supp_fig:Ram_EAM_neb_H_TT_strain}
    \end{subfigure}
    
    \begin{subfigure}{.48\textwidth}
        \includegraphics[width=\linewidth]{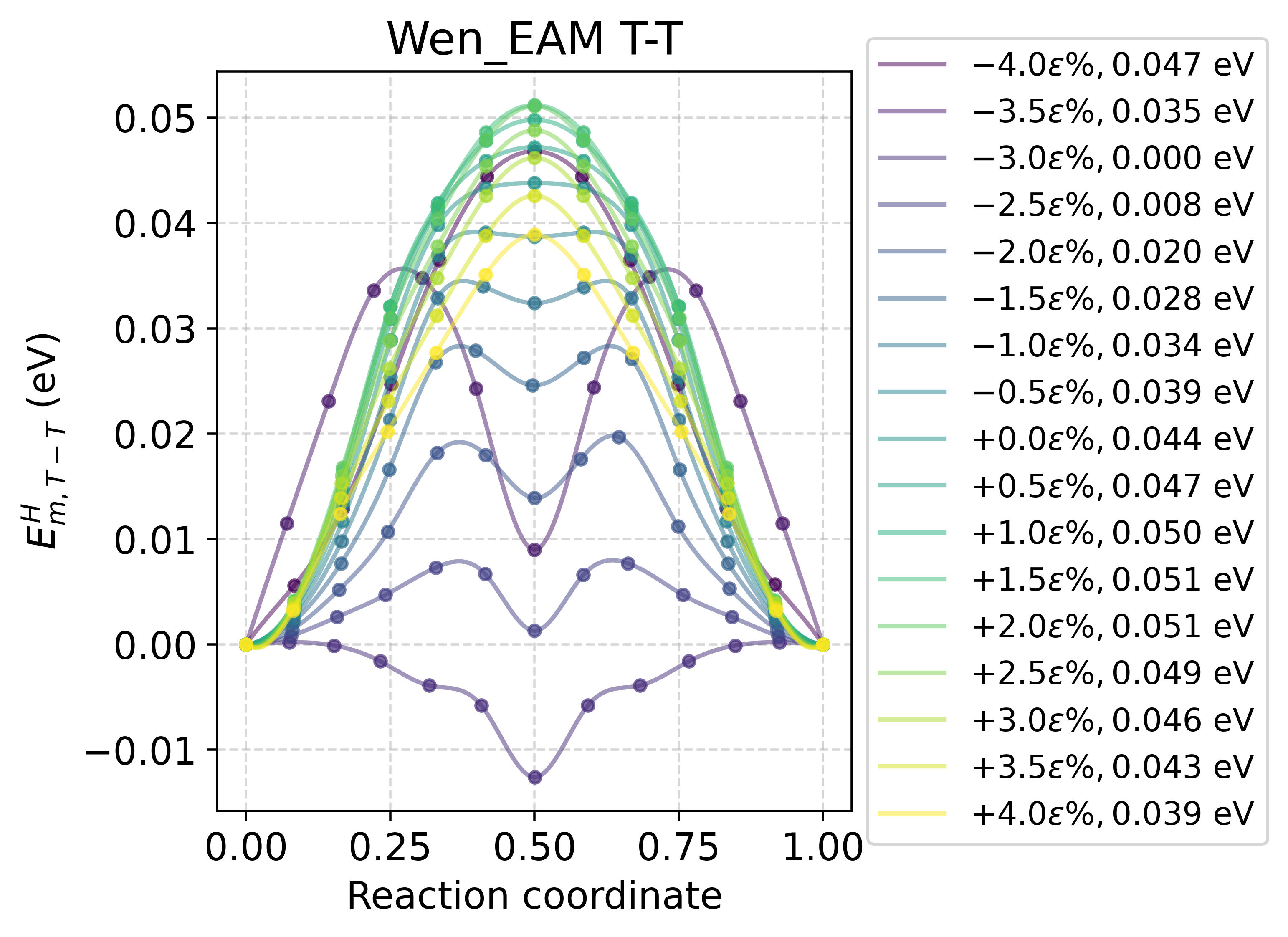}
        \caption{}
        \label{supp_fig:Wen_EAM_neb_H_TT_strain}
    \end{subfigure}%
    \begin{subfigure}{.48\textwidth}
        \includegraphics[width=\linewidth]{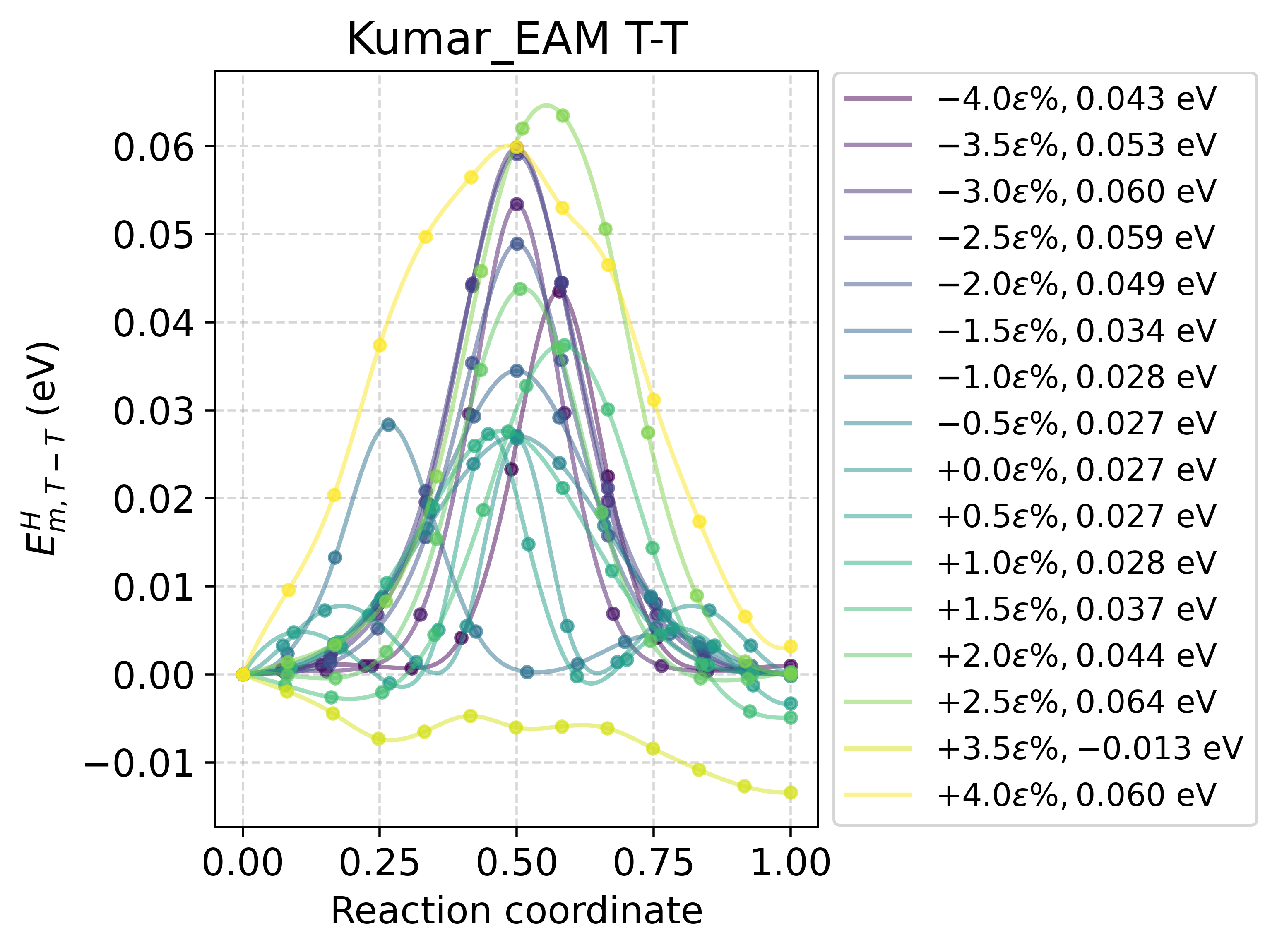}
        \caption{}
        \label{supp_fig:Kumar_EAM_neb_H_TT_strain}
    \end{subfigure}
    \begin{subfigure}{.48\textwidth}
        \includegraphics[width=\linewidth]{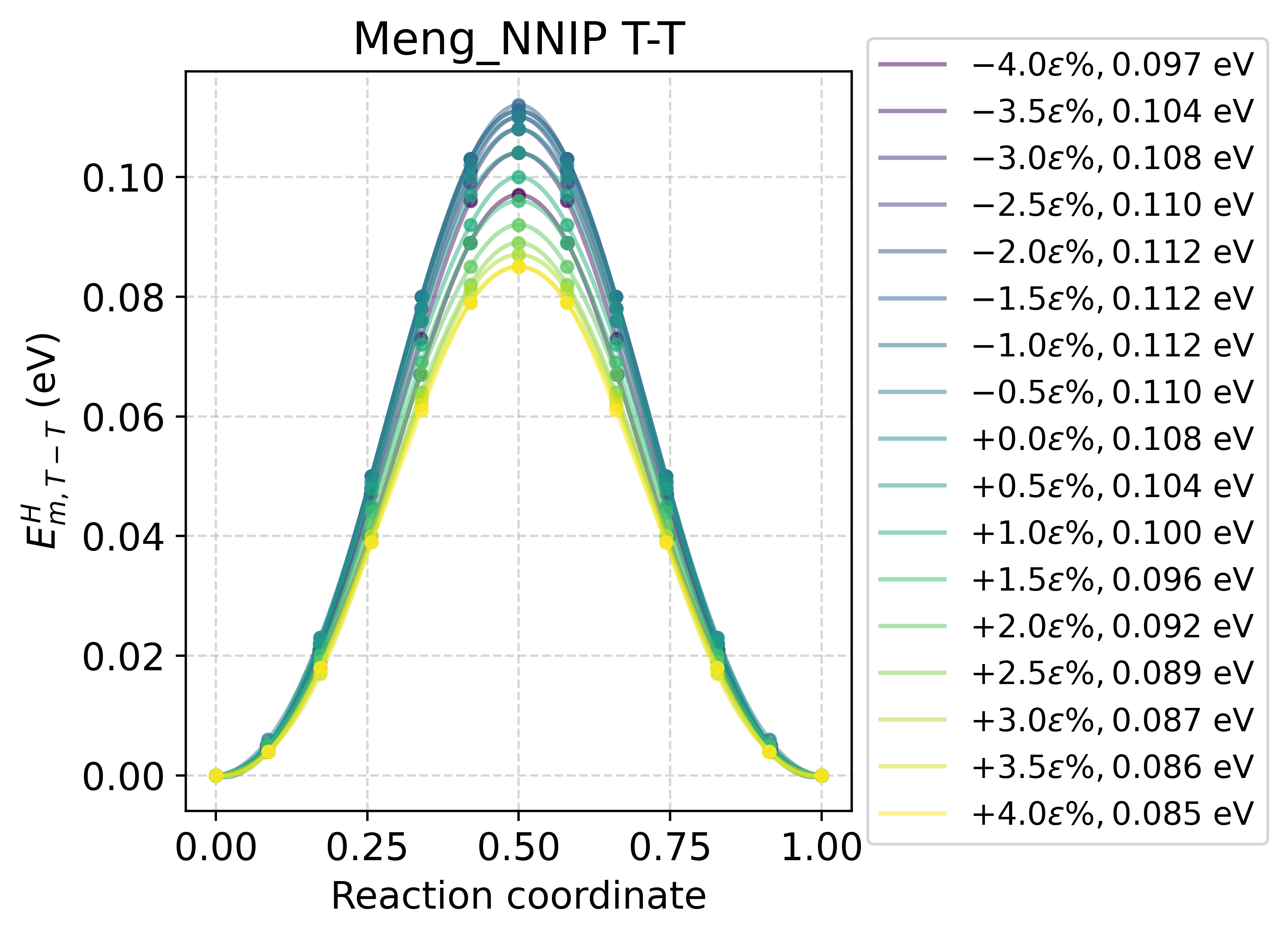}
        \caption{}
        \label{supp_fig:Meng_NNIP_neb_H_TT_strain}
    \end{subfigure}
    \caption{The migration energy barriers of the T-T transition for the H atom as a function of hydrostatic strain applied to the surrounding \aFe-lattice. The maximum barrier energy at each strain is shown in the legend.}
    \label{supp_fig:neb_H_TT_strain}
\end{figure*}

\begin{figure*}
\centering
    \begin{subfigure}{.48\textwidth}
        \includegraphics[width=\linewidth]{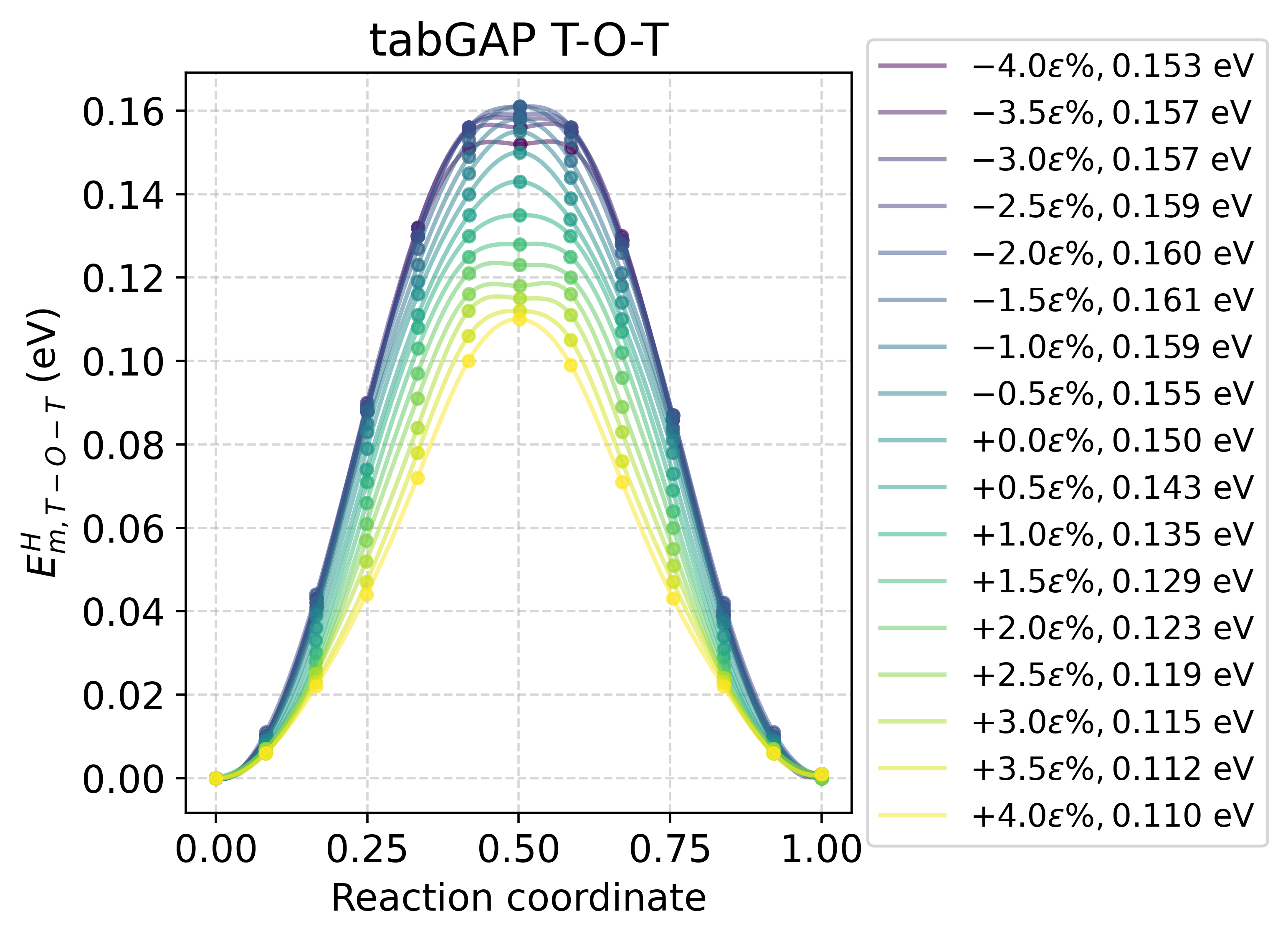}
        \caption{}
        \label{supp_fig:tabGAP_neb_H_TOT_strain}
    \end{subfigure}%
    \begin{subfigure}{.48\textwidth}
        \includegraphics[width=\linewidth]{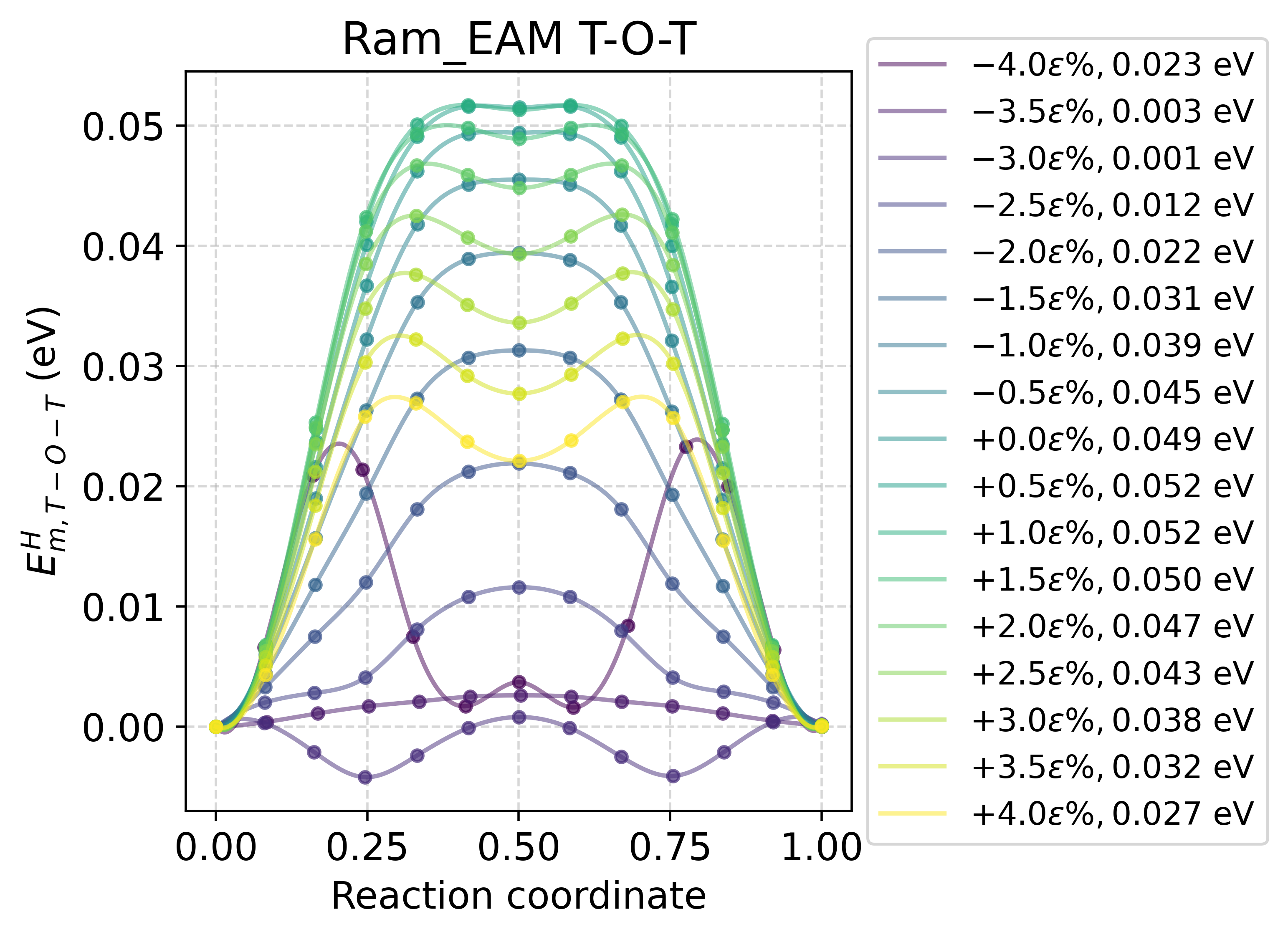}
        \caption{}
        \label{supp_fig:Ram_EAM_neb_H_TOT_strain}
    \end{subfigure}

    \begin{subfigure}{.48\textwidth}
        \includegraphics[width=\linewidth]{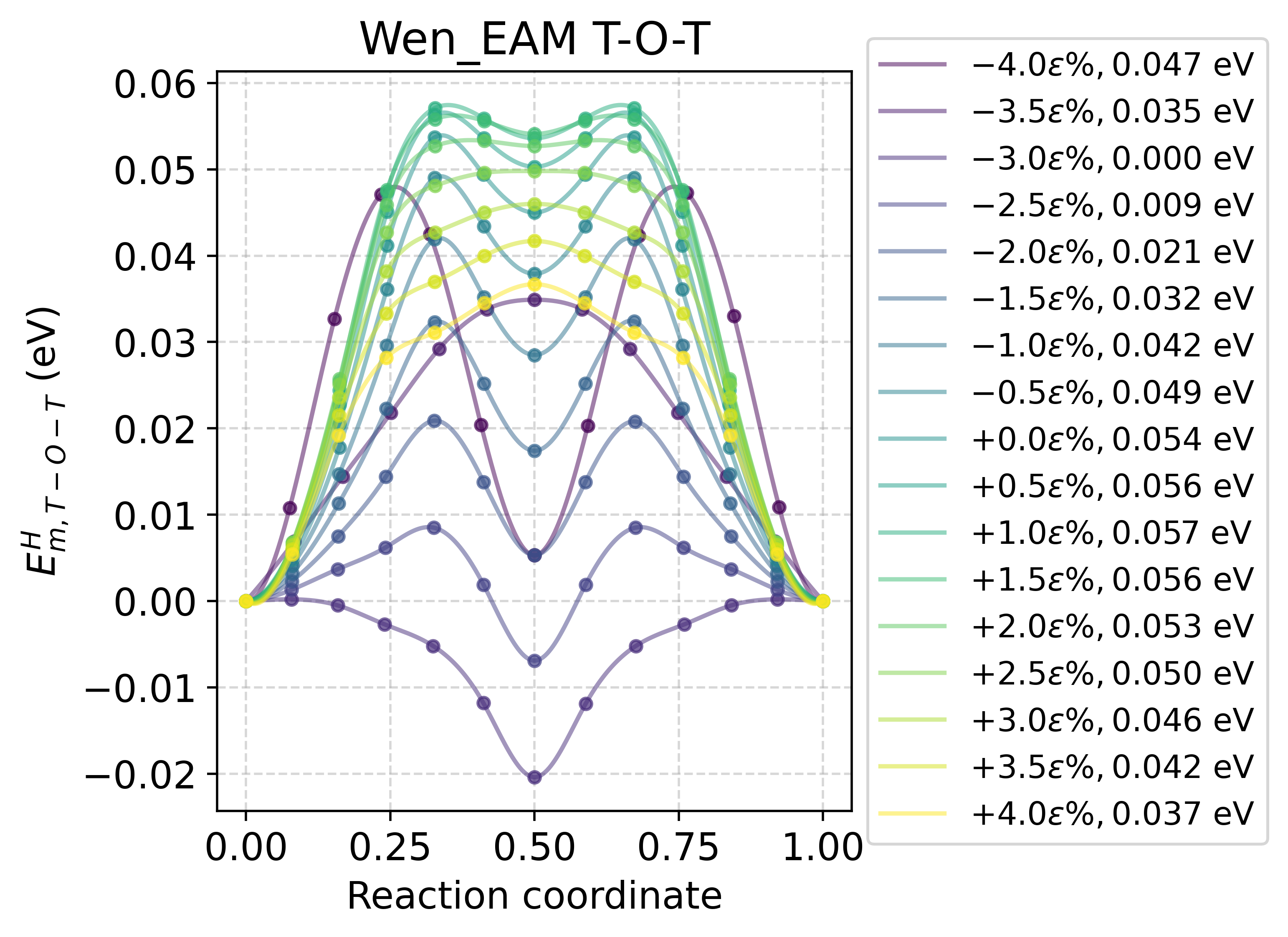}
        \caption{}
        \label{supp_fig:Wen_EAM_neb_H_TOT_strain}
    \end{subfigure}%
    \begin{subfigure}{.48\textwidth}
        \includegraphics[width=\linewidth]{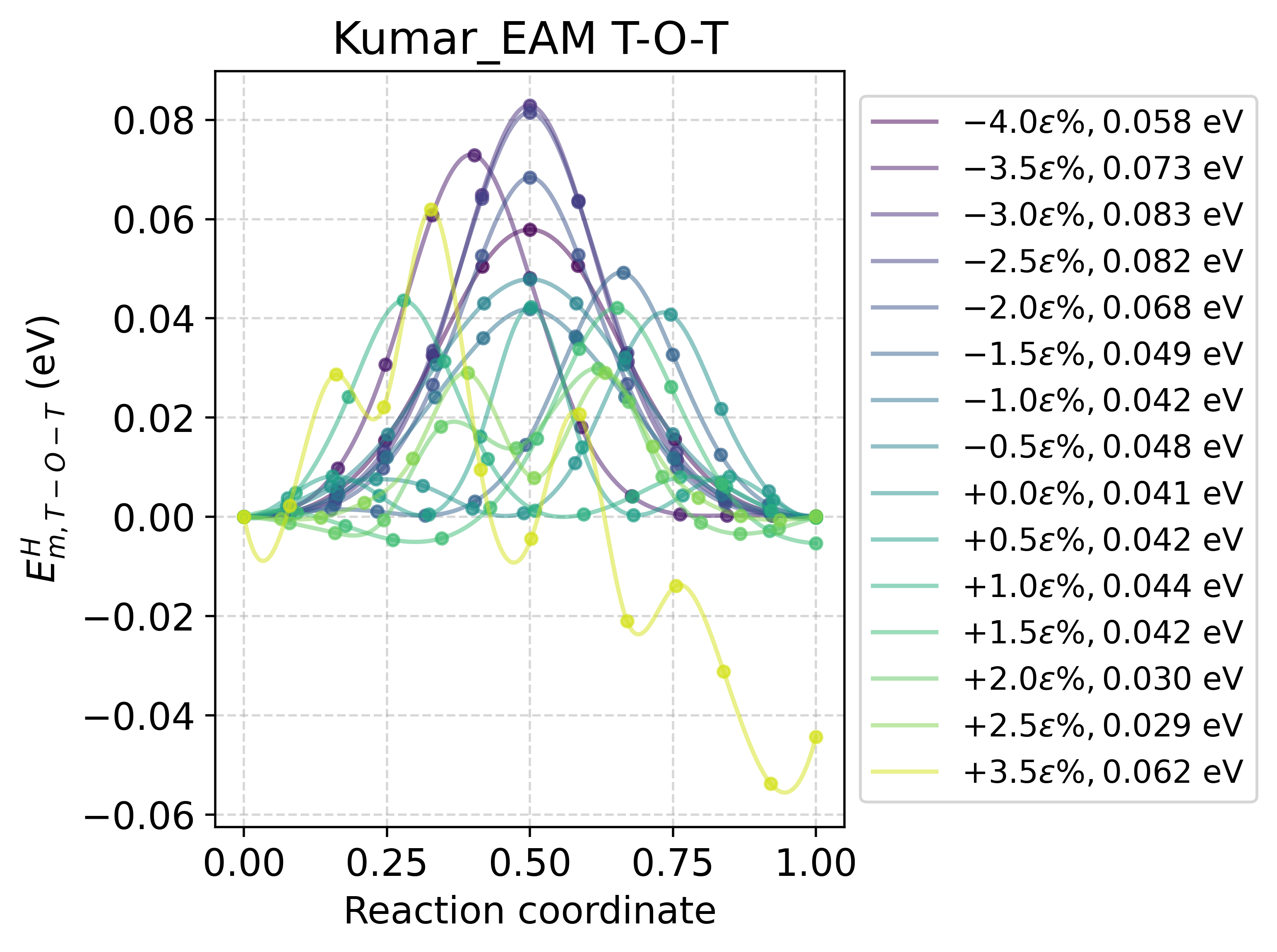}
        \caption{}
        \label{supp_fig:Kumar_EAM_neb_H_TOT_strain}
    \end{subfigure}
    \begin{subfigure}{.48\textwidth}
        \includegraphics[width=\linewidth]{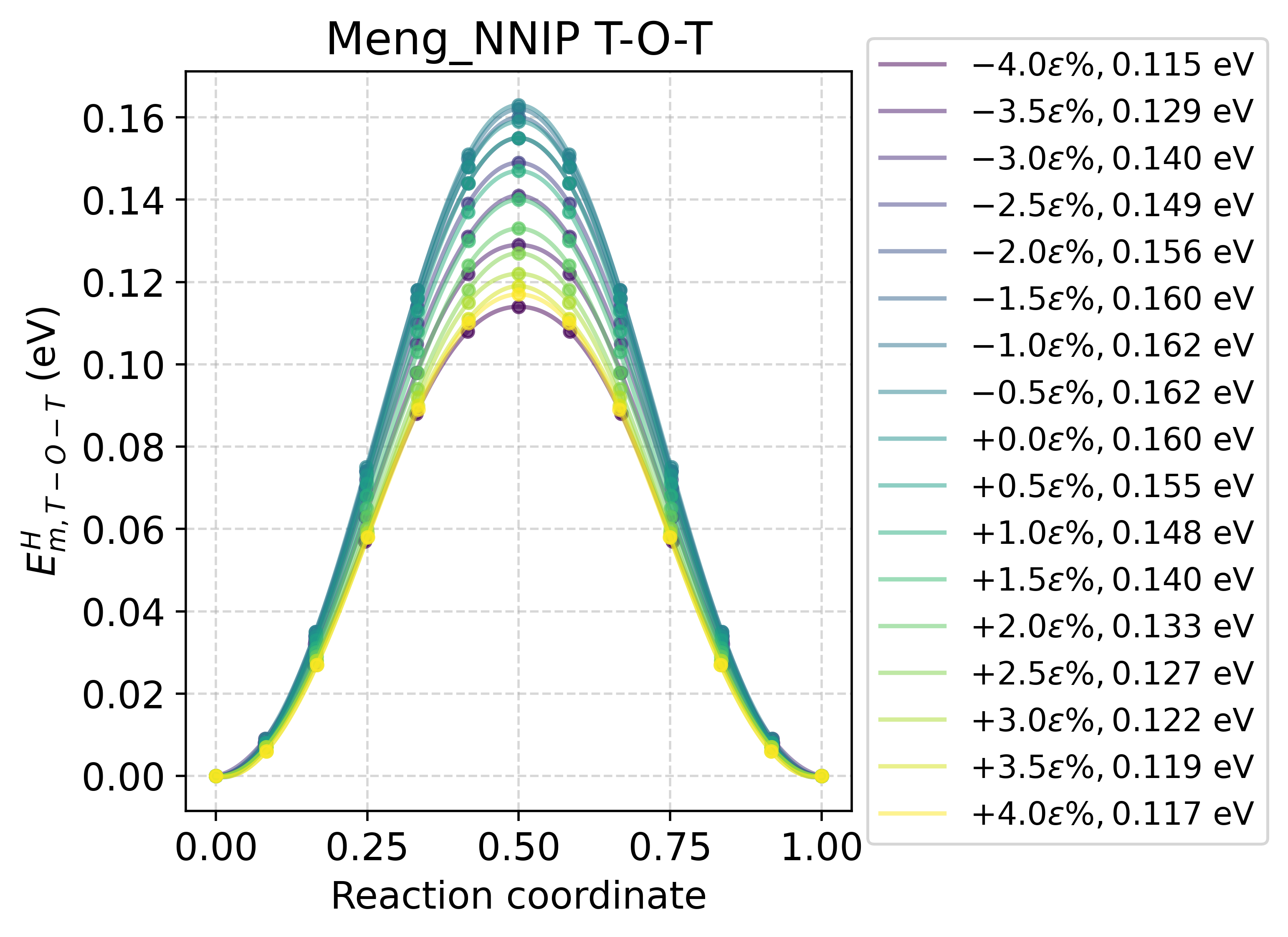}
        \caption{}
        \label{supp_fig:Meng_NNIP_neb_H_TOT_strain}
    \end{subfigure}
    \caption{The migration energy barriers of the T-O-T transition for the H atom as a function of hydrostatic strain applied to the surrounding \aFe-lattice. The maximum barrier energy at each strain is shown in the legend.}
    \label{supp_fig:neb_H_TOT_strain}
\end{figure*}

The migration barriers for the movement of an H atom through the \aFe-lattice were calculated for the Fe-H tabGAP presented in this work and for the other tested IAPs from literature. The two migration paths, T-T and T-O-T, were investigated. A hydrostatic strain ($\varepsilon$) in the range of $-4$ to $\SI{4}{\%}$ was applied to the surrounding \aFe lattice, and CI-NEB was used with the specifications outlined in Sec. II D of the main text. The results are shown in Figs. \ref{supp_fig:neb_H_TT_strain} and \ref{supp_fig:neb_H_TOT_strain} for the T-T and the T-O-T paths, respectively.

The tabGAP model shows a very regular and a linear trend of the migration barriers for both the T-T and T-O-T transitions (Figs. \ref{supp_fig:tabGAP_neb_H_TT_strain} and \ref{supp_fig:tabGAP_neb_H_TOT_strain}), with the barrier height decreasing as the Fe-lattice is expanded. Only at very high, $<-2.5\varepsilon\%$, compressive strains do we see this trend changing, with the migration barrier either plateauing for the T-T path (\ref{supp_fig:tabGAP_neb_H_TT_strain}), or going down slightly for the T-O-T(\ref{supp_fig:tabGAP_neb_H_TOT_strain}). The Meng\_NNIP model \cite{mengGeneralpurposeNeuralNetwork2021} shows a very similar trend, just with the compressive strain behavior being accentuated.

The three classical EAMs show varied responses to hydrostatic strain. The \aFe-lattice did not always relax to a regular crystalline structure at high strains when the Kumar\_EAM \cite{kumarEffectHydrogenPlasticity2023} was used, explaining the irregularity in the migration barriers given by the model. In addition, at the two strains of $+3.0$ and $+4.0\varepsilon\%$ the NEB calculation failed to converge with the Kumar\_EAM. The Wen\_EAM \cite{wenNewInteratomicPotential2021} reports the T-T saddle point and the O-site to be stable sites at various strains. Ram\_EAM \cite{ramasubramaniamInteratomicPotentialsHydrogen2009} shows this behavior as well for the O-site, but less so for the T-T saddle point. These two models both show a decrease in the migration barriers at mild compressive strains in disagreement with the ML-based non-ZPE-corrected potentials.

\section{Finite-temperature elastic constants of \aFe-H}

In Tab. \ref{supp_tab:elastic_constants} we provide the numerical data from the simulations presented in the main text for evaluating the elastic constants of the \aFe-H system with the tabGAP model. The table contains the elastic constants computed with the tabGAP model for two volume-matched lattices, which we named \aFe-H and \aFevol, with the latter being a H-free \aFe structure with a lattice expansion applied that matches the equilibrium condition of the former H-loaded lattice. The Eq. (11) of the main text was used for this. The volumes of the lattices were kept constant with the NVT ensemble during the finite-temperature sampling of the stress tensor.

\begin{table*}
\centering
\caption{The elastic constants predicted by the tabGAP model for the Fe-H system within a range of $\myqtyrange{0.0}{5.0}{\atp}$ of interstitial H content for the temperature range of $\myqtyrange{20}{500}{K}$. The \aFevol-columns contain the volume-matched values for pure \aFe structures.}
    \begin{threeparttable}
    \begin{tabular}{@{}c cccc | cccc | cccc@{}}
    \toprule
    & \multicolumn{4}{c}{$C_{11} (\si{GPa})$} & \multicolumn{4}{c}{$C_{12} (\si{GPa})$} & \multicolumn{4}{c}{$C_{44} (\si{GPa})$} \\
    \midrule
        $\si{\atp}$ H & \multicolumn{2}{c}{$0.00$} & \multicolumn{2}{c}{$0.08$} & \multicolumn{2}{c}{$0.00$} & \multicolumn{2}{c}{$0.08$} & \multicolumn{2}{c}{$0.00$} & \multicolumn{2}{c}{$0.08$} \\
        \midrule
        Temperature $(\si{K})$ & \aFe-H & \aFevol & \aFe-H & \aFevol & \aFe-H & \aFevol & \aFe-H & \aFevol & \aFe-H & \aFevol & \aFe-H & \aFevol \\
        \midrule
        $20$ & 269.23 & 269.23 & 269.02 & 268.79 & 143.66 & 143.67 & 143.67 & 143.40 & 108.73 & 108.74 & 108.61 & 108.61 \\
        $50$ & 266.89 & 266.93 & 266.67 & 266.44 & 142.37 & 142.39 & 142.36 & 142.09 & 108.31 & 108.33 & 108.22 & 108.19 \\
        $100$ & 263.13 & 263.13 & 262.90 & 262.67 & 140.35 & 140.30 & 140.33 & 140.04 & 107.67 & 107.69 & 107.54 & 107.56 \\
        $200$ & 255.43 & 255.50 & 255.22 & 255.03 & 136.47 & 136.60 & 136.58 & 136.20 & 106.50 & 106.47 & 106.29 & 106.38 \\
        $300$ & 248.05 & 248.34 & 248.02 & 247.74 & 133.10 & 133.42 & 133.44 & 133.01 & 105.30 & 105.31 & 105.15 & 105.14 \\
        $400$ & 241.10 & 241.19 & 240.80 & 240.68 & 130.27 & 130.26 & 130.39 & 129.97 & 104.21 & 104.28 & 104.13 & 104.02 \\
        $500$ & 234.78 & 234.52 & 234.55 & 233.93 & 127.83 & 127.63 & 127.76 & 127.27 & 103.26 & 102.84 & 103.12 & 103.04 \\
        \midrule
        & \multicolumn{2}{c}{$0.19$} & \multicolumn{2}{c}{$0.56$} & \multicolumn{2}{c}{$0.19$} & \multicolumn{2}{c}{$0.56$} & \multicolumn{2}{c}{$0.19$} & \multicolumn{2}{c}{$0.56$} \\
        \midrule
        $20$ & 268.68 & 268.08 & 267.55 & 265.76 & 143.66 & 142.99 & 143.60 & 141.63 & 108.41 & 108.40 & 107.80 & 107.71 \\
        $50$ & 266.35 & 265.77 & 265.26 & 263.47 & 142.37 & 141.70 & 142.34 & 140.35 & 108.02 & 107.99 & 107.39 & 107.28 \\
        $100$ & 262.59 & 262.02 & 261.48 & 259.69 & 140.37 & 139.70 & 140.34 & 138.24 & 107.38 & 107.36 & 106.69 & 106.68 \\
        $200$ & 254.90 & 254.51 & 253.55 & 252.22 & 136.52 & 135.86 & 136.67 & 134.56 & 106.13 & 106.14 & 105.48 & 105.50 \\
        $300$ & 247.43 & 247.18 & 246.22 & 244.74 & 133.37 & 132.68 & 133.59 & 131.01 & 104.97 & 104.93 & 104.39 & 104.34 \\
        $400$ & 240.41 & 240.03 & 239.14 & 237.91 & 130.33 & 129.61 & 130.63 & 128.10 & 103.99 & 103.92 & 103.30 & 103.33 \\
        $500$ & 233.83 & 233.48 & 232.93 & 232.19 & 127.67 & 126.95 & 128.26 & 126.12 & 102.72 & 102.90 & 102.37 & 102.22 \\
        \midrule
        & \multicolumn{2}{c}{$1.00$} & \multicolumn{2}{c}{$1.53$} & \multicolumn{2}{c}{$1.00$} & \multicolumn{2}{c}{$1.53$} & \multicolumn{2}{c}{$1.00$} & \multicolumn{2}{c}{$1.53$} \\
        \midrule
        $20$ & 266.20 & 262.92 & 264.82 & 259.79 & 143.55 & 139.95 & 143.55 & 138.11 & 107.04 & 106.87 & 106.22 & 105.94 \\
        $50$ & 263.90 & 260.64 & 262.53 & 257.52 & 142.27 & 138.68 & 142.31 & 136.84 & 106.63 & 106.47 & 105.85 & 105.55 \\
        $100$ & 260.11 & 256.94 & 258.77 & 253.93 & 140.37 & 136.63 & 140.38 & 134.88 & 106.05 & 105.86 & 105.26 & 104.94 \\
        $200$ & 252.10 & 249.62 & 250.25 & 246.71 & 136.90 & 133.05 & 137.09 & 131.26 & 104.88 & 104.73 & 103.99 & 103.80 \\
        $300$ & 244.63 & 242.52 & 242.88 & 239.72 & 133.95 & 129.76 & 134.31 & 128.08 & 103.80 & 103.69 & 102.90 & 102.77 \\
        $400$ & 237.62 & 235.65 & 235.88 & 232.99 & 130.96 & 126.98 & 131.83 & 125.07 & 102.59 & 102.71 & 101.70 & 101.77 \\
        $500$ & 231.63 & 229.23 & 230.06 & 227.02 & 128.46 & 124.34 & 128.93 & 122.99 & 101.79 & 101.75 & 100.81 & 101.08 \\
        \midrule
        & \multicolumn{2}{c}{$3.01$} & \multicolumn{2}{c}{$5.00$} & \multicolumn{2}{c}{$3.01$} & \multicolumn{2}{c}{$5.00$} & \multicolumn{2}{c}{$3.01$} & \multicolumn{2}{c}{$5.00$} \\
        \midrule
        $20$ & 260.94 & 250.51 & 256.84 & 238.45 & 143.42 & 132.64 & 143.66 & 125.53 & 103.83 & 103.21 & 100.82 & 99.68 \\
        $50$ & 258.63 & 248.35 & 254.64 & 236.45 & 142.29 & 131.43 & 142.63 & 124.37 & 103.48 & 102.84 & 100.51 & 99.36 \\
        $100$ & 254.95 & 244.81 & 250.67 & 233.23 & 140.43 & 129.40 & 141.02 & 122.63 & 102.91 & 102.31 & 99.95 & 98.91 \\
        $200$ & 245.88 & 238.03 & 240.48 & 226.75 & 137.91 & 126.08 & 138.90 & 119.26 & 101.75 & 101.34 & 98.70 & 98.04 \\
        $300$ & 237.88 & 231.44 & 232.17 & 220.91 & 135.53 & 123.08 & 136.61 & 116.77 & 100.60 & 100.40 & 97.84 & 97.22 \\
        $400$ & 231.17 & 225.07 & 225.78 & 214.99 & 132.75 & 120.45 & 134.11 & 114.33 & 99.59 & 99.58 & 96.87 & 96.52 \\
        $500$ & 226.05 & 219.66 & 221.21 & 209.86 & 130.37 & 118.63 & 131.58 & 112.45 & 98.78 & 98.70 & 96.03 & 95.97 \\
    \bottomrule
    \end{tabular}
    \end{threeparttable}
\label{supp_tab:elastic_constants}
\end{table*}



\section{H-screw dislocation properties}

\begin{figure}
\centering
    \begin{subfigure}{.35\textwidth}
        \captionsetup{margin={0.11\textwidth, 0pt}}
        \includegraphics[width=\linewidth]{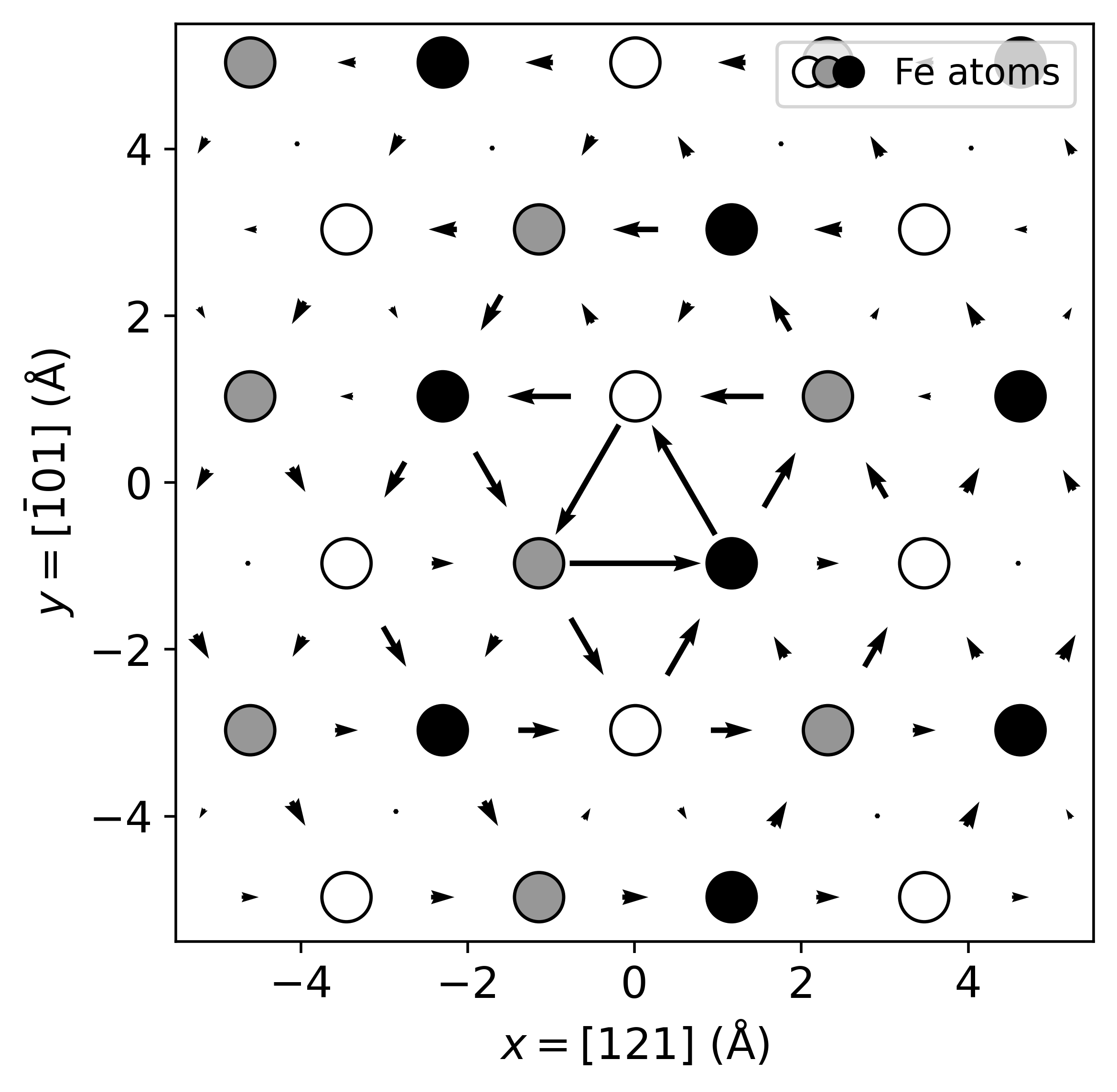}
    \caption{Pure Fe}
    \label{supp_fig:pure_Fe_ddm}
    \end{subfigure}\\
    \begin{subfigure}{.35\textwidth}
        \includegraphics[width=\linewidth]{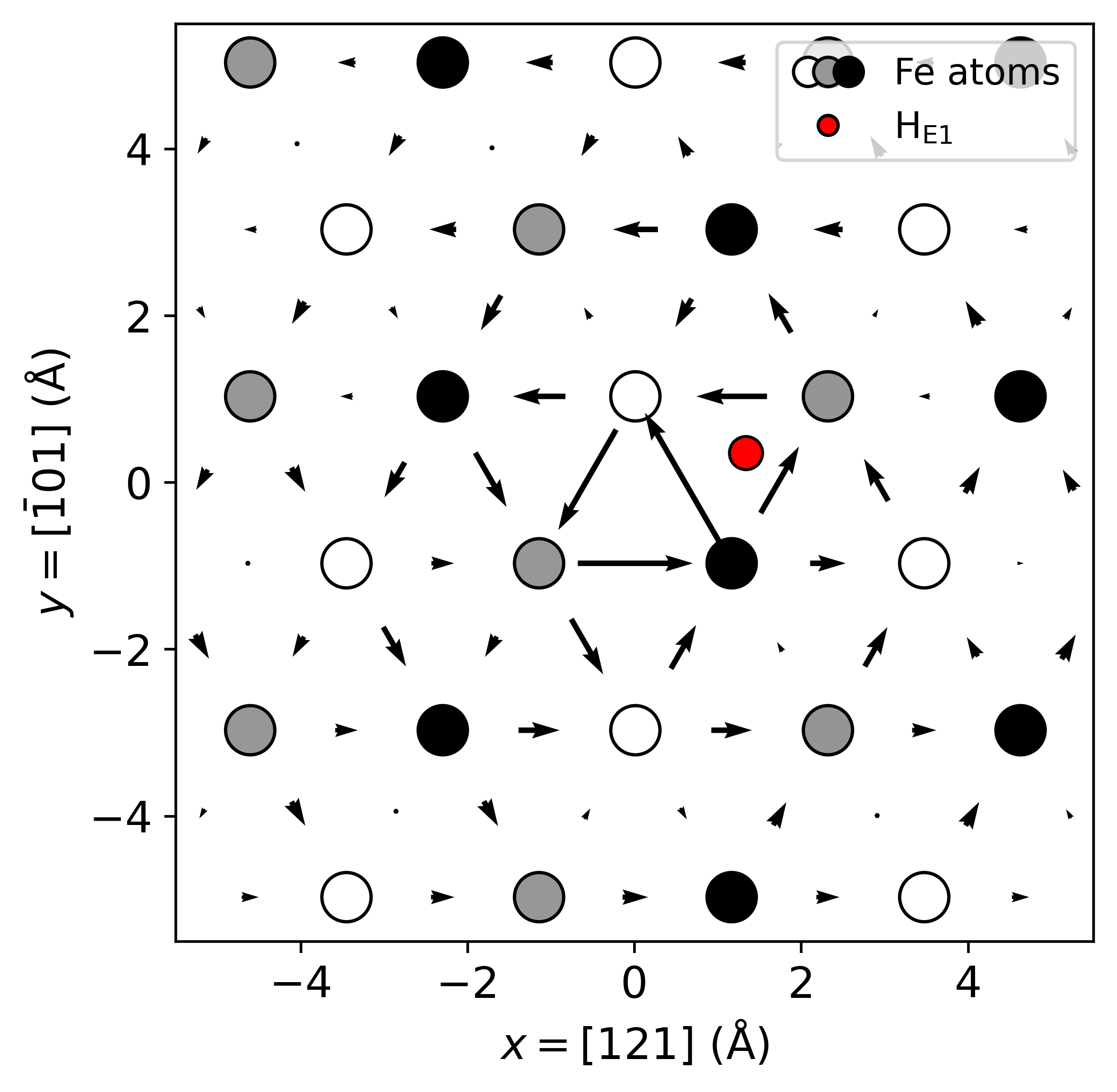}
    \caption{$1/40$ H/$b$}
    \label{supp_fig:Fe_H_ddm_40b}
    \end{subfigure}
    \begin{subfigure}{.35\textwidth}
        \includegraphics[width=\linewidth]{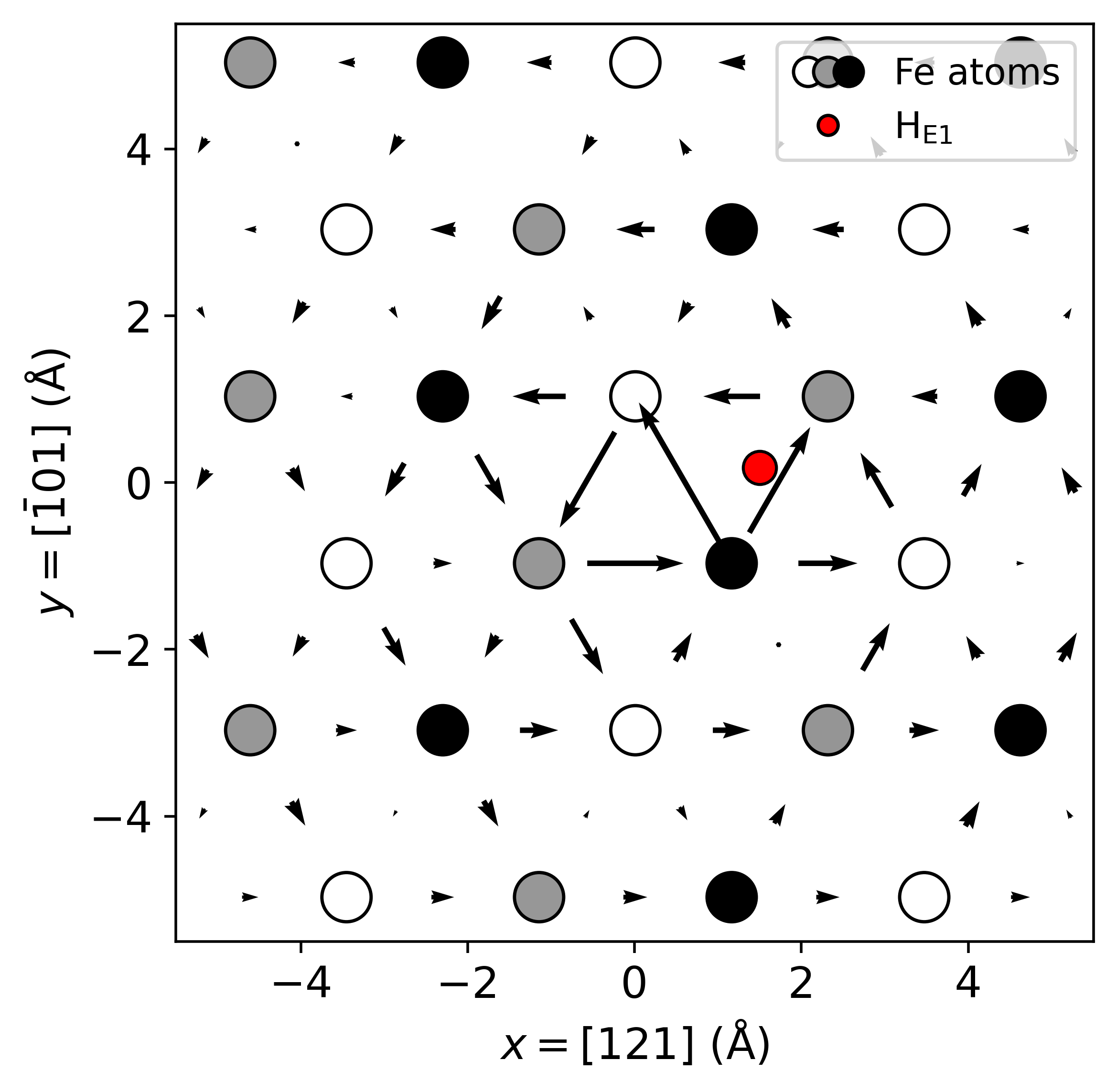}
    \caption{$1$ H/$b$}
    \label{supp_fig:Fe_H_ddm_1b}
    \end{subfigure}\\
    \begin{subfigure}{.35\textwidth}
        \includegraphics[width=\linewidth]{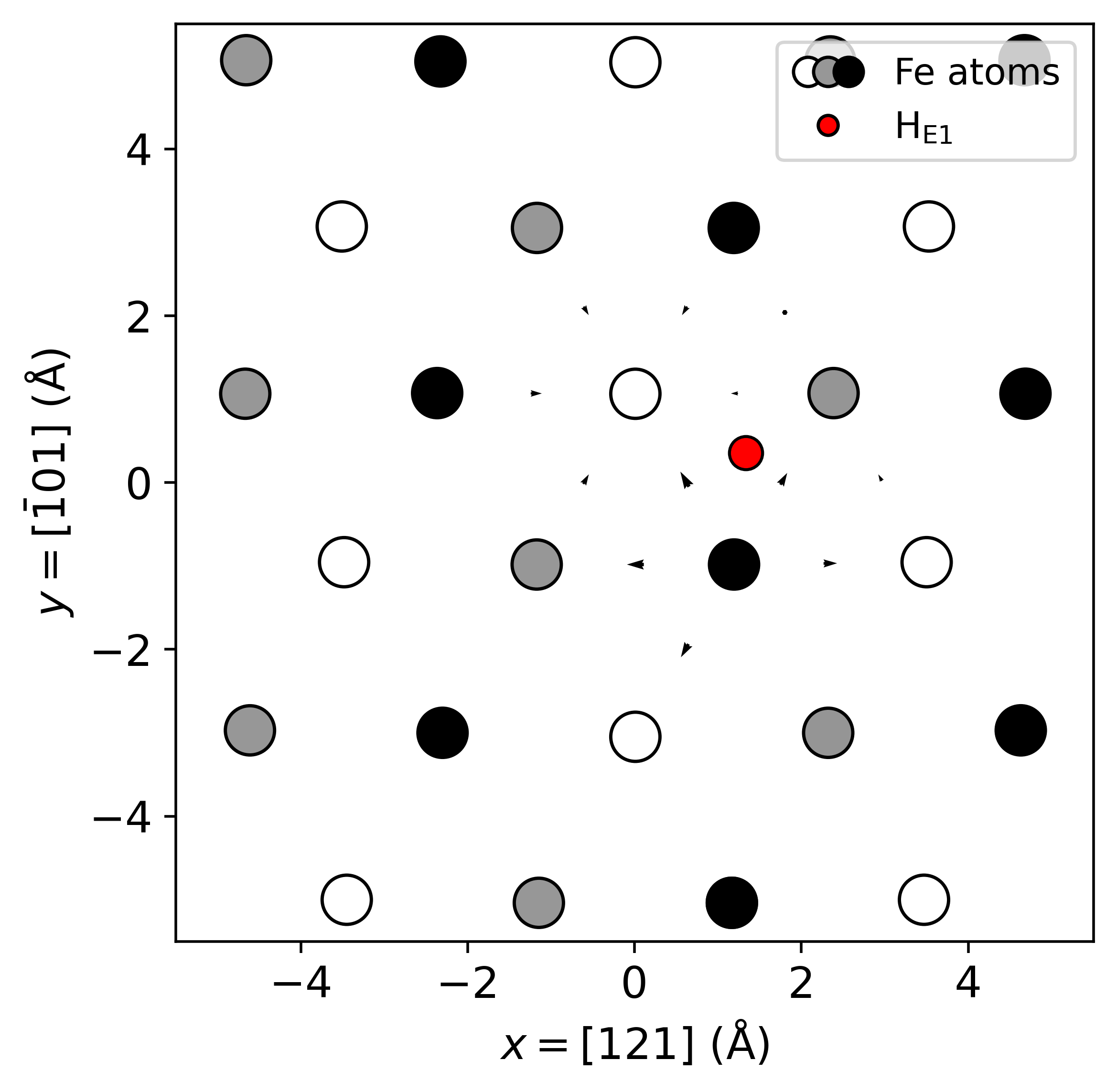}
    \caption{$\Delta$DDM of pure Fe vs. $1/40$ H/$b$}
    \label{supp_fig:Fe_H_diff_ddm_40b}
    \end{subfigure}
    \begin{subfigure}{.35\textwidth}
        \includegraphics[width=\linewidth]{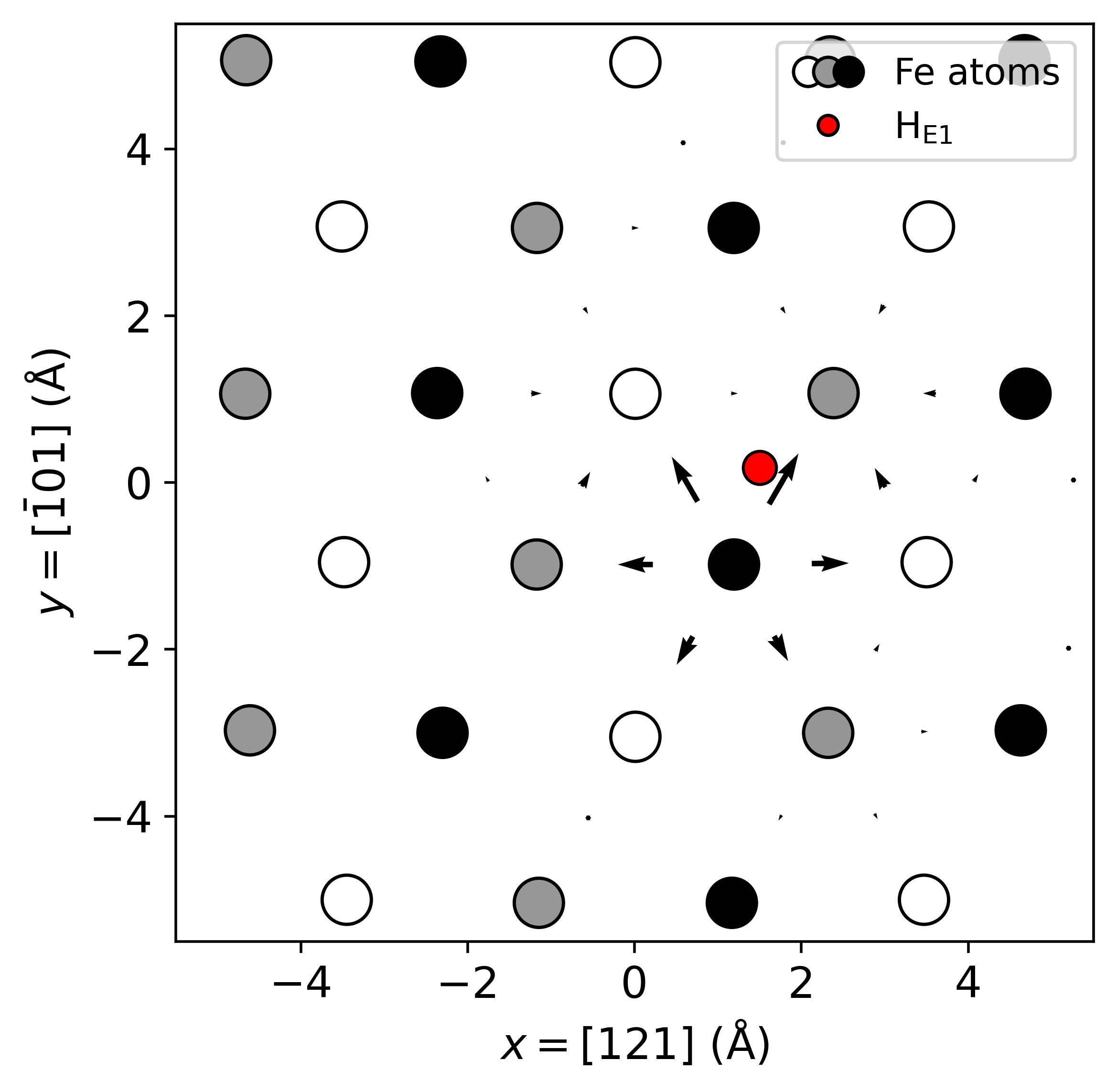}
    \caption{$\Delta$DDM of pure Fe vs. $1$ H/$b$}
    \label{supp_fig:Fe_H_diff_ddm_1b}
    \end{subfigure}
\caption{The differential displacement maps given by the tabGAP for a $1/2\hkl<111>\hkl{110}$ screw dislocation. (\subref{supp_fig:pure_Fe_ddm}) is pure Fe. (\subref{supp_fig:Fe_H_ddm_40b})-(\subref{supp_fig:Fe_H_ddm_1b}) show maps with $1/40$ H/$b$ and $1$ H/$b$ of H in the E1 site, respectively. (\subref{supp_fig:Fe_H_diff_ddm_40b})-(\subref{supp_fig:Fe_H_diff_ddm_1b}) show the difference between the differential displacement maps of the pure Fe and H-occupied structures. Only the arrows within $|b/2|$ of the z-coordinate of the H atom are shown for clarity, and the arrows are scaled by $b/2$. The Fe atoms are colored according to their position in the $z=\hkl[11-1]$ direction.}
\label{supp_fig:ddmaps}
\end{figure}

In Fig. \ref{supp_fig:ddmaps} we show the core structure of the $1/2\hkl<111>\hkl{110}$ screw dislocation with the differential displacement map (DDM) \cite{vitekCoreStructure1/21111970}. The dislocation structure was created as explained in Sec. II D of the main text. Here we show the two extreme cases of H concentration we examined, namely $1/40$ (left-hand side figures) and $1$ H/$b$ (left-hand side figures) which were achieved by varying the dislocation line length, and inserting a single H atom into a E1 site adjacent to the dislocation core. The lower row of figures shows the difference between the pure Fe and the Fe-H DDMs of corresponding H line density. When there's less than $1$ H/$b$ in the E1 site, we see a weak and localized split-like reconstruction of the dislocation core toward the H atom. When the density is $\SI{1.0}{H/b}$, the core undergoes a much stronger split-like reconstruction. The exact site in which the H relaxes into also moves slightly to the positive x, and negative y directions.

\begin{figure}
    \includegraphics[width=0.48\textwidth]{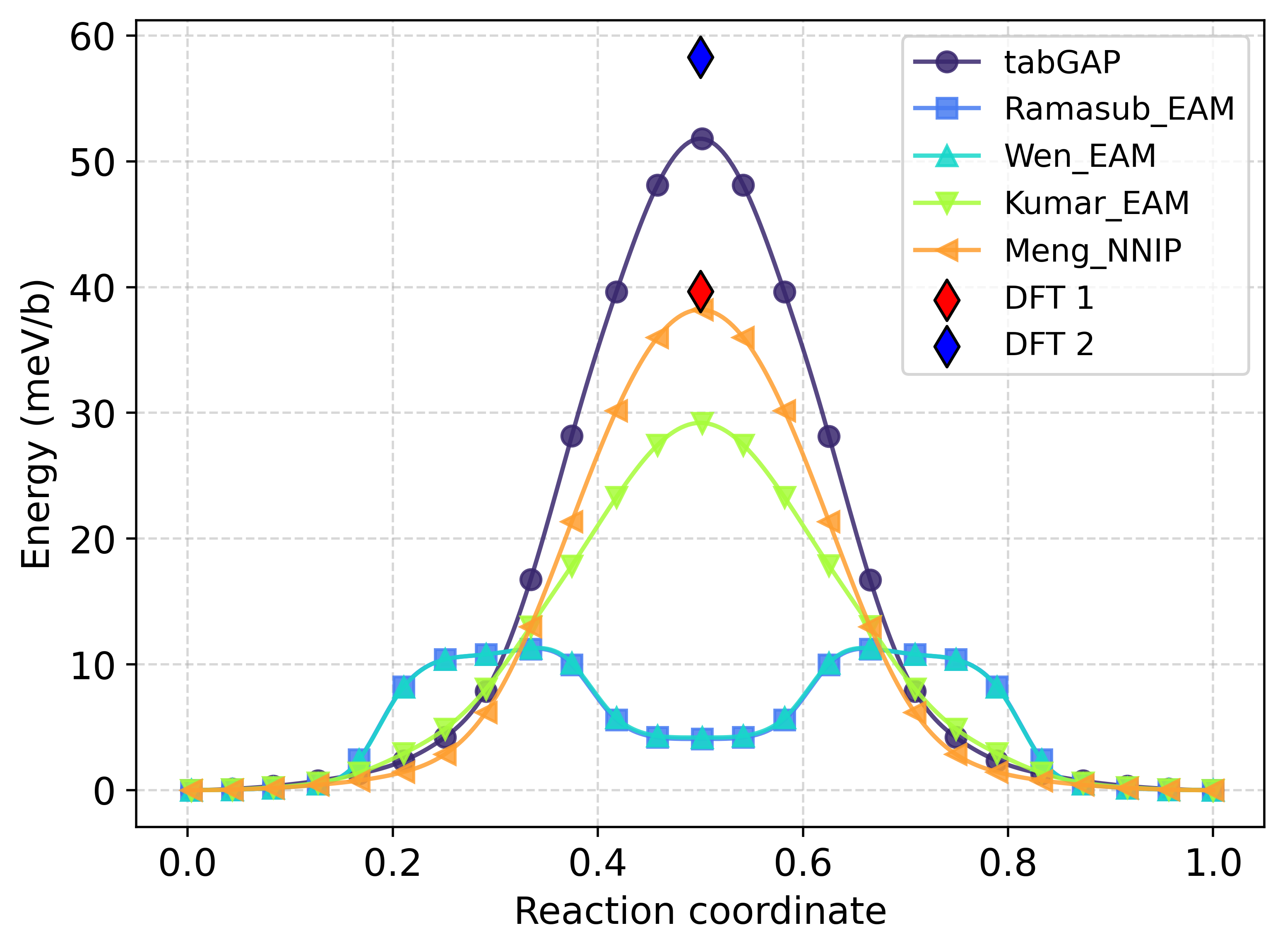}
    \caption{The Peierls barrier for a $1/2\hkl<111>\hkl{110}$ screw dislocation moving in the \hkl<112> direction from CI-NEB calculations with the present tabGAP model and the other tested Fe-H IAPs. The DFT results are from Refs. \cite{ventelonInitioInvestigationPeierls2013} and \cite{dragoniAchievingDFTAccuracy2018} for DFT 1 and DFT 2, respectively.}
    \label{supp_fig:pure_Fe_peierls}
\end{figure}

An important problem that has been noted for the two earlier classical EAM Fe-H IAPs, the Ram\_EAM and the Wen\_EAM, is that they predict a dual-hump shaped Peierls barrier for the $1/2\hkl<111>\hkl{110}$ screw dislocation motion, while the consensus in DFT calculations is that this is single-hump shaped \cite{kumarEffectHydrogenPlasticity2023, ITAKURA20123698, ventelonInitioInvestigationPeierls2013, dragoniAchievingDFTAccuracy2018}. This discrepancy has important implications in the kink-pair nucleation, creating a discontinuity in the nucleation enthalpy - shear stress dependency \cite{gordonScrewDislocationMobility2010, gordonScrewDislocationMobility2011}. In Fig. \ref{supp_fig:pure_Fe_peierls} we compare the Peierls barriers for the screw dislocation motion given by the presented Fe-H tabGAP model, demonstrating its single-humped shape, and the other tested Fe-H IAPs. The tabGAP model gives a Peierls barrier height of $\SI{51.8}{meV/b}$ which falls in between the two DFT results of $\SI{40}{meV/b}$ and $\SI{58}{meV/b}$ from Refs. \cite{ventelonInitioInvestigationPeierls2013, dragoniAchievingDFTAccuracy2018}. The barrier is known to vary up to $\SI{20}{meV/b}$ depending on the details of the calculation method \cite{ventelonInitioInvestigationPeierls2013}. Both the Meng\_NNIP and the Kumar\_EAM also give a single-humped barrier with heights of $\SI{38.2}{meV/b}$ and $\SI{29.2}{meV/b}$. In comparison, the double-humped shape with a metastable intermediate split core configuration given by the Ram\_EAM and Wen\_EAM is shown as well. The barrier given by these two IAPs is $\SI{11.2}{meV/b}$.

\section{Interpolation of the kink-pair nucleation path}

\begin{figure}
    \centering
    \setlength{\fboxsep}{0pt}
    \setlength{\fboxrule}{0.4pt}
    \setkeys{Gin}{keepaspectratio}
    \begin{subfigure}{.32\textwidth}
        \fbox{\includegraphics[width=\dimexpr\linewidth-2\fboxrule\relax]{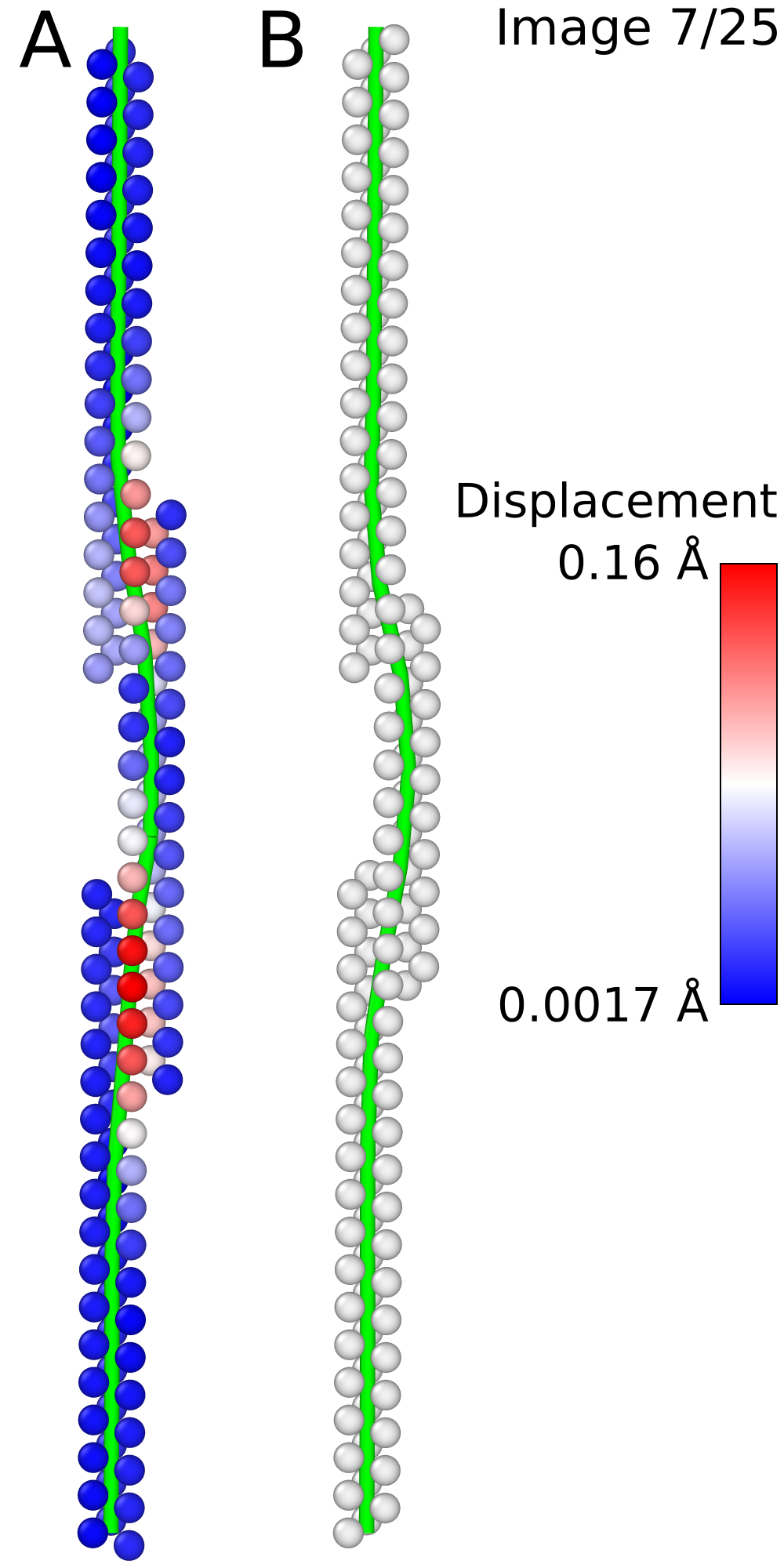}}
    \caption{}
    \label{supp_fig:kp_inter1}
    \end{subfigure}
    \begin{subfigure}{.32\textwidth}
        \fbox{\includegraphics[width=\dimexpr\linewidth-2\fboxrule\relax]{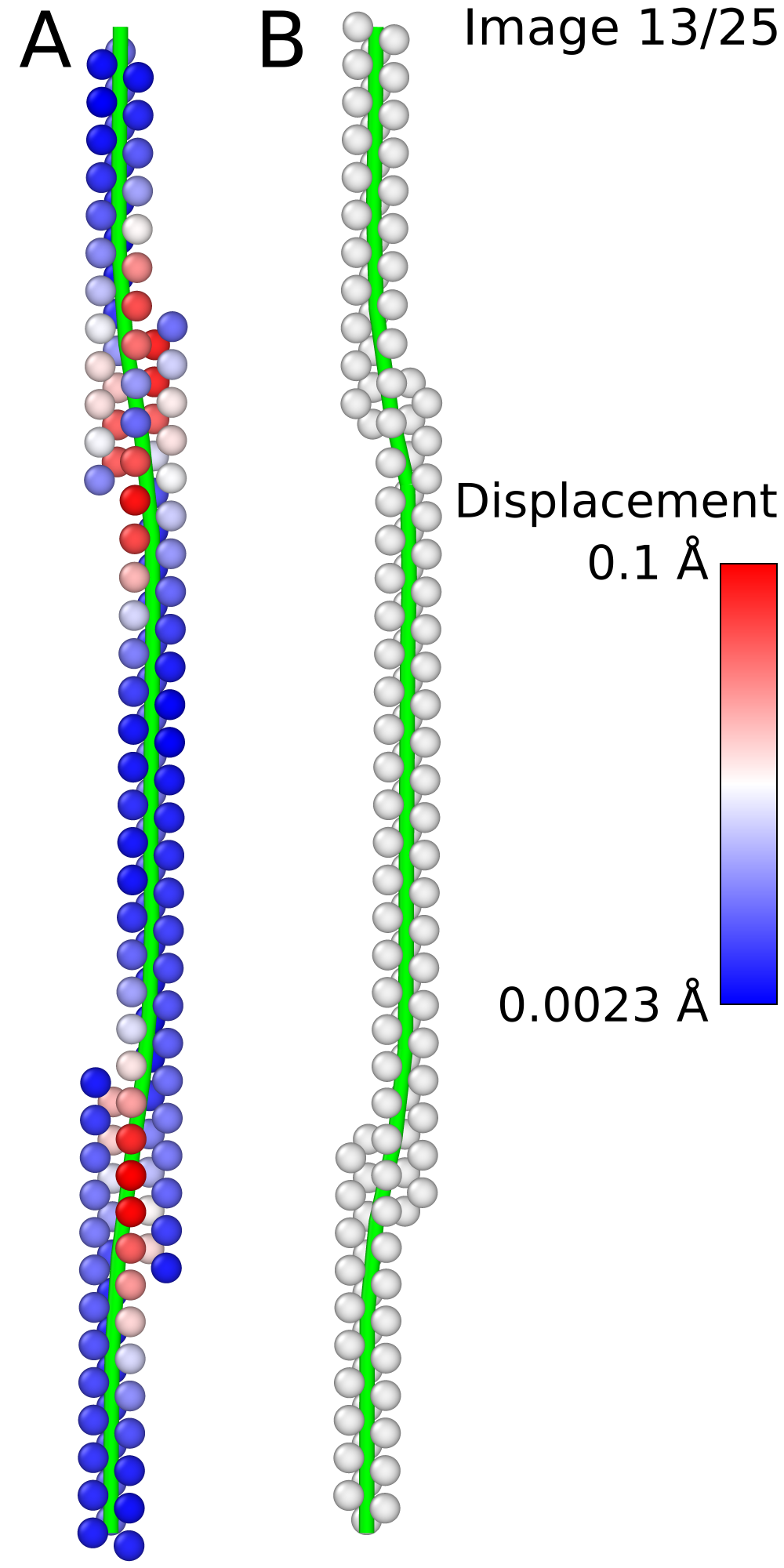}}
    \caption{}
    \label{supp_fig:kp_inter2}
    \end{subfigure}
    \begin{subfigure}{.32\textwidth}
        \fbox{\includegraphics[width=\dimexpr\linewidth-2\fboxrule\relax]{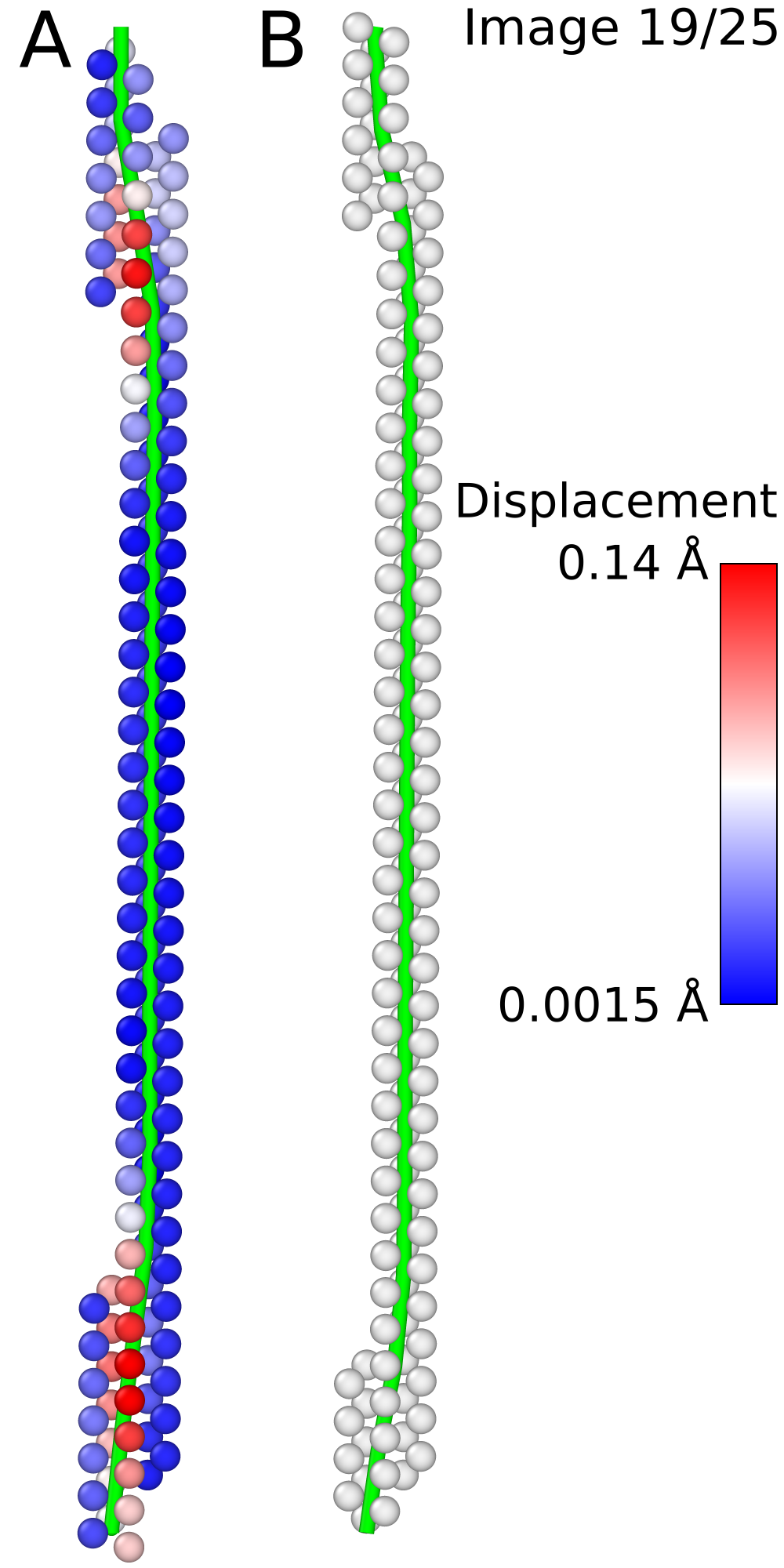}}
    \caption{}
    \label{supp_fig:kp_inter3}
    \end{subfigure}
\caption{The difference between the interpolated kink-pair structure (dislocations labeled A) and the result kink-pair from the converged CI-NEB calculation (dislocations labeled B). Three sets of images along the kink-pair nucleation pathway are shown in (\subref{supp_fig:kp_inter1}-\subref{supp_fig:kp_inter3}). the Fe dislocation core atoms in the A-structures are colored according to their displacement magnitudes from the positions of the B-structure Fe atoms.}
\label{supp_fig:kp_interpolation}
\end{figure}

In our CI-NEB simulations of the screw dislocation kink-pair nucleation mechanism we constructed the initial reaction path with a spatially progressing interpolation scheme, where the interpolation between the initial and final dislocation structures ($\SI{40}{b}$ long screw dislocations on neighboring Peierls valleys) is initiated from the center of the dislocation line, and propagated outwards along the dislocation line. A simple binary on-off interpolation like what we just outlined would induce large strains to the system. To alleviate this we applied a $\tanh$-function based weight to the interpolated coordinates. This adjusts the size of the interpolation step by the distance between the atoms and the propagating kink-pair, giving a smooth change in the atomic positions. The resulting interpolated kink-pair structure is visualized in Fig. \ref{supp_fig:kp_interpolation} alongside the optimized structure given by the CI-NEB calculation with the Fe-H tabGAP model. We show the displacement magnitudes of the dislocation core atoms between the interpolated and optimized kink-pair structures to be within $\SI{0.16}{Å}$ which indicates that the aforementioned interpolation scheme constructs a kink-pair structure that is reasonably close to what the tabGAP model predicts. 

\section{H binding to an edge dislocation}\label{supp_sec:edge_dislocation_binding}

\begin{figure}
    \centering
    \includegraphics[width=0.48\textwidth]{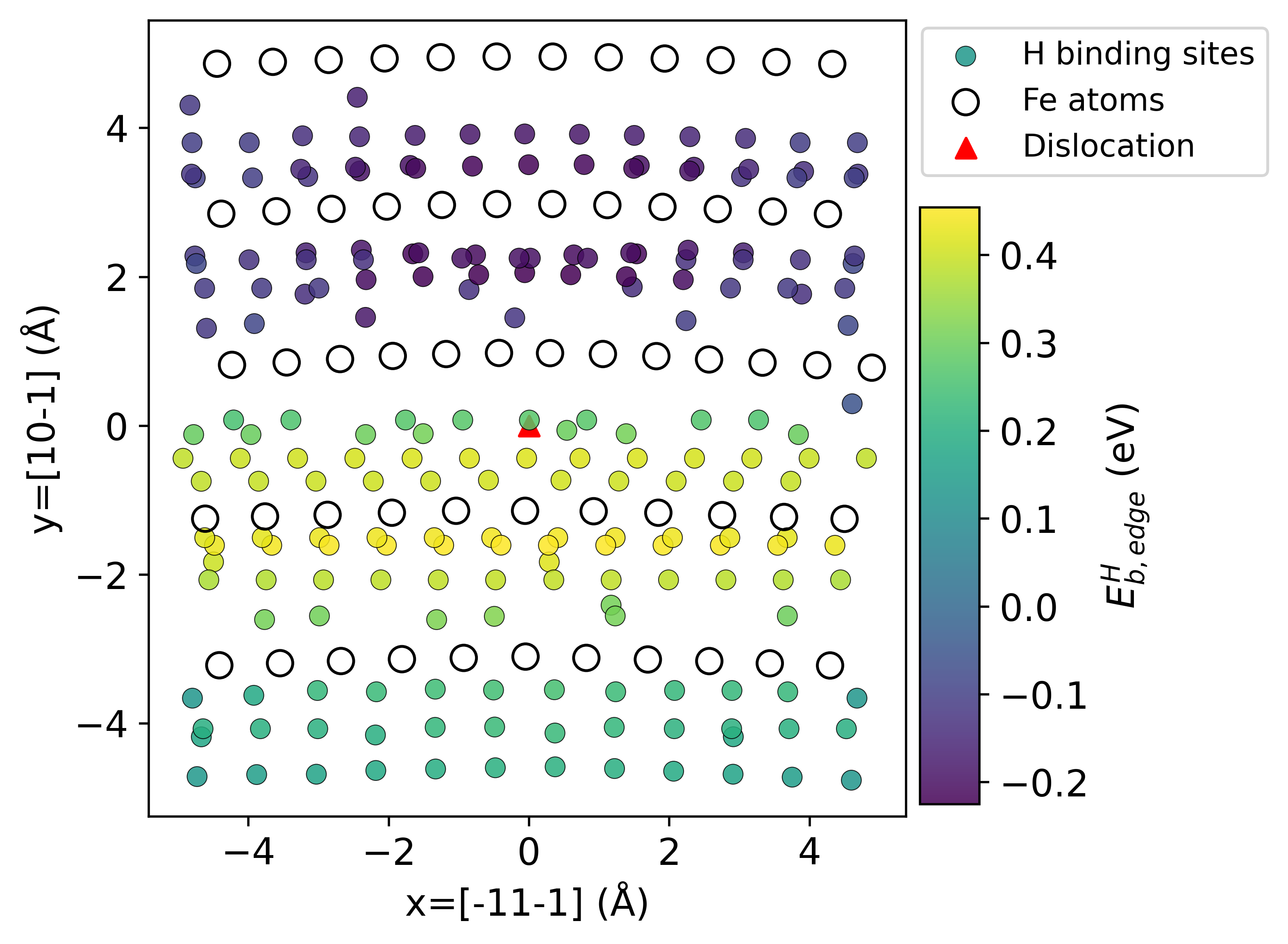}
    \caption{The binding sites and energies for a H atom near a $1/2\hkl<111>\hkl{110}$ edge dislocation calculated with the tabGAP.}
    \label{supp_fig:H_Eb_edge}
\end{figure}

In Fig. \ref{supp_fig:H_Eb_edge} we show the binding sites near a $1/2\hkl<111>\hkl{110}$ edge dislocation. Here, the compressive stress and the tensile stress regions above and below the dislocation, respectively, can be seen directly from the H binding energies in the upper (non-binding) and lower (binding) parts of the figure. The red triangle in the middle indicates the location of the dislocation core, which is perpendicular to the plane of the figure. The glide plane of the dislocation separates the regions with negative and positive binding energy H sites. An H atom strongly binds within the tensile stress region with a maximum binding energy of $\SI{0.45}{eV}$ just below the glide plane layer of Fe atoms. This agrees well with Zhao and Lu's \cite{zhaoQMMMStudy2011a} QM/MM method calculation, which reported a maximum binding energy of $\SI{0.47}{eV}$ for the edge dislocation.

Strong binding sites are found throughout the tensile part of the search area. This is explained by the movement of the edge dislocation toward the inserted H atom during relaxation. The dislocation moved toward H atoms inserted in the tensile stress region and away from H atoms inserted in the compressive stress region. The $1/2\hkl<111>\hkl{110}$ edge dislocation is known to be highly mobile, with a Peierls barrier of $\sim\SI{0.05}{meV/b}$ \cite{zhangEfficiencyAccuracyTransferability2024}, which explains the observed movement of the dislocation.

\FloatBarrier

\bibliography{bibliography}